\documentclass[%
preprint
 amsmath, amssymb, %
 aip,cha,%
]{revtex4-1}

\usepackage{CJKutf8}

\usepackage{amsmath}
\usepackage{amssymb}
\usepackage{docs}%
\usepackage{bm}%
\usepackage{xcolor}
\usepackage{multirow}
\usepackage{dcolumn}% Align table columns on decimal point
\usepackage{graphicx}% Include figure files
\usepackage{epstopdf}
\usepackage[export]{adjustbox}
\usepackage[colorlinks=true,linkcolor=blue]{hyperref}%
\usepackage{algorithm}
\usepackage{algpseudocode}
\usepackage{subcaption}
\usepackage{color}
\usepackage{scalerel,stackengine}

\stackMath % \widecheck symbol required
\newcommand\widecheck[1]{%
	\savestack{\tmpbox}{\stretchto{%
			\scaleto{%
				\scalerel*[\widthof{\ensuremath{#1}}]{\kern-.6pt\bigwedge\kern-.6pt}%
				{\rule[-\textheight/2]{1ex}{\textheight}}%WIDTH-LIMITED BIG WEDGE
			}{\textheight}% 
		}{0.5ex}}%
	\stackon[1pt]{#1}{\scalebox{-1}{\tmpbox}}%
}

\expandafter\ifx\csname package@font\endcsname\relax\else
 \expandafter\expandafter
 \expandafter\usepackage
 \expandafter\expandafter
 \expandafter{\csname package@font\endcsname}%
\fi
\begin{document}
		
\begin{CJK*}{UTF8}{gbsn}
\title{Self-scaling tensor basis neural network for Reynolds stress modeling of wall-bounded turbulence}

\author{Zelong Yuan}%
\author{Yuzhu Pearl Li}%
\email[Email address for correspondence:\;]{pearl.li@nus.edu.sg}
\affiliation{\small Department of Civil and Environmental Engineering, National University of Singapore, Singapore 117576, Republic of Singapore }

%\author{}
%\affiliation{}
%\author{}
%\affiliation{}
%\author{}
%\affiliation{}

\date{\today}% It is always \today, today,
             %  but any date may be explicitly specified

\begin{abstract}
Recent advances in data-driven turbulence modeling have established tensor basis neural networks (TBNN) as a physically grounded framework for Reynolds-stress closure in Reynolds-averaged Navier–Stokes (RANS) simulations. However, their robustness in wall-bounded turbulent flows remains limited across Reynolds numbers and geometries due to the lack of an intrinsic scaling mechanism.
In this work, we propose a self-scaling tensor basis neural network (STBNN) for Reynolds-stress modeling of wall-bounded turbulence. The model incorporates an invariant velocity-gradient normalization derived from the first two invariants of the velocity-gradient tensor, providing an intrinsic and geometry-independent scale that balances strain and rotation effects without relying on empirical coefficients or wall-distance inputs. Owing to its frame-indifferent formulation, the approach preserves Galilean and rotational invariance while maintaining a physically interpretable representation of Reynolds-stress anisotropy.
STBNN is evaluated through \emph{a priori} and \emph{a posteriori} studies using direct numerical simulation (DNS) data of canonical wall-bounded flows, including plane channel and periodic hill flows. In \emph{a priori} tests, the model accurately reproduces Reynolds-stress anisotropy, with correlation coefficients exceeding 99\% and relative errors below 10\%, while capturing near-wall scaling and logarithmic-layer behavior. In \emph{a posteriori} RANS simulations, STBNN predicts mean velocity profiles in close agreement with DNS and improves prediction of separation and reattachment compared with linear and quadratic eddy-viscosity models and the baseline TBNN. Notably, a model trained at low Reynolds numbers generalizes to higher Reynolds numbers and unseen geometries for canonical flows. These results demonstrate the effectiveness of the proposed framework for data-driven Reynolds-stress modeling in wall-bounded turbulent flows.
\end{abstract}

%\pacs{47.20.Qr,  47.65.-d}% PACS, the Physics and Astronomy
                             % Classification Scheme.
%\keywords{Suggested keywords}%Use showkeys class option if keyword
                              %display desired
\maketitle
\end{CJK*}

\section{Introduction}\label{sec:level1}

High-fidelity prediction of turbulent flows remains a fundamental challenge in computational fluid dynamics (CFD), owing to the prohibitive computational expense of direct numerical simulation (DNS) and large-eddy simulation (LES) for practical engineering applications.\cite{speziale1990analytical, durbin2018, pope2000} Consequently, Reynolds-averaged Navier–Stokes (RANS) approaches remain the workhorse of industrial CFD simulations and are expected to retain this role for the foreseeable future.\cite{slotnick2014cfd,duraisamy2019turbulence,duraisamy2021perspectives} The fundamental challenge of RANS modelling lies in accurately closing the Reynolds-stress tensor arising from the averaging of the Navier–Stokes equations. Over the past decades, a wide range of RANS turbulence closures have been developed, ranging from linear eddy-viscosity models (LEVM) such as Spalart–Allmaras\cite{spalart1992one,spalart2000strategies},  $k-\varepsilon$ \cite{jones1972prediction,launder1974application,yakhot1986renormalization,yakhot1992development} and $k-\omega$ families \cite{wilcox1988reassessment,menter1994two,wilcox2008formulation,menter2025generalized} to more elaborate Reynolds-stress transport models (RSM).\cite{wilcox2006turbulence,li2022reynolds,li2022turbulence} Eddy-viscosity models offer computational efficiency and numerical robustness but their accuracy often deteriorates in flows with strong anisotropy and separation, whereas RSMs improve physical fidelity at an increased computational cost and reduced robustness.

In recent years, machine-learning-based methods have gained increasing attention as promising tools for improving turbulence closures.\cite{tracey2015machine,singh2016using,pan2018data,zhu2019machine,gamahara2017searching,yang2019predictive,xie2020artificial,Xie2021Artificial,yuan2020,yuan2021,yuan2021a,yuan2022,novati2021automating,bae2022scientific,kutz2017deep,Milani2021ScalarTBNN,zhang2022b,Cai2024TBNN}  Early studies demonstrated that data-driven correction of Reynolds stresses can improve mean-flow predictions over those obtained from the baseline two-equation $k-\varepsilon$ model.\cite{wang2017physics,wu2018physics,xiao2019} A major advance was the tensor-basis neural network (TBNN) proposed by Ling \emph{et al.}\cite{ling2016reynolds,ling2016machine}, which embeds Galilean and rotational invariance into the constitutive representation of the anisotropy tensor. Subsequent developments incorporated physics-informed learning strategies\cite{wu2018physics,jiang2021interpretable}, symbolic-regression frameworks \cite{weatheritt2016novel,zhao2020rans,Ji2025SymbolicTBNN}, realizability constraints\cite{McConkey2025Realisability}, historical effects \cite{Myklebust2025UnsteadyTBNN, kawabata2026machine} and improved coupling algorithms \cite{Boureima2022DynamicCalibration,Liu2023ExtrapolationMLRANS} to enhance robustness when integrating learned stresses into RANS solvers. Previous studies on Reynolds-stress closure modelling have also shown that the adequacy of the tensor basis is crucial to predictive accuracy and robustness, particularly when representing strongly anisotropic turbulence.\cite{panda2018representation,panda2020reliable,panda2022evaluation}

Despite these significant advances, a fundamental limitation persists: even invariant-preserving models show limited generalization to Reynolds numbers and geometries outside the training conditions.\cite{xiao2019,duraisamy2019turbulence} A possible explanation is that the local mean-velocity gradient does not uniquely determine the turbulence state in wall-bounded flows. Moreover, frameworks such as the TBNN commonly employs the turbulent time scale $k/\varepsilon$ for normalization, where $k$ is the turbulent kinetic energy and $\varepsilon$ is the dissipation rate. The difficulty of accurately predicting $\varepsilon$ near walls introduces regime-dependent uncertainty and consequently degrades generalization.

The present work addresses this issue by introducing a self-scaling tensor-basis neural network (STBNN) closure. Instead of modifying network architecture or increasing training data, the approach reformulates the input representation using an invariant-based normalization derived from the first two invariants of the velocity-gradient tensor. This normalization provides an intrinsic local time scale that balances strain and rotation effects and may help dynamically similar turbulence states be represented more consistently over the Reynolds-number range and geometrical variations examined in this study. The developed model preserves frame invariance while improving the generalization capability within the tested canonical RANS cases.

The proposed STBNN is investigated using both \emph{a priori} and \emph{a posteriori} assessments to examine its predictive capability and practical applicability in RANS computations. Canonical channel flows spanning a range of Reynolds numbers and the periodic-hill configuration with varying geometrical complexity are considered, enabling assessment across Reynolds-number variation within a canonical flow and applicability to flows with distinct underlying physics. Comparisons with conventional linear and quadratic eddy-viscosity models (QEVM) and the baseline TBNN are used to elucidate the impact of the physically consistent self-scaling formulation, particularly in terms of robustness and transferability across flow conditions.

The remainder of this paper is structured as follows. The governing equations and turbulence models, including the LEVM, QEVM, TBNN and the proposed STBNN, are introduced in Sec.~\ref{sec:level2}. The flow configurations and associated DNS databases are described in Sec.~\ref{sec:level3}. Secs.~\ref{sec:level4} and \ref{sec:level5} present the \emph{a priori} and \emph{a posteriori} analyses, respectively. Concluding remarks are given in Sec.~\ref{sec:level6}.

\section{Governing equations and turbulence models}\label{sec:level2}
The incompressible Navier–Stokes equations for a Newtonian fluid read \citep{pope2000}, 
\begin{equation}
	\frac{{\partial {u_i}}}{{\partial {x_i}}} = 0,
	\label{ns1}
\end{equation}
\begin{equation}
	\frac{{\partial {u_i}}}{{\partial t}} + \frac{{\partial \left( {{u_i}{u_j}} \right)}}{{\partial {x_j}}} =  - \frac{{\partial p}}{{\partial {x_i}}} + \nu\frac{{{\partial ^2}{u_i}}}{{\partial {x_j}\partial {x_j}}},
	\label{ns2}
\end{equation}
where $u_i$, $x_i$, $t$, and $p$ denote the velocity components, spatial coordinates, time, and the pressure divided by the constant density, respectively, and $\nu$ represents the kinematic viscosity. The summation convention for repeated indices is adopted throughout this paper for simplicity.

By decomposing the velocity and pressure fields into mean and fluctuating components via Reynolds averaging ($u_i = \bar{u}_i + u_i'$, $p = \bar{p} + p'$) and applying ensemble averaging to the governing equations, the Reynolds-averaged Navier–Stokes (RANS) equations are derived\cite{Reynolds1895}:
\begin{equation}
	\frac{{\partial {{\bar u}_i}}}{{\partial {x_i}}} = 0,
	\label{RANS1}
\end{equation}
\begin{equation}
	\frac{{\partial \left( {{{\bar u}_i}{{\bar u}_j}} \right)}}{{\partial {x_j}}} =  - \frac{{\partial \bar p}}{{\partial {x_i}}} + \nu\frac{{{\partial ^2}{{\bar u}_i}}}{{\partial {x_j}\partial {x_j}}} - \frac{{\partial {R_{ij}}}}{{\partial {x_j}}},
	\label{RANS2}
\end{equation}
where $R_{ij} = \overline{u_i^\prime u_j^\prime}$ denotes the unclosed Reynolds stress tensor.  The primary challenge in the RANS framework lies in modelling the unknown Reynolds stresses to close the momentum equations. In RANS modeling, only the anisotropic part of the Reynolds stress requires explicit closure, since the isotropic contribution can be absorbed into the mean pressure. Accordingly, the deviatoric part of the Reynolds stress tensor is defined by\cite{pope2000}
 \begin{equation}
	R_{ij}^d=R_{ij}-\frac{1}{3}R_{kk}\delta_{ij}=R_{ij}-\frac{2}{3}k\delta_{ij},
	\label{Rij_decomp}
\end{equation}
where $k = \overline{u_i^\prime u_i^\prime}/2$ is the turbulent kinetic energy per unit mass and $\delta_{ij}$ denotes the Kronecker delta.
\subsection{Linear eddy-viscosity model (LEVM)}\label{sec:level2_LEVM}
A widely used approach is the Boussinesq eddy-viscosity hypothesis,\cite{pope2000} which assumes an analogy between the Reynolds stresses and molecular viscous stresses, postulating a linear relationship with the mean velocity gradients. This leads to the linear eddy-viscosity model (LEVM),\cite{pope2000}
\begin{equation}
	 R_{ij}^d =  - 2{\nu _T}{S_{ij}},
	\label{Rij_LEVM}
\end{equation}
where ${S_{ij}} = \left( {\partial {{\bar u}_i}/\partial {x_j} + \partial {{\bar u}_j}/\partial {x_i}} \right)/2$ is the mean strain-rate tensor and $\nu_T$ represents the turbulent eddy viscosity to be modeled separately. Following previous studies on data-driven Reynolds-stress augmentation\cite{wang2017physics,wu2018physics}, the Launder–Sharma $k-\varepsilon $ model is employed as the baseline for comparison and evaluation of the predictive accuracy of the proposed approach. In the baseline $k-\varepsilon $ model, the eddy viscosity is prescribed as\cite{launder1974application}
\begin{equation}
	{\nu _T} = {C_\mu }{f_\mu }\frac{{{k^2}}}{\varepsilon },
	\label{nut_ke}
\end{equation}
where $C_\mu = 0.09$ and $f_\mu = \exp \left[ { - 3.4/{{\left( {1 + {{{\mathop{ Re}\nolimits} }_t}}/50 \right)}^2}} \right]$ is a near-wall damping function. Here, $\varepsilon $ denotes the dissipation rate of turbulent kinetic energy and ${{\mathop{ Re}\nolimits} _t} = {k^2}/\left( {\nu \varepsilon } \right)$ is turbulent Reynolds number. 

The linear eddy-viscosity model has demonstrated satisfactory performance in a wide range of canonical wall-bounded flows, such as flat-plate boundary layers and fully developed channel flows, where turbulence production and dissipation are approximately balanced. However, the LEVM model inherently assumes that the principal axes of the Reynolds stress anisotropy tensor remain aligned with those of the mean strain-rate tensor. Although this co-alignment is appropriate for simple shear flows under near-equilibrium conditions, it frequently fails in more complex turbulent flows involving flow separation, streamline curvature, or secondary strains. In these scenarios, non-equilibrium dynamics and rotational effects induce misalignment between the principal directions of stress and strain, leading to substantial predictive inaccuracies. To overcome these shortcomings, more general representations of the Reynolds stress anisotropy have been proposed. In these approaches, the anisotropy tensor $b_{ij}=R_{ij}^d/2k$ is expressed as a nonlinear function of the mean strain-rate tensor $S_{ij}$ and the mean rotation-rate tensor ${\Omega_{ij}} = \left( {\partial {{\bar u}_i}/\partial {x_j} - \partial {{\bar u}_j}/\partial {x_i}} \right)/2$.

\begin{figure}\centering
	\includegraphics[width=0.7\textwidth]{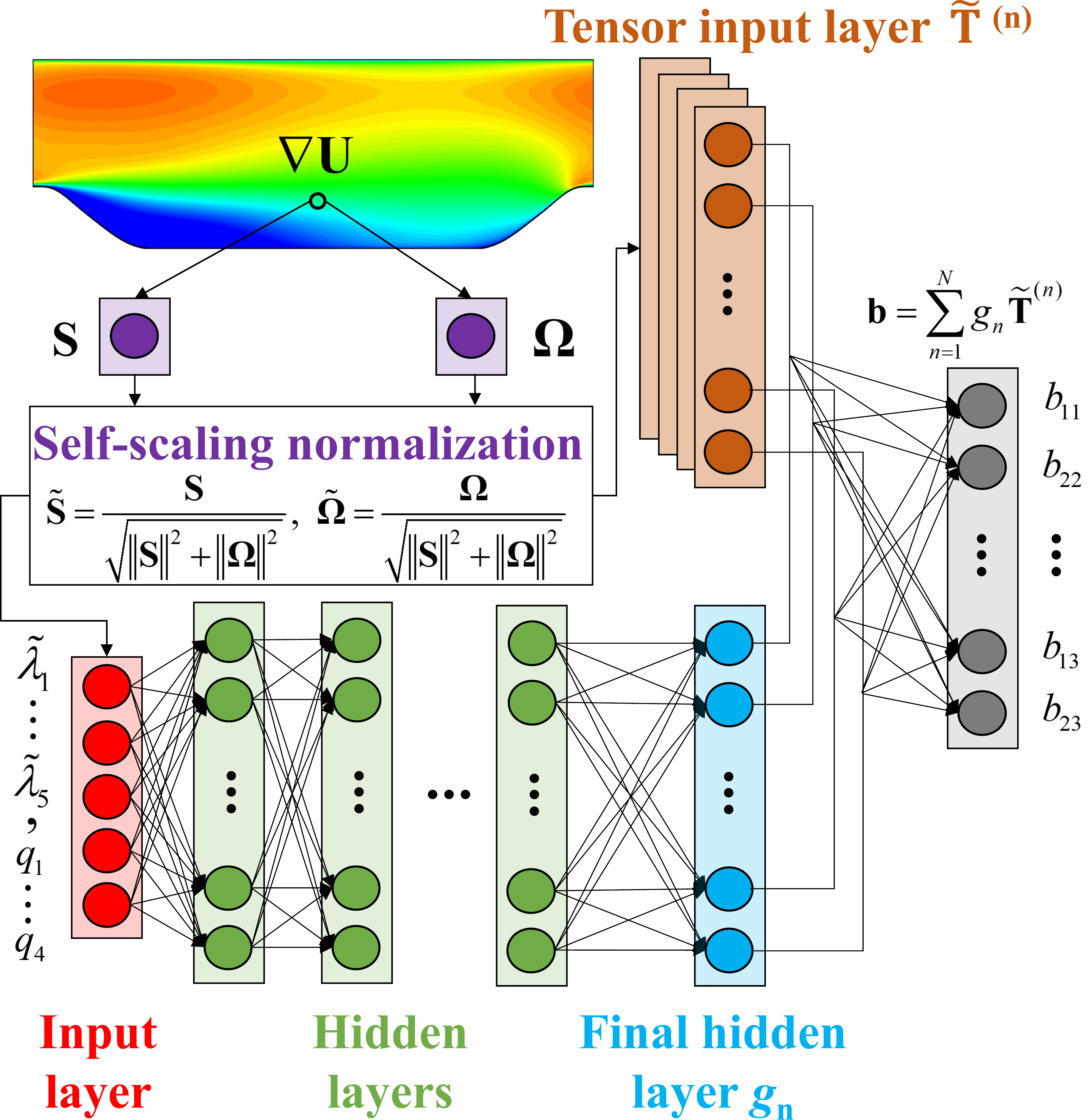}
	\caption{Schematic diagram of the self-scaling tensor basis neural network (STBNN).}\label{fig:1}
\end{figure}

\subsection{General eddy-viscosity framework and quadratic eddy-viscosity model (QEVM)}\label{sec:level2_QEVM}
One of the most general frameworks for nonlinear eddy-viscosity modeling was proposed by Pope\cite{pope1975more} to enhance the universality of RANS closures. In this framework, the anisotropy tensor of Reynolds stress is assumed to be a function of the nondimensional mean strain-rate and the rotation-rate tensors. Building on the theory of integrity bases for isotropic tensor functions and the Cayley–Hamilton theorem, Pope\cite{pope1975more} demonstrated that the Reynolds stress anisotropy tensor (${b_{ij}} = R_{ij}^d/2k$) can generally be expressed as a linear combination of ten finite tensor polynomials,
\begin{equation}
	{\bf{b}}\left( {{\bf{\hat S}},{\bf{\hat \Omega }}} \right) = \sum\limits_{n = 1}^{10} {{g_n}\left( {{{\hat \lambda }_1},{{\hat \lambda }_2}, \cdots ,{{\hat \lambda }_5}} \right){{{\bf{\hat T}}}^{\left( n \right)}}},
	\label{bij_Pope}
\end{equation}
with five invariants
\begin{equation}
	{{\hat \lambda }_1} = {\rm{tr}}\left( {{{{\bf{\hat S}}}^2}} \right),\;\;{{\hat \lambda }_2} = {\rm{tr}}\left( {{{{\bf{\hat \Omega }}}^2}} \right),\;\;{{\hat \lambda }_3} = {\rm{tr}}\left( {{{{\bf{\hat S}}}^3}} \right),\;\;{{\hat \lambda }_4} = {\rm{tr}}\left( {{{{\bf{\hat \Omega }}}^2}{\bf{\hat S}}} \right),\;\;{{\hat \lambda }_5} = {\rm{tr}}\left( {{{{\bf{\hat \Omega }}}^2}{{{\bf{\hat S}}}^2}} \right),
	\label{lambda_GEVM}
\end{equation} 
and ten independent isotropic basis tensors
\begin{equation}
	\left\{ {\begin{array}{*{20}{l}}
			{{\hat{\bf{T}}^{(1)}} = \widehat {\bf{S}},}&{{\hat{\bf{T}}^{(6)}} = {{\widehat {\bf{\Omega }}}^2}\widehat {\bf{S}} + \widehat {\bf{S}}{{\widehat {\bf{\Omega }}}^2} - \frac{2}{3}{\mkern 1mu} {\rm{tr}}\left( {\widehat {\bf{S}}{{\widehat {\bf{\Omega }}}^2}} \right){{\bf{I}}},}\\
			{{\hat{\bf{T}}^{(2)}} = \widehat {\bf{S}}\widehat {\bf{\Omega }} - \widehat {\bf{\Omega }}\widehat {\bf{S}},}&{{\hat{\bf{T}}^{(7)}} = \widehat {\bf{\Omega }}\widehat {\bf{S}}{{\widehat {\bf{\Omega }}}^2} - {{\widehat {\bf{\Omega }}}^2}\widehat {\bf{S}}\widehat {\bf{\Omega }},}\\
			{{\hat{\bf{T}}^{(3)}} = {{\widehat {\bf{S}}}^2} - \frac{1}{3}{\mkern 1mu} {\rm{tr}}\left( {{{\widehat {\bf{S}}}^2}} \right){{\bf{I}}},}&{{\hat{\bf{T}}^{(8)}} = \widehat {\bf{S}}\widehat {\bf{\Omega }}{{\widehat {\bf{S}}}^2} - {{\widehat {\bf{S}}}^2}\widehat {\bf{\Omega }}\widehat {\bf{S}},}\\
			{{\hat{\bf{T}}^{(4)}} = {{\widehat {\bf{\Omega }}}^2} - \frac{1}{3}{\mkern 1mu} {\rm{tr}}\left( {{{\widehat {\bf{\Omega }}}^2}} \right){{\bf{I}}},}&{{\hat{\bf{T}}^{(9)}} = {{\widehat {\bf{\Omega }}}^2}{{\widehat {\bf{S}}}^2} + {{\widehat {\bf{S}}}^2}{{\widehat {\bf{\Omega }}}^2} - \frac{2}{3}{\mkern 1mu} {\rm{tr}}\left( {{{\widehat {\bf{S}}}^2}{{\widehat {\bf{\Omega }}}^2}} \right){{\bf{I}}},}\\
			{{\hat{\bf{T}}^{(5)}} = \widehat {\bf{\Omega }}{{\widehat {\bf{S}}}^2} - {{\widehat {\bf{S}}}^2}\widehat {\bf{\Omega }},}&{{\hat{\bf{T}}^{(10)}} = \widehat {\bf{\Omega }}{{\widehat {\bf{S}}}^2}{{\widehat {\bf{\Omega }}}^2} - {{\widehat {\bf{\Omega }}}^2}{{\widehat {\bf{S}}}^2}\widehat {\bf{\Omega }}.}
	\end{array}} \right.
	\label{Tij_GEVM}
\end{equation}
Here,  $\hat{\boldsymbol{S}} = \hat{\tau}\boldsymbol{S}$ and $\hat{\boldsymbol{\Omega}} = \hat{\tau}\boldsymbol{\Omega}$ are the mean strain-rate and rotation-rate tensors nondimensionalized using the turbulent time scale $\hat{\tau} = k/\varepsilon$, respectively. $\bf{I}$ represents the identity tensor. The coefficients $g_n$ are scalar functions of the associated invariants $\hat \lambda_i$. The symbol ``$\mathrm{tr}(\cdot)$" denotes the trace of a second-order tensor, and tensor products represent standard matrix multiplication with summation implied over repeated indices; namely, ${\bf{AB}} = {A_{ik}}{B_{kj}}$ for arbitrary second-order tensors ${\bf{A}}$ and ${\bf{B}}$. 

By retaining only the first four tensorial terms in Pope’s integrity basis, Shih \emph{et al.}\cite{Shih1993Realizable} proposed a quadratic eddy-viscosity model (QEVM) that preserves the dominant nonlinear effects. The deviatoric Reynolds stress tensor of QEVM model is expressed as\cite{Shih1993Realizable}
\begin{equation}
R_{ij}^d =  - 2{C_\mu }k\hat T_{ij}^{\left( 1 \right)} + \frac{{{k^3}}}{{{\varepsilon ^2}\left[ {1000 + {{\left( {2\left| {{{\hat \lambda }_1}} \right|} \right)}^{3/2}}} \right]}}\left[ {{C_1}\hat T_{ij}^{\left( 2 \right)} + {C_2}\hat T_{ij}^{\left( 3 \right)} + {C_3}\hat T_{ij}^{\left( 4 \right)}} \right],
	\label{Rij_d_QEVM}
\end{equation}
where model coefficients are ${C_\mu } = \frac{{2/3}}{{1.25 + \sqrt {2\left| {{{\hat \lambda }_1}} \right|}  + 0.9\sqrt {2\left| {{{\hat \lambda }_2}} \right|} }}$, $C_1 = 15$, $C_2 = 3$, and $C_3 = -19$.

\subsection{Tensor basis neural network (TBNN)}\label{sec:level2_TBNN}
The tensor basis neural network (TBNN), proposed by Ling \emph{et al.}\cite{ling2016reynolds} as an extension of Pope’s generalized eddy-viscosity Model (Eqs.~\ref{bij_Pope}-\ref{Tij_GEVM})\cite{pope1975more}, represents the Reynolds stress anisotropy tensor as a linear combination of isotropic tensor basis functions constructed from the mean velocity gradient tensor, with the corresponding scalar coefficients learned by a neural network as functions of a set of scalar invariants.

Within the TBNN framework, the basis coefficients associated with the tensor bases are parameterized as nonlinear functions of the invariants through a neural network. The Reynolds stress anisotropy tensor is modeled by\cite{ling2016reynolds,Cai2024TBNN}
\begin{equation}
	{\bf{b}} = \sum\limits_{n = 1}^N {{g_n}} ({\boldsymbol{\hat{\lambda}}},{\bf{q}};{\bf{\theta }}){\widehat {\bf{T}}^{(n)}},
	\label{bij_TBNN}
\end{equation}
where \(\boldsymbol{\hat{\lambda}} = \{\hat \lambda_1,\hat \lambda_2,\ldots,\hat\lambda_5\}\) denotes the invariant set defined in Eq.~\ref{lambda_GEVM}, \(\mathbf q = \{q_1,q_2,\ldots,q_m\}\) represents a set of auxiliary scalar features, and \(\boldsymbol\theta\) are the trainable parameters of the neural network. Following the work of Wu \emph{et al.}\cite{wu2018physics}, auxiliary scalar features are introduced to encode additional physical information beyond the invariant set. 
\begin{equation}
	{q_1} = \ln \left( {1 + \frac{{\sqrt k {d_w}}}{\nu }} \right),\;\;{q_2} = \ln \left( {1 + \frac{{{k^2}}}{{\nu \varepsilon }}} \right),\;\;{q_3} = \frac{{{d_w}}}{{{L_{{\rm{ref}}}}}},\;\;{q_4} = \frac{k}{\varepsilon }\left\| {\bf{S}} \right\|,
	\label{aux_qi}
\end{equation}
with $d_w$ the distance to the nearest wall, $L_{\mathrm{ref}}$ a reference length scale, and $\left\| {\bf{S}} \right\| = \sqrt {{S_{ij}}{S_{ij}}}$ the magnitude of the strain-rate tensor. Here, $q_1$ represents a wall-distance-based Reynolds number, $q_2$ corresponds to the turbulent Reynolds number, $q_3$ characterizes a nondimensional  wall-normal geometric scale, and $q_4$ denotes the ratio of the turbulent timescale to the mean strain timescale. By embedding Galilean and rotational invariance directly into the neural-network architecture, TBNN ensures physically consistent predictions and provides improved robustness compared with unconstrained data-driven models.

The Reynolds stress anisotropy tensor $b_{ij}$ is a symmetric second-order tensor and thus possesses six independent components in three-dimensional space. In addition, $b_{ij}$ is subject to the trace-free constraint $b_{ii}= 0$, which reduces the number of independent degrees of freedom to five. In non-degenerate flows, these five degrees of freedom can be fully spanned by five linearly independent basis tensors. Furthermore, Lund and Novikov \cite{lund1992} showed that representing basis coefficients $g_n$ as ratios of polynomials in the corresponding integrity invariants allows the number of required basis tensors to be reduced from ten to five. Accordingly, in the present study, the deviatoric Reynolds stresses reconstructed by the TBNN models are expressed as a linear combination of the first five basis tensors ($N=5$). 

\subsection{Self-scaling tensor basis neural network (STBNN)}\label{sec:level2_STBNN}
In standard TBNN formulations, the mean strain-rate and rotation-rate tensors are nondimensionalized using a turbulence time scale $ \hat{\tau}=k/\varepsilon $. However, the dissipation rate $ \varepsilon$ is not directly resolvable near walls and typically requires empirical damping or wall-function corrections, introducing geometry dependence and limiting model generalizability. To overcome this limitation, the present STBNN model adopts an invariant velocity-gradient normalization constructed from the first two invariants of the velocity-gradient tensor. This yields an intrinsic, geometry-independent scale that naturally balances strain- and rotation-rate effects without relying on empirical coefficients or wall-distance information.

In the proposed STBNN framework illustrated in Fig.~\ref{fig:1}, the strain-rate and rotation-rate tensors are nondimensionalized using an invariant velocity-gradient magnitude. Specifically, the self-scaled strain-rate and rotation-rate tensors are defined as
\begin{equation}
{\bf{\tilde S}} = \frac{{\bf{S}}}{{\sqrt {\left| {{\lambda _1}} \right| + \left| {{\lambda _2}} \right|} }} = \frac{{\bf{S}}}{{\sqrt {{{\left\| {\bf{S}} \right\|}^2} + {{\left\| {\bf{\Omega }} \right\|}^2}} }},\;\;{\bf{\tilde \Omega }}  = \frac{{\bf{\Omega}}}{{\sqrt {\left| {{\lambda _1}} \right| + \left| {{\lambda _2}} \right|} }} = \frac{{\bf{\Omega }}}{{\sqrt {{{\left\| {\bf{S}} \right\|}^2} + {{\left\| {\bf{\Omega }} \right\|}^2}} }},
	\label{selfScale_SR}
\end{equation}
where ``$\lVert \cdot \rVert$" denotes the Frobenius norm. The normalization factor represents a scalar magnitude constructed from the first two invariants ($\lambda_1 = {\rm{tr}}\left( {{{{\bf{S}}}^2}} \right)$ and $\lambda_2 = {\rm{tr}}\left( {{{{\bf{\Omega}}}^2}} \right)$) of the velocity-gradient tensor, providing an intrinsic local deformation scale. Unlike the standard TBNN formulation based on the turbulence time scale $ \hat{\tau}=k/\varepsilon $, the present scaling is purely kinematic, frame-indifferent, and independent of wall distance or empirical damping functions. To distinguish the self-scaled quantities from those used in the standard TBNN formulation, a tilde is employed to denote the self-scaling tensors (${{\tilde \tau }^{ - 1}} = {\sqrt {{{\left\| {\bf{S}} \right\|}^2} + {{\left\| {\bf{\Omega }} \right\|}^2}} }$).

The proposed self-scaling normalization is constructed from invariants of the velocity-gradient tensor, and therefore reflects the local kinematic balance between the strain-rate and rotation-rate tensors. In particular, it can be interpreted in terms of normalized quantities,
	$\tilde{\mathbf{S}} = \mathbf{S} / \sqrt{\|\mathbf{S}\|^2 + \|\mathbf{\Omega}\|^2}$ and 
	$\tilde{\mathbf{\Omega}} = \mathbf{\Omega} / \sqrt{\|\mathbf{S}\|^2 + \|\mathbf{\Omega}\|^2}$,
	which provide bounded and physically interpretable measures of the relative contributions of strain and rotation. This normalization enables a consistent treatment of both strain-dominated and rotation-dominated regions without introducing empirical coefficients.
	
From a physical perspective, this interpretation is qualitatively consistent with vortex identification concepts such as the $\lambda_2$ criterion\cite{jeong1995identification}, which characterize flow topology through the interplay between strain and rotation in the velocity-gradient tensor. It should be noted, however, that the present formulation does not explicitly invoke any vortex identification procedure, but instead employs invariant-based normalization to provide a compact and frame-invariant representation of local flow kinematics.
Compared with commonly used normalization strategies based on turbulence quantities (e.g., $k$ and $\varepsilon$) or wall-distance scaling, the present formulation depends solely on local velocity-gradient information and avoids reliance on modeled quantities.

The Reynolds stress anisotropy tensor is then expressed in the tensor basis form as
\begin{equation}
	{\bf{b}} = \sum\limits_{n = 1}^N {{g_n}} ({\boldsymbol{\tilde{\lambda}}},{\bf{q}};{\bf{\theta }}){\widetilde {\bf{T}}^{(n)}},
	\label{bij_STBNN}
\end{equation}
where $g_n$ are the basis coefficients predicted by the neural network, and \(\boldsymbol{\tilde{\lambda}} = \{\tilde \lambda_1,\tilde \lambda_2,\ldots,\tilde\lambda_5\}\) and $\widetilde{\mathbf T}^{(n)}$ respectively denote the scalar invariants and tensor bases constructed from the self-scaled tensors $\tilde{\mathbf S}$ and $\tilde{\boldsymbol\Omega}$. The five invariants are expressed as
\begin{equation}
	{{\tilde \lambda }_1} = {\rm{tr}}\left( {{{{\bf{\tilde S}}}^2}} \right),\;\;{{\tilde \lambda }_2} = {\rm{tr}}\left( {{{{\bf{\tilde \Omega }}}^2}} \right),\;\;{{\tilde \lambda }_3} = {\rm{tr}}\left( {{{{\bf{\tilde S}}}^3}} \right),\;\;{{\tilde \lambda }_4} = {\rm{tr}}\left( {{{{\bf{\tilde \Omega }}}^2}{\bf{\tilde S}}} \right),\;\;{{\tilde \lambda }_5} = {\rm{tr}}\left( {{{{\bf{\tilde \Omega }}}^2}{{{\bf{\tilde S}}}^2}} \right).
	\label{lambda_STBNN}
\end{equation} 
In the present work, the first five tensor bases are retained and defined by
\begin{equation}
	\left\{ \begin{array}{l}
		{{{\bf{\widetilde T}}}^{(1)}} = {\bf{\tilde S}},\\
		{{{\bf{\widetilde T}}}^{(2)}} = {\bf{\tilde S\tilde \Omega }} - {\bf{\tilde \Omega \tilde S}},\\
		{{{\bf{\widetilde T}}}^{(3)}} = {{{\bf{\tilde S}}}^2} - \frac{1}{3}{\rm{tr}}\left( {{{{\bf{\tilde S}}}^2}} \right){\bf{I}},\\
		{{{\bf{\widetilde T}}}^{(4)}} = {{{\bf{\tilde \Omega }}}^2} - \frac{1}{3}{\rm{tr}}\left( {{{{\bf{\tilde \Omega }}}^2}} \right){\bf{I}},\\
		{{{\bf{\widetilde T}}}^{(5)}} = {\bf{\tilde \Omega }}{{{\bf{\tilde S}}}^2} - {{{\bf{\tilde S}}}^2}{\bf{\tilde \Omega }}.
	\end{array} \right.
	\label{Tij_STBNN}
\end{equation}
The resulting tensor bases are symmetric and traceless and inherit Galilean and rotational invariance from the underlying self-scaled strain-rate and rotation-rate tensors.

Since the self-scaling normalization is constructed exclusively from the scalar quantities $\lVert \mathbf S \rVert$ and $\lVert \boldsymbol\Omega \rVert$, which correspond to the square roots of the first two invariants of the velocity gradient tensor, the introduction of the self-scaled tensors does not violate the invariance properties of the original TBNN framework. These quantities remain invariant under coordinate rotations and Galilean transformations, and therefore the proposed normalization preserves both rotational invariance and Galilean invariance. At the same time, it provides a unified and adaptive scaling of the tensor bases ($\widetilde{\mathbf T}^{(n)} \sim O(1)$) across different flow regimes.

\section {Case set-up for plane channel and periodic hill flows}\label{sec:level3}
Two canonical wall-bounded turbulent flows are considered to examine the performance of the proposed STBNN framework, namely the fully developed plane channel flows and the flows over periodic hills. These configurations are widely used benchmarks in turbulence modeling, providing complementary challenges associated with near-wall behavior, Reynolds-number dependence, or strong flow separation. 

\begin{figure}\centering
	\begin{subfigure}[c]{0.49\textwidth}
		\centering
		\parbox{0.9\linewidth}{\centering ($a$)}\vspace{-0.2em}
		\includegraphics[width=0.9\linewidth]{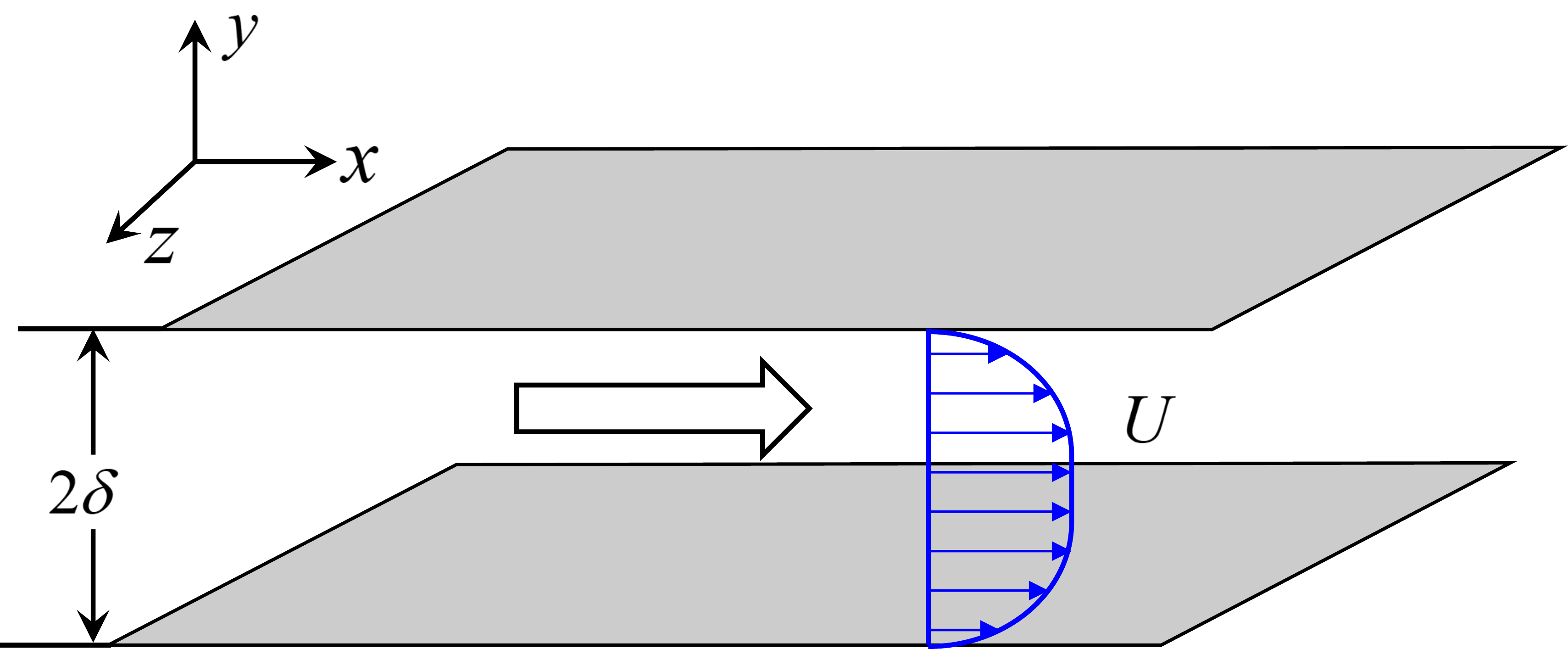}
	\end{subfigure}
	\hfill
	\begin{subfigure}[c]{0.49\textwidth}
		\centering
		\parbox{0.9\linewidth}{\centering ($b$)}\vspace{-0.2em}
		\includegraphics[width=0.9\linewidth]{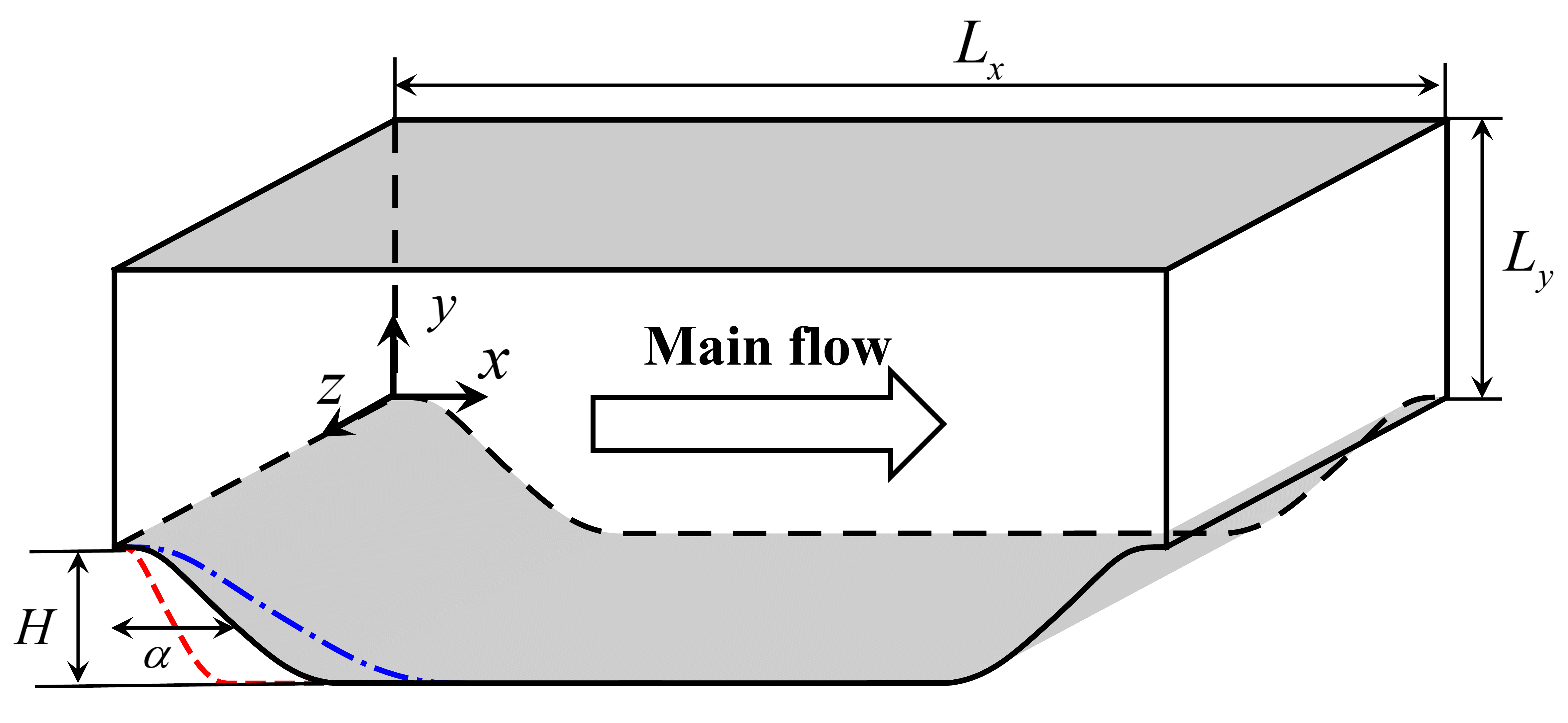}
		
		\vspace{0.5em}
		
		\includegraphics[width=0.9\linewidth]{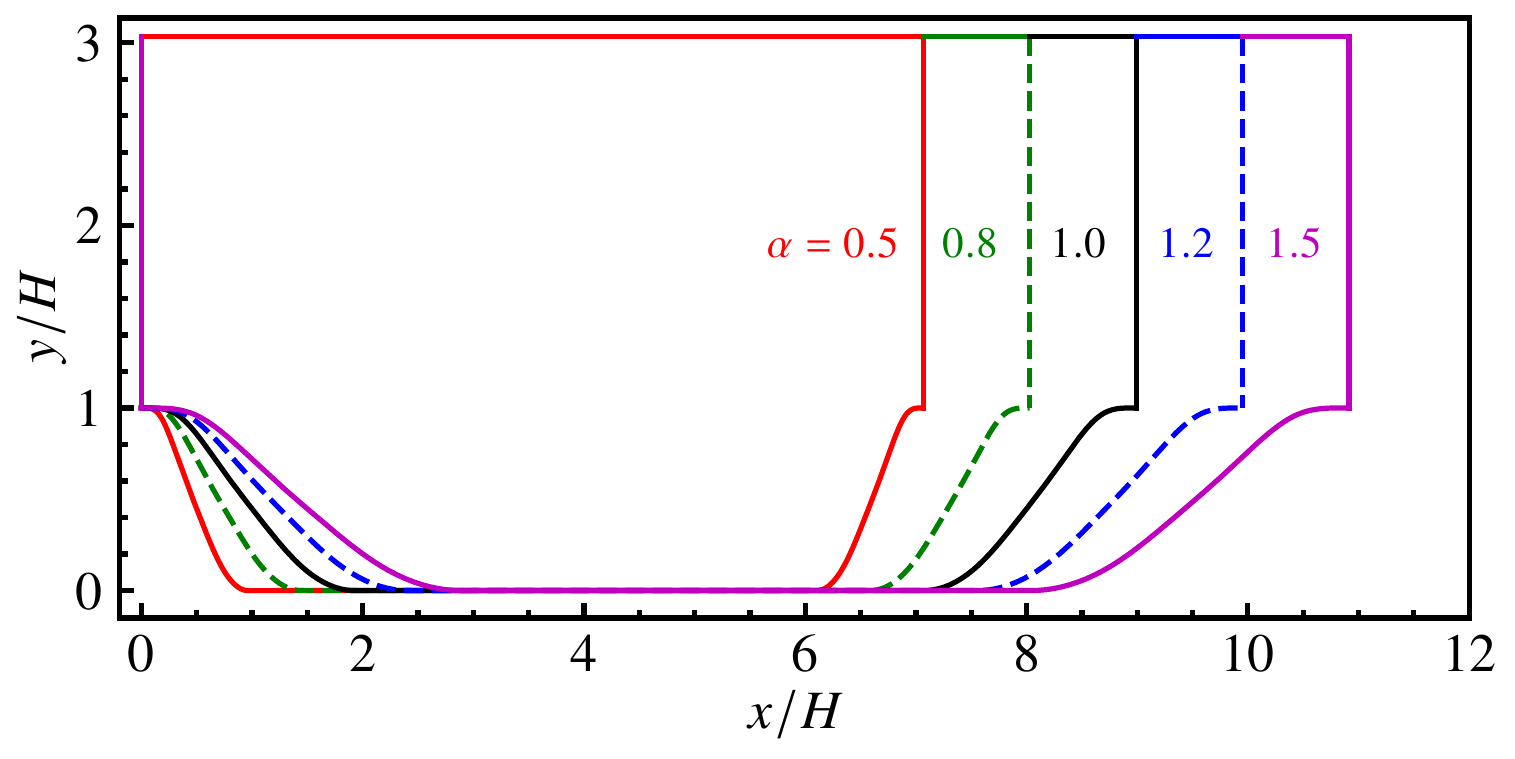}
	\end{subfigure}
	\caption{Configurations of plane channel and periodic hill flows: (a) plane channel; and (b) periodic hill with varying hill steepness factor $\alpha$.}
	\label{fig:2}
\end{figure}

\subsection {Plane channel flows}

The plane channel flow consists of an incompressible turbulent flow between two infinite parallel plates separated by a distance $2h$ shown in Fig.~\ref{fig:2}a. The streamwise, wall-normal, and spanwise directions are aligned with the $x_1$ ($x$), $x_2$ ($y$), and $x_3$ ($z$) axes, respectively. Owing to its statistical one-dimensionality in the wall-normal direction and well-established properties, plane channel flow serves as a benchmark for validating turbulence closures, including recent machine-learning-assisted RANS models.

This configuration plays a fundamental role in the development and evaluation of turbulence models, as it exhibits well-defined near-wall asymptotic behaviour, a clear separation of flow regions, and a wealth of high-fidelity reference data. Comprehensive direct numerical simulation (DNS) databases spanning a wide range of friction Reynolds numbers ($Re_\tau = u_\tau \delta/\nu$, where $u_\tau=\sqrt{\tau_w/\rho}$ is the friction velocity, $\delta$ the channel half-height, and $\tau_w$ the mean wall shear stress) are available in the literature, from which the present study employs the statistical profiles of mean velocity and Reynolds stresses.\cite{Lee2015Channel5200,kaneda2021velocity,hoyas2022wall} In the present study, plane channel flows covering the range $Re_\tau$= 550-10000 are employed to examine the ability of the proposed model to capture near-wall asymptotic behaviour as well as its robustness across Reynolds numbers.

To explicitly examine Reynolds-number generalization, the neural network is trained using channel-flow data at a subset of friction Reynolds numbers, ${{\mathop{Re}\nolimits} _\tau } \in \left\{ {1000,2000,4000,8000} \right\}$, while the remaining cases ${{\mathop{Re}\nolimits} _\tau } \in \left\{ {550,5200,10000} \right\}$ are reserved exclusively for validation. This design enables a systematic evaluation of model performance at Reynolds numbers not encountered during training. 

\subsection {Periodic hill flows with varying geometries}
The second test case is turbulent flow over periodic hills (Fig.~\ref{fig:2}), a canonical separated-flow configuration widely used for turbulence-model assessment due to the presence of strong adverse-pressure-gradient separation and subsequent reattachment. The bulk Reynolds number is defined based on the hill crest height $H$ and the bulk velocity $U_b$ at the domain inlet, ${{\mathop{Re}\nolimits}_H} = {U_b}H/\nu $. In the present study, the configuration at $Re_H = 5600$ is considered, based on DNS data reported by Xiao \emph{et al.}\cite{xiao2020flows}. These high-fidelity data were subsequently incorporated into the database compiled by McConkey \emph{et al.}\cite{mcconkey2021curated}, which serves as the primary data source for this work. 

The computational domain consists of a hill geometry that repeats periodically in the streamwise direction. Periodic boundary conditions are imposed at the inlet and outlet, while no-slip conditions are applied at the top and bottom walls. The domain height is constant $L_y/H=3.036$. The streamwise length varies with the steepness parameter $\alpha$, while the flat section length is constant for all geometries. The total horizontal length is given by $L_x/H=3.858\alpha+5.142$\cite{xiao2020flows}. Training is performed using cases with $\alpha  \in \left\{ {0.5,1.0,1.5} \right\}$, whereas intermediate geometries with $\alpha  \in \left\{ {0.8,1.2} \right\}$ are reserved for validation. This setting examines the applicability of the model to additional geometries within the canonical periodic-hill family.

Table~\ref{tab:1} summarizes the datasets and varying parameters used for training and validation in plane channel and periodic hill cases. For all training cases, the data are randomly split into 70\% training and 30\% testing subsets, providing a statistically robust evaluation of model performance.

\subsection {Architecture of the TBNN and STBNN models}
In the present study, both the baseline tensor-basis neural network (TBNN) and the self-scaling tensor-basis neural network (STBNN) employ an identical neural network architecture and training configuration. The difference between two models lies solely in the nondimensionalization and scaling of the tensor bases.

A fully connected feed-forward neural network with five hidden layers is adopted for both TBNN and STBNN models, with twenty neurons in each hidden layer. The Gaussian Error Linear Unit (GELU)\cite{hendrycks2016gaussian} is adopted for all hidden layers, and a linear function is applied at the output layer. In the artificial neural network, the output $X_i^{(l)}$ at layer $l$ is obtained from the outputs $X_j^{(l-1)}$ of the previous layer through a linear mapping followed by a nonlinear activation. The forward propagation of the neural network is given by
\begin{equation}
	X_i^l = \sigma \left( {\sum\limits_j {W_{ij}^lX_j^{l - 1}}  + b_i^l} \right),
	\label{ANN_mapping}
\end{equation} 
where $W_{ij}^{(l)}$, $b_i^{(l)}$ and $\sigma$ denote the weight matrix, bias vector and activation function  at layer $l$, respectively. The GELU activation function is defined as\cite{hendrycks2016gaussian}
\begin{equation}
	\sigma \left( x \right) = x\Phi \left( x \right),
	\label{ANN_GELU}
\end{equation} 
with $\Phi \left(  \cdot  \right)$ being the cumulative distribution function of standard Gaussian distribution.  

GELU activation function provides a smooth and continuously differentiable mapping, and has been reported to exhibit stable behavior in regression tasks involving continuous physical quantities \cite{hendrycks2016gaussian}. The influence of different activation functions within the present framework is examined in Appendix~\ref{sec:appendixA}.

The input feature vector consists of nine nondimensional variables, including four auxiliary feature parameters $q_1$–$q_4$ (defined in Eq.~\ref{aux_qi}) and five scalar invariants of the velocity gradient tensor. For the TBNN model, the invariants are denoted by $\hat\lambda_1$–$\hat\lambda_5$ (see Eq.~\ref{lambda_GEVM}), while the STBNN model employs $\tilde\lambda_1$–$\tilde\lambda_5$ (see Eq.~\ref{lambda_STBNN}). Based on the tensor-basis formulation, the first five tensor bases are retained (see in Eqs.~\ref{Tij_GEVM} and \ref{Tij_STBNN}), and the neural network predicts the corresponding five basis coefficients. Accordingly, the output dimension of the neural network is five. The influence of the underlying neural network architecture is examined in Appendix~\ref{sec:appendixB}, where replacing the ANN with alternative mappings (namely, ResNet and KAN) yields comparable performance under the same invariant-based tensor formulation.

For the TBNN model, the basis tensors are constructed from the strain-rate and rotation-rate tensors nondimensionalized by the conventional time scale $\hat \tau=k/\varepsilon$, i.e. $\hat{\mathbf S}=\hat \tau\mathbf S$ and $\hat{\boldsymbol\Omega}=\hat \tau\boldsymbol\Omega$. The resulting tensor bases are denoted by $\hat{\mathbf T}^{(n)}$, and the network-predicted basis coefficients are directly used to reconstruct the Reynolds stress anisotropy tensor.

\begin{table}[t]
	\centering
	\caption{Overview of datasets and varying parameters used for training, testing, and validation in plane channel and periodic hill cases.}
	\label{tab:1}
	\begin{tabular}{lcccccc}
		\hline\hline
		Case & Data points & Parameter varied & Training set & Validation set & ${Re}_b$ (training) & ${Re}_b$ (validation) \\
		\hline
		Plane channel
		& 5716
		& ${Re}_\tau$
		& 1000, 2000
		& 550, 5200
		& 20000, 43480
		& 10000, 125000 \\
		& 
		& 
		& 4000, 8000
		& 10000
		& 93510, 199680
		& 290190 \\
		Periodic hill
		& 73755
		& Steepness ($\alpha$)
		& 0.5, 1.0, 1.5
		& 0.8, 1.2
		& 5600
		& 5600 \\
		\hline\hline
	\end{tabular}
\end{table}

For the STBNN model, an adaptive self-scaling normalization strategy is introduced by rescaling the strain-rate and rotation-rate tensors with their combined magnitude. Specifically, $\tilde {\mathbf S}=\tilde \tau\mathbf S$ and $\tilde {\mathbf \Omega}=\tilde \tau\mathbf \Omega$ with the inverse scaling factor  ${{\tilde \tau }^{ - 1}} = {\sqrt {{{\left\| {\bf{S}} \right\|}^2} + {{\left\| {\bf{\Omega }} \right\|}^2}} }$. The basis tensors built from $(\tilde{\mathbf S},\tilde{\boldsymbol\Omega})$ are denoted by $\widetilde{\mathbf T}^{(n)}$. With the same network architecture and outputs, the Reynolds stress anisotropy tensor is reconstructed. The distinction between TBNN and STBNN models lies solely in the self-scaled basis tensors used in the tensor-basis representation.

The TBNN and STBNN models are trained by minimizing the mean-squared error (MSE) between the predicted and DNS deviatoric Reynolds-stress tensors. The loss function is defined by
\begin{equation}
	L\!\left(\mathbf R^{d,\mathrm{mod}},\,\mathbf R^{d,\mathrm{DNS}}\right)
	=
	\frac{1}{N_s}
	\sum_{i=1}^{N_s}
	\left\|
	\mathbf R_i^{d,\mathrm{DNS}}
	-
	2k_i\,\mathbf b_i^{\mathrm{mod}}
	\right\|^2.
	\label{loss_Rij}
\end{equation}
This loss directly measures the error in the modeled deviatoric Reynolds stress that enters the RANS momentum equations (Eq.~\ref{RANS2}), rather than in the normalized anisotropy tensor. The network parameters are optimized using the AdamW optimizer\cite{kingma2014adam,loshchilov2017decoupled} with an initial learning rate $\gamma=10^{-3}$. The same network architecture, input features, and training strategy are applied to both the plane channel flow and the periodic hill flow cases. The input and output variables of the models and their parameter settings are listed in Table~\ref{tab:2}. 

\begin{table}[t]
	\centering
	\caption{Parameters of the TBNN and STBNN models.}
	\label{tab:2}
	\begin{tabular}{lcccccc}
		\hline\hline
		Model & Learning rate & Epoch & Hidden layers & Neurons per layer & Inputs & Outputs \\
		\hline
		TBNN 
		& $10^{-3}$ 
		& 10000 
		& 5 
		& 20 
		& $\left(\begin{array}{c}
			\hat{\lambda}_1,\hat{\lambda}_2,\hat{\lambda}_3,\hat{\lambda}_4,\hat{\lambda}_5,\\
			q_1,q_2,q_3,q_4
		\end{array}\right)$
		& $R_{11}^d, R_{22}^d, R_{33}^d, R_{12}^d$ \\
		
		STBNN 
		& $10^{-3}$ 
		& 10000 
		& 5 
		& 20 
		& $\left(\begin{array}{c}
			\tilde{\lambda}_1,\tilde{\lambda}_2,\tilde{\lambda}_3,\tilde{\lambda}_4,\tilde{\lambda}_5,\\
			q_1,q_2,q_3,q_4
		\end{array}\right)$
		& $R_{11}^d, R_{22}^d, R_{33}^d, R_{12}^d$ \\
		\hline\hline
	\end{tabular}
\end{table}

\section {\emph{A priori} tests of the STBNN models}\label{sec:level4}
In this section, we perform \emph{a priori} tests to examine the predictive accuracy of the proposed STBNN model in comparison with the baseline closures, including LEVM, QEVM, and TBNN models. The assessment is conducted using high-fidelity DNS data for wall-bounded turbulent flows under different flow conditions, namely plane channel flows at various friction Reynolds numbers and periodic hill flows with varying hill steepness. 

The model performance is evaluated through the prediction of Reynolds stresses. Two indicators are employed to quantify the discrepancy between the reference value ($Q^{\rm{ref}}$) and the model prediction ($Q^{\rm{model}}$) for arbitrary targeted quantity $Q$.  These metrics are the correlation coefficient ${\rm C}(Q)$ and the relative error ${\rm E_r}(Q)$, defined respectively as\cite{yuan2020,yuan2021,yuan2021a,yuan2022}
\begin{equation}
	{\rm C}\left( Q \right) = \frac{{\left\langle {\left( {{Q^{\rm{ref}}} - \left\langle {{Q^{\rm{ref}}}} \right\rangle } \right)\left( {{Q^{\rm{model}}} - \left\langle {{Q^{\rm{model}}}} \right\rangle } \right)} \right\rangle }}{{{{\left\langle {{{\left( {{Q^{\rm{ref}}} - \left\langle {{Q^{\rm{ref}}}} \right\rangle } \right)}^2}} \right\rangle }^{1/2}}{{\left\langle {{{\left( {{Q^{\rm{model}}} - \left\langle {{Q^{\rm{model}}}} \right\rangle } \right)}^2}} \right\rangle }^{1/2}}}},
	\label{CorrCoeff}
\end{equation}
\begin{equation}
	{\rm E_r}\left( Q \right) = \frac{{{{\left\langle {{{\left( {{Q^{\rm{ref}}} - {Q^{\rm{model}}}} \right)}^2}} \right\rangle }^{1/2}}}}{{{{\left\langle {{{\left( {{Q^{\rm{ref}}}} \right)}^2}} \right\rangle }^{1/2}}}},
	\label{relativeErr}
\end{equation}
where ``$\left\langle  \right\rangle $" represents the spatial averaging over the entire domain. High correlation coefficients combined with low relative errors indicate improved \emph{a priori} predictive accuracy of turbulence models. 

\subsection {Plane channel flows}

The correlation coefficients and relative errors of the predicted deviatoric Reynolds stresses in plane channel flow for both training and validation datasets are summarized in Tables~\ref{tab:3} and~\ref{tab:4}. For all Reynolds stress components, the STBNN model achieves correlation coefficients essentially equal to unity (greater than 99.9\%) and relative errors close to zero (below 2\%). The small discrepancy between the training and validation sets indicates that the model does not suffer from overfitting and maintains high predictive accuracy at untrained Reynolds numbers.

\begin{table}[t]
	\centering
	\caption{Correlation coefficients of the deviatoric Reynolds stress for plane channel flow.}
	\label{tab:3}
	
	\begin{tabular}{c l c c c c}
		\hline\hline
		${\rm{Case \slash C}}\left( {R_{ij}^d} \right)$  & Model & $R_{11}^d$ & $R_{22}^d$ & $R_{33}^d$ & $R_{12}^d$ \\
		\hline
		
		\multirow{4}{*}{%
			\begin{tabular}{c}
				Training set \\[3pt]
				${{Re}}_\tau \in \left\{ \begin{array}{l}
					1000,{\mkern 1mu} 2000,{\mkern 1mu} \\
					4000,{\mkern 1mu} 8000
				\end{array} \right\}$
		\end{tabular}}
		& LEVM  & 0      & 0      & 0      & 0.5766 \\
		& QEVM  & 0.9739 & 0.957 & 0.9725 & 0.9212 \\
		& TBNN  & 0.7836 & 0.747 & 0.7258 & 0.6001 \\
		& \textbf{STBNN} & \textbf{1.0} & \textbf{1.0} & \textbf{0.9999} & \textbf{1.0} \\
		\hline
		
		\multirow{4}{*}{%
			\begin{tabular}{c}
				Validation set \\[3pt]
				${{Re}}_\tau \in \left\{ \begin{array}{l}
					550,{\mkern 1mu} 5200,{\mkern 1mu} \\
					10000
				\end{array} \right\}$
		\end{tabular}}
		& LEVM  & 0      & 0      & 0      & 0.5917 \\
		& QEVM  & 0.9718 & 0.9593 & 0.9665 & 0.9036 \\
		& TBNN  & 0.8028 & 0.7586 & 0.7869 & 0.6141 \\
		& \textbf{STBNN} & \textbf{0.9995} & \textbf{0.9999} & \textbf{0.996} & \textbf{0.9998} \\
		\hline\hline
	\end{tabular}
\end{table}

\begin{table}[t]
	\centering
	\caption{Relative errors of the deviatoric Reynolds stress for plane channel flow.}
	\label{tab:4}
	
	\begin{tabular}{c l c c c c}
		\hline\hline
		${\rm Case \slash Er}(R_{ij}^d)$ & Model & $R_{11}^d$ & $R_{22}^d$ & $R_{33}^d$ & $R_{12}^d$ \\
		\hline
		
		\multirow{4}{*}{%
			\begin{tabular}{c}
				Training set \\[3pt]
				${{Re}}_\tau \in \left\{ \begin{array}{l}
					1000,{\mkern 1mu} 2000,{\mkern 1mu} \\
					4000,{\mkern 1mu} 8000
				\end{array} \right\}$
		\end{tabular}}
		& LEVM  & 1.0 & 1.0 & 1.0 & 1.6152 \\
		& QEVM  & 0.3799 & 0.4753 & 0.2533 & 1.3855 \\
		& TBNN  & 0.9875 & 0.9875 & 0.9879 & 0.9887 \\
		& \textbf{STBNN} & \textbf{0.0017} & \textbf{0.0013} & \textbf{0.0052} & \textbf{0.001} \\
		\hline
		
		\multirow{4}{*}{%
			\begin{tabular}{c}
				Validation set \\[3pt]
				${{Re}}_\tau \in \left\{ \begin{array}{l}
					550,{\mkern 1mu} 5200,{\mkern 1mu} \\
					10000
				\end{array} \right\}$
		\end{tabular}}
		& LEVM  & 1.0 & 1.0 & 1.0 & 1.6381 \\
		& QEVM  & 0.3884 & 0.4726 & 0.2875 & 1.3447 \\
		& TBNN  & 0.9887 & 0.9889 & 0.9887 & 0.9905 \\
		& \textbf{STBNN} & \textbf{0.0251} & \textbf{0.0103} & \textbf{0.0674} & \textbf{0.0103} \\
		\hline\hline
	\end{tabular}
\end{table}

By contrast, the LEVM fails to reproduce the anisotropic normal stresses, leading to vanishing correlations and large errors. The QEVM improves the prediction of the normal components but still exhibits noticeable inaccuracies in the shear Reynolds stress. The TBNN model further enhances the correlations. However, its errors remain significantly larger than those of STBNN (exceeding 90\%), particularly in the validation set. These results demonstrate the superior accuracy of the STBNN model and its improved predictive transferability to the untrained Reynolds numbers examined in the present study.

\begin{figure}\centering
	\begin{subfigure}{0.5\textwidth}
		\centering
		{($a$)}
		\includegraphics[width=0.9\linewidth,valign=t]{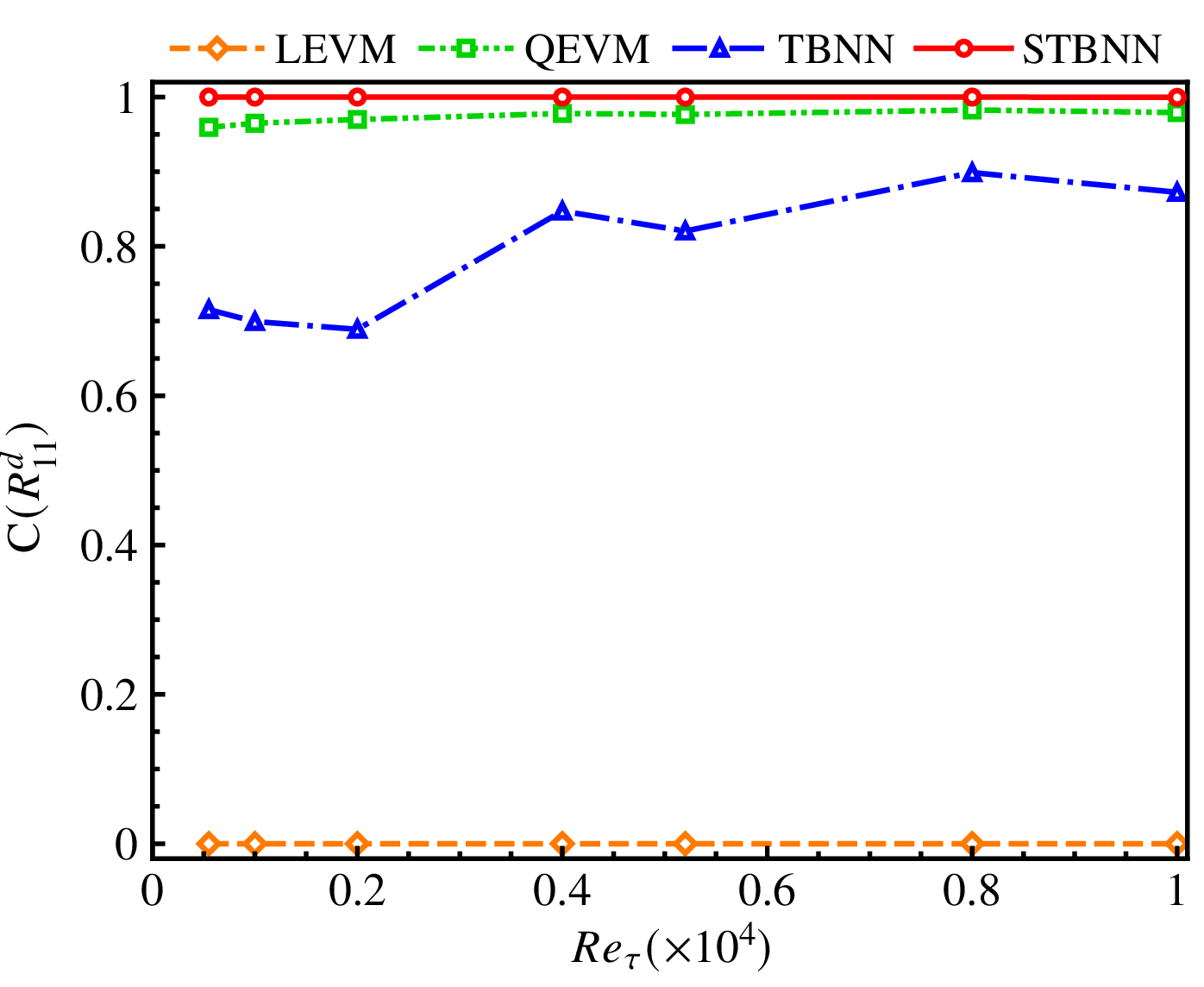}
	\end{subfigure}%
	\begin{subfigure}{0.5\textwidth}
		\centering
		{($b$)}
		\includegraphics[width=0.9\linewidth,valign=t]{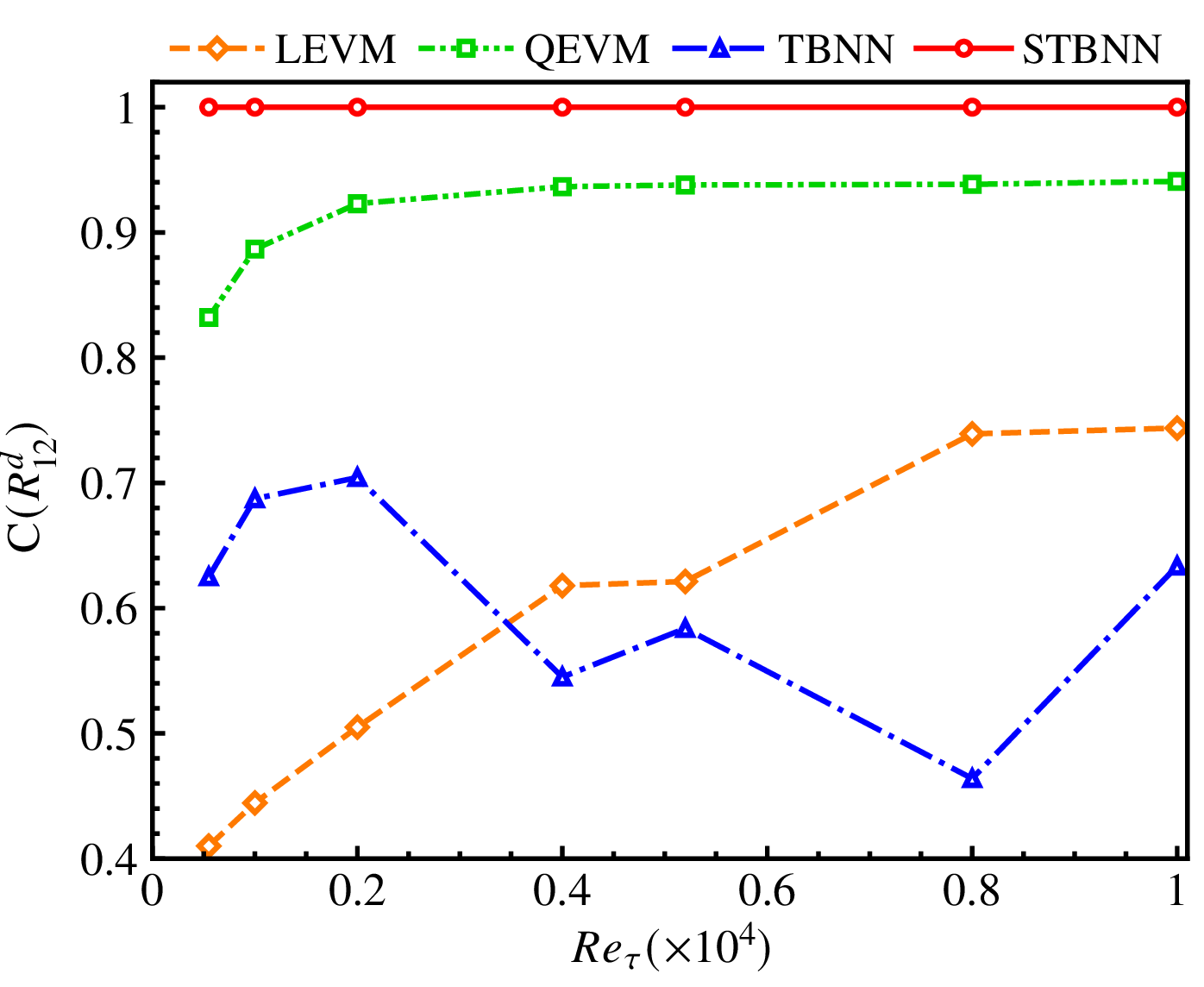}
	\end{subfigure}
	\begin{subfigure}{0.5\textwidth}
		\centering
		{($c$)}
		\includegraphics[width=0.9\linewidth,valign=t]{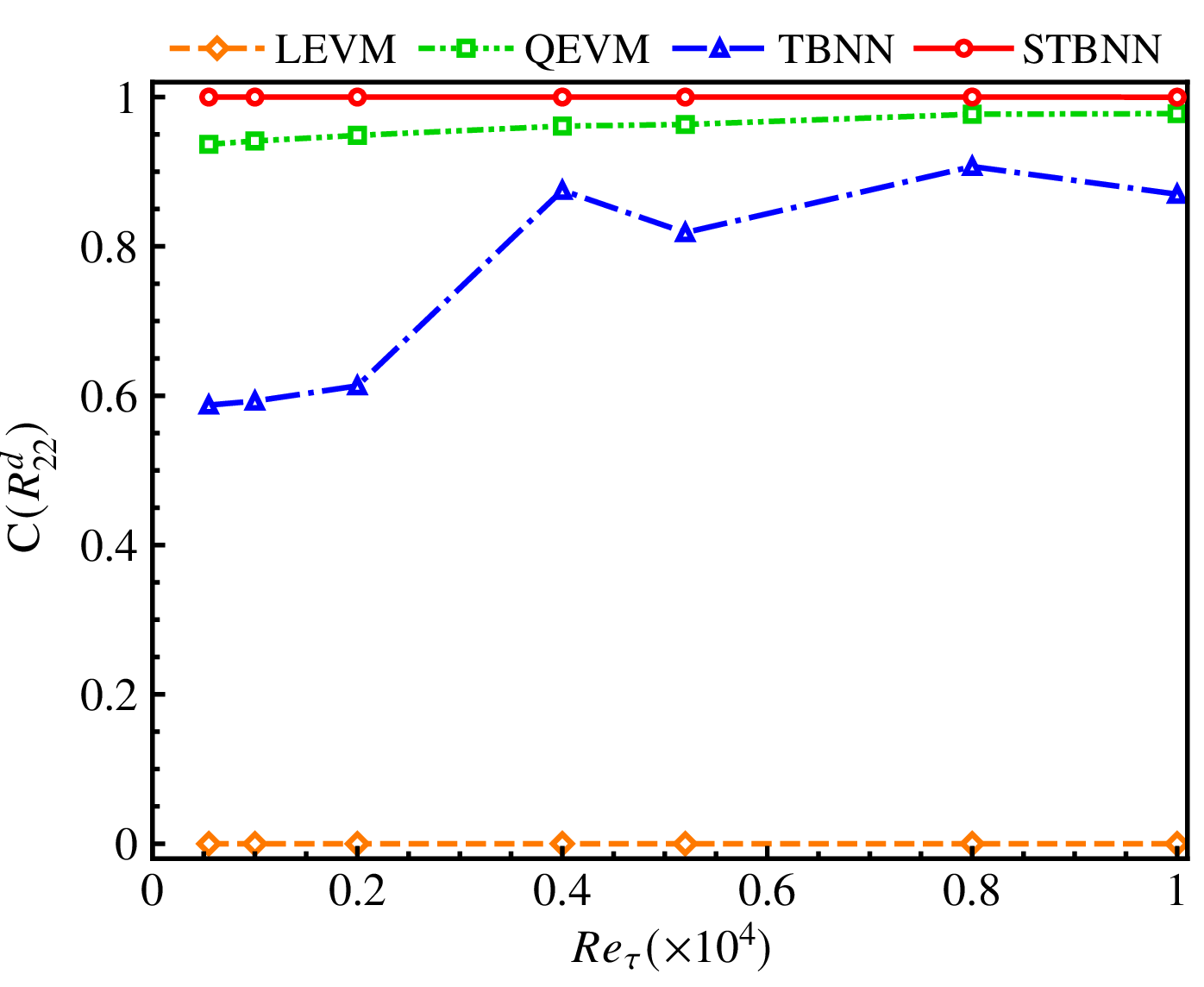}
	\end{subfigure}%
	\begin{subfigure}{0.5\textwidth}
		\centering
		{($d$)}
		\includegraphics[width=0.9\linewidth,valign=t]{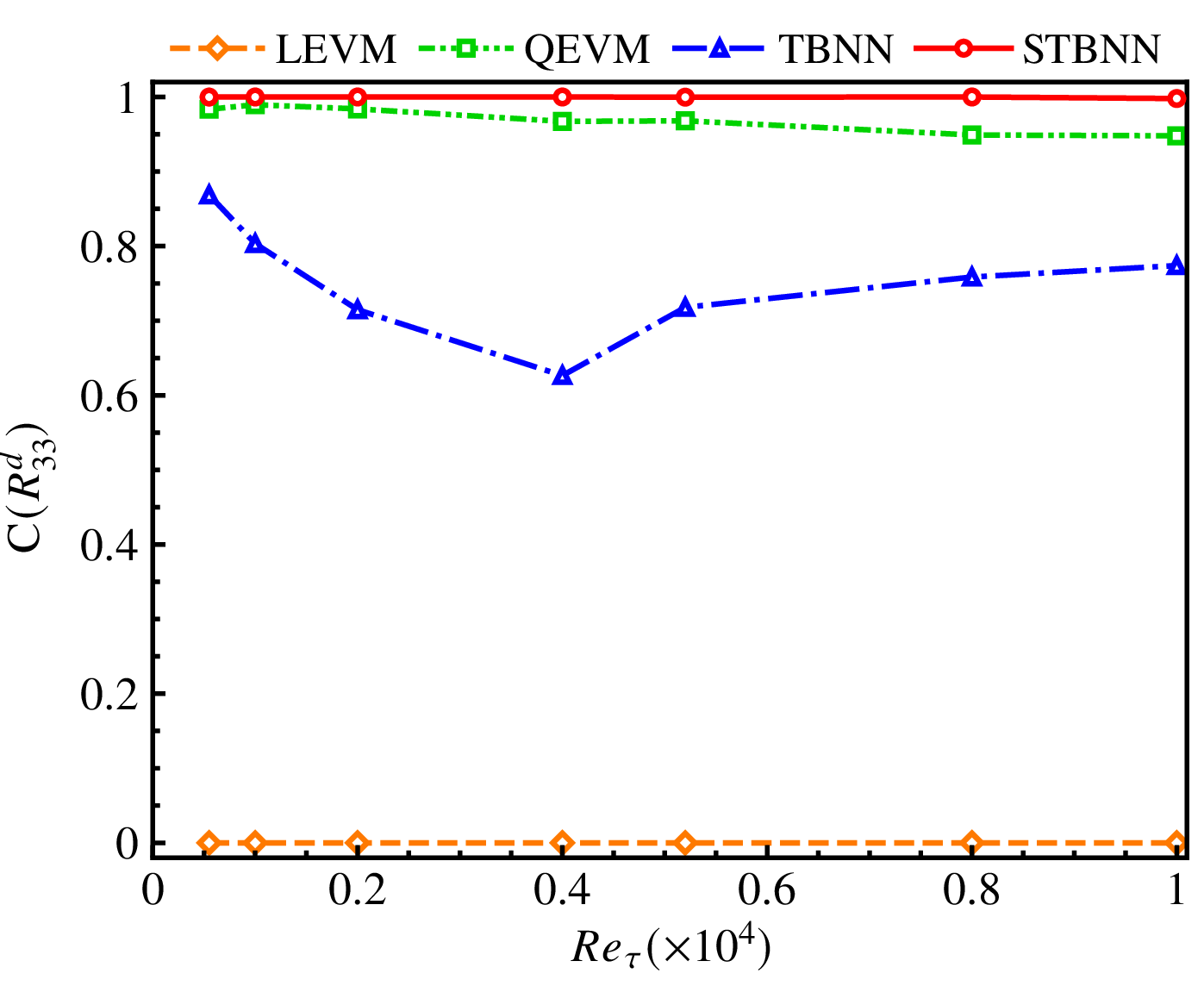}
	\end{subfigure}
	\caption{Correlation coefficients of modeled deviatoric Reynolds stress components in plane channel flow across friction Reynolds numbers (${ Re}_\tau$): (a) $R_{11}^d$; (b) $R_{12}^d$; (c) $R_{22}^d$; (d) $R_{33}^d$.}
	
	\label{fig:3}
\end{figure}

\begin{figure}\centering
	\begin{subfigure}{0.5\textwidth}
		\centering
		{($a$)}
		\includegraphics[width=0.9\linewidth,valign=t]{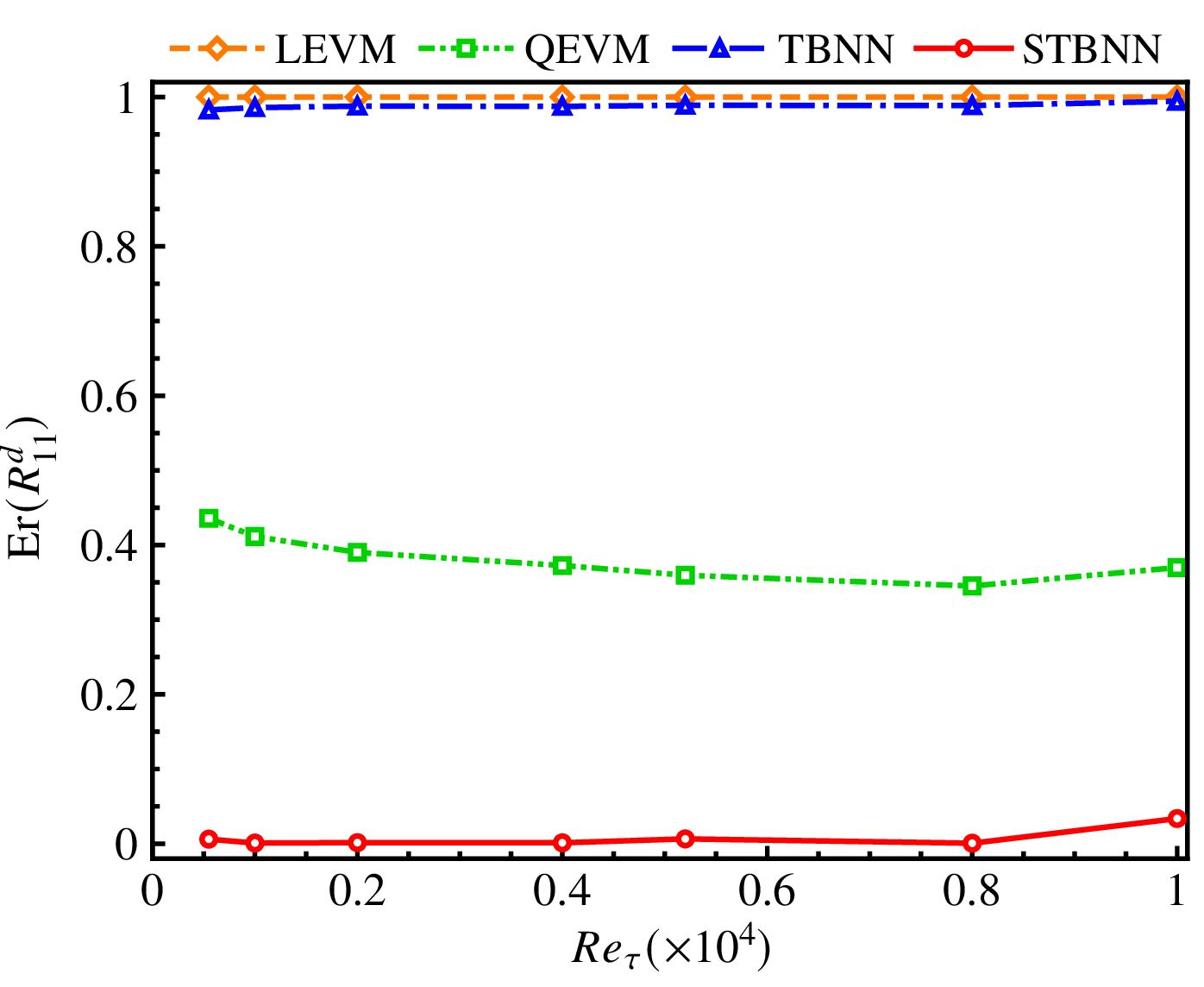}
	\end{subfigure}%
	\begin{subfigure}{0.5\textwidth}
		\centering
		{($b$)}
		\includegraphics[width=0.9\linewidth,valign=t]{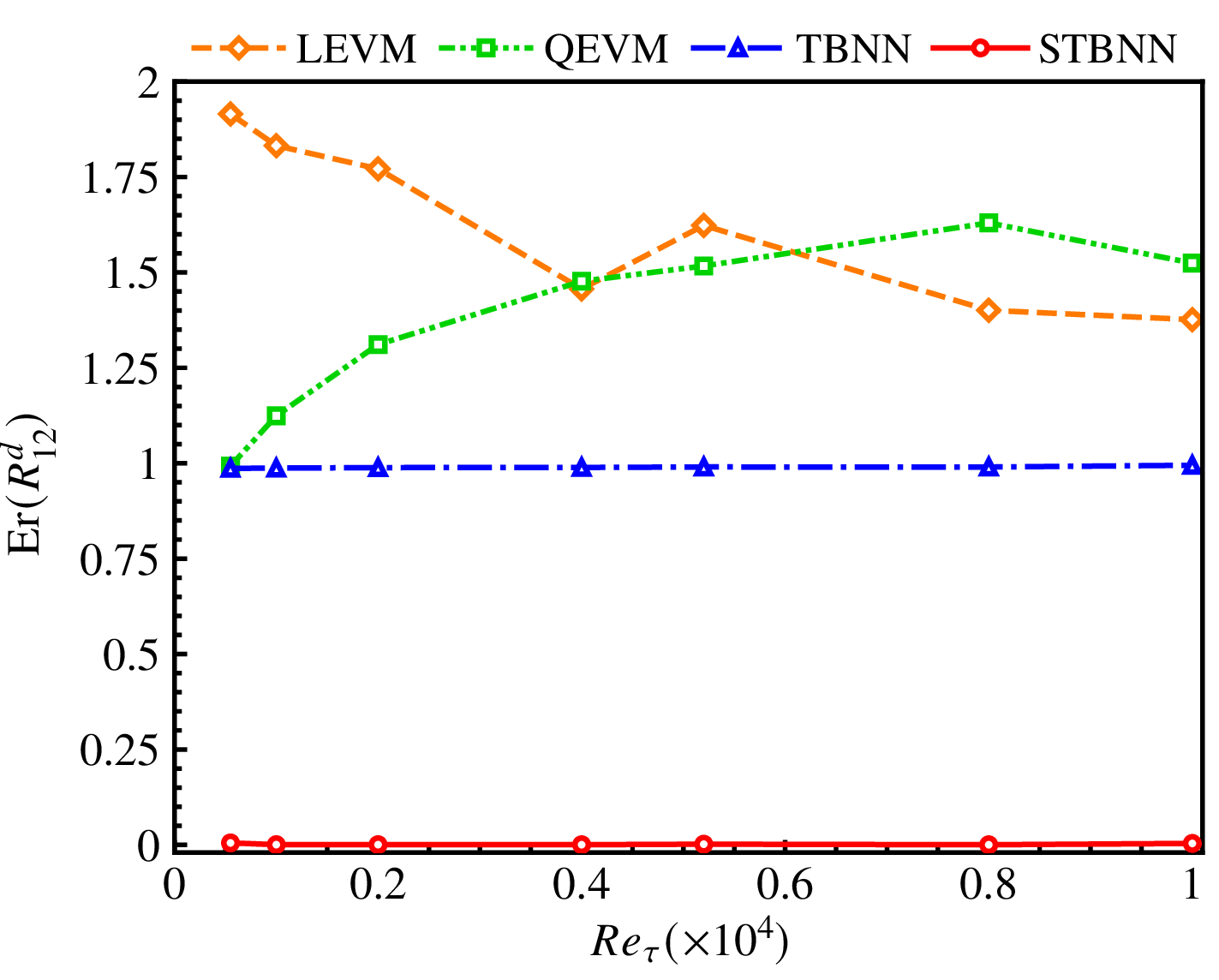}
	\end{subfigure}
	\begin{subfigure}{0.5\textwidth}
		\centering
		{($c$)}
		\includegraphics[width=0.9\linewidth,valign=t]{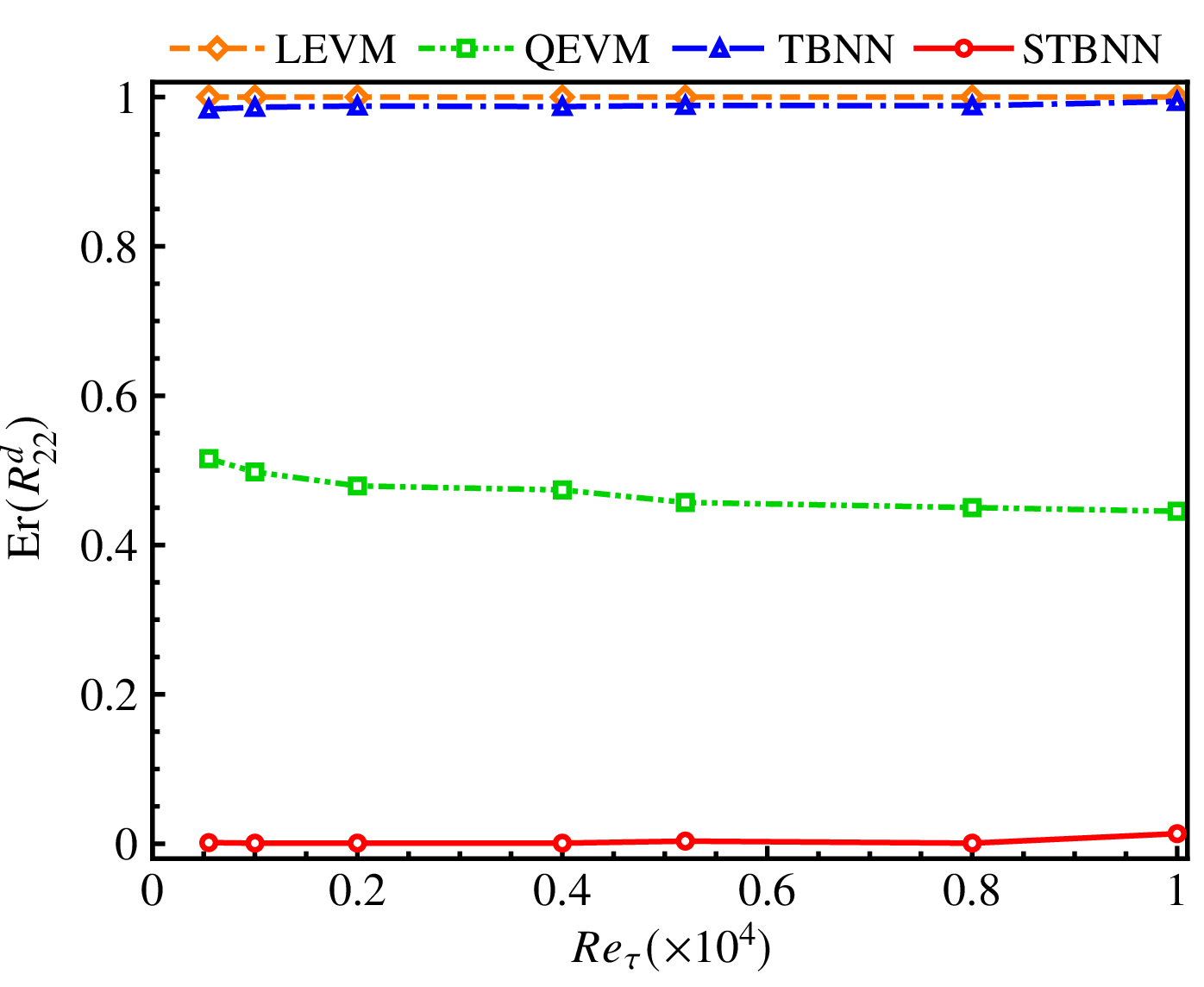}
	\end{subfigure}%
	\begin{subfigure}{0.5\textwidth}
		\centering
		{($d$)}
		\includegraphics[width=0.9\linewidth,valign=t]{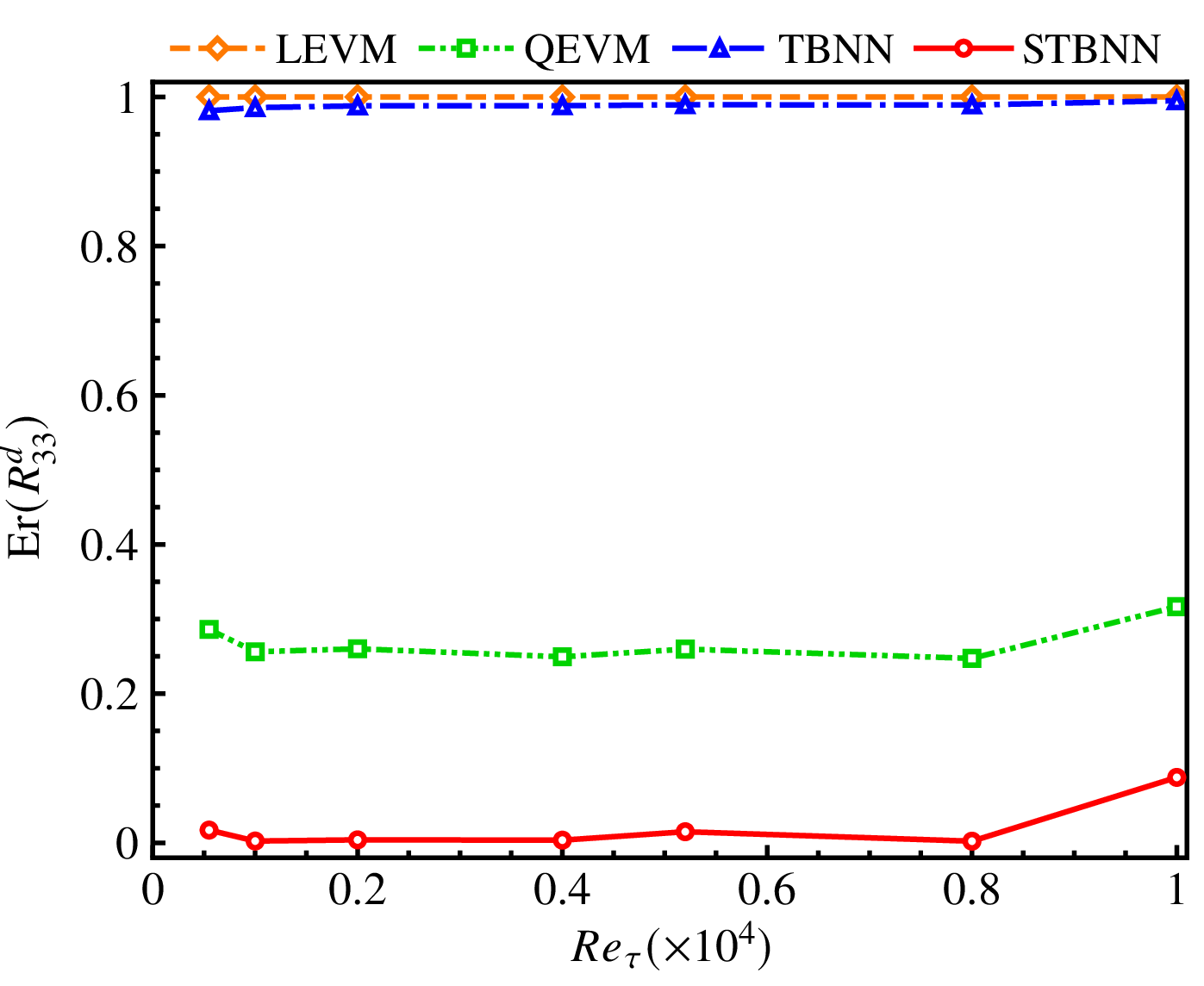}
	\end{subfigure}
	\caption{Relative errors of modeled deviatoric Reynolds stress components in plane channel flow across friction Reynolds numbers (${Re}_\tau$): (a) $R_{11}^d$; (b) $R_{12}^d$; (c) $R_{22}^d$; (d) $R_{33}^d$.}
	\label{fig:4}
\end{figure}

The Reynolds-number dependence of the predictions is further illustrated in Figs.~\ref{fig:3} and~\ref{fig:4}. The STBNN model maintains nearly constant correlation coefficients close to unity and consistently small relative errors over the entire range $550 \le Re_\tau \le 10^4$. In contrast, the TBNN predictions exhibit intermediate accuracy between LEVM and QEVM, whereas the STBNN model consistently provides the most accurate predictions across all Reynolds numbers. The weak Reynolds-number dependence of the STBNN predictions indicates that the model captures the scaling behaviour of the stress anisotropy more consistently over the tested Reynolds-number range.

\begin{figure}
	\centering
	
	\begin{subfigure}{0.32\linewidth}
		\includegraphics[width=\linewidth]{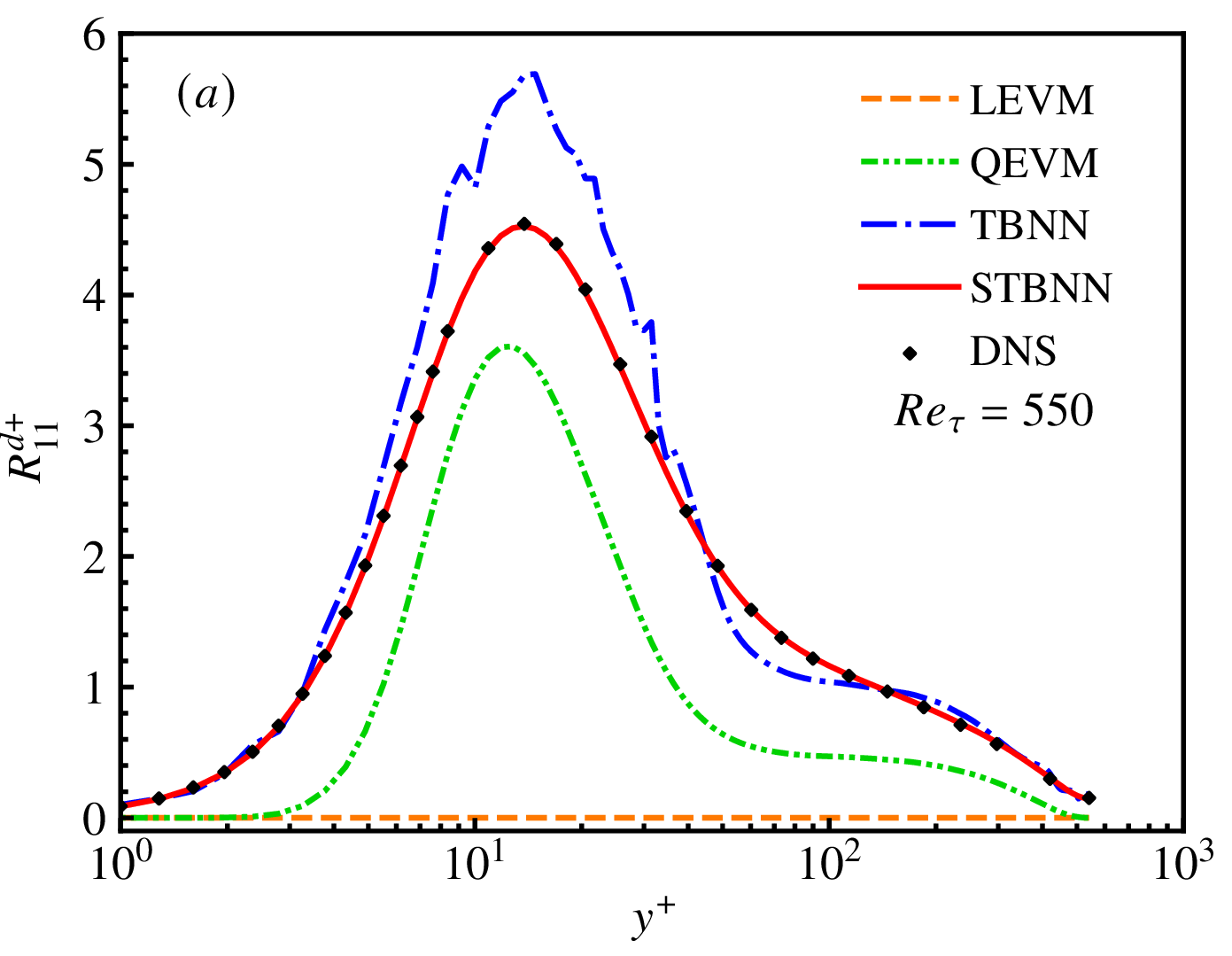}
	\end{subfigure}
	\begin{subfigure}{0.32\linewidth}
		\includegraphics[width=\linewidth]{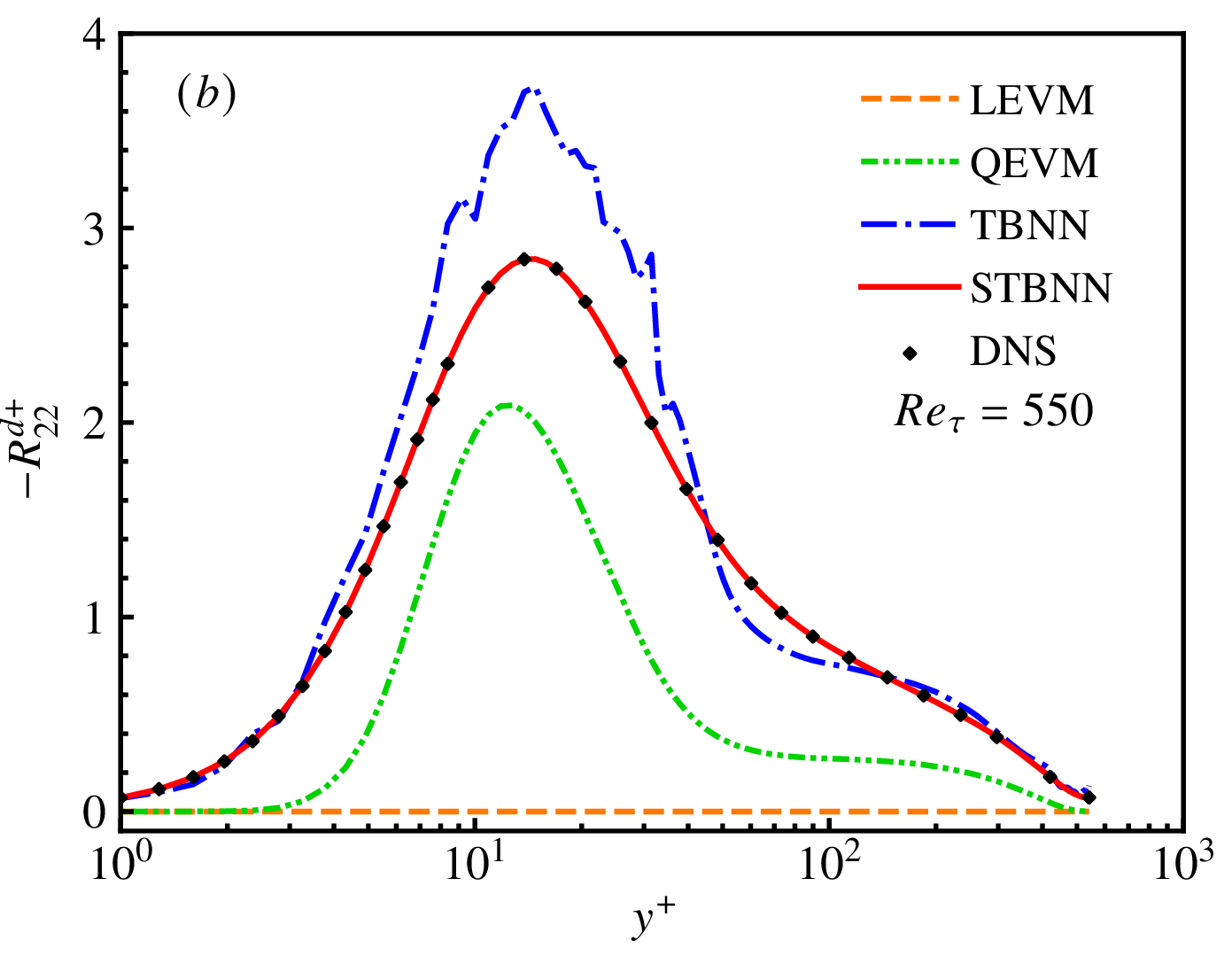}
	\end{subfigure}
	\begin{subfigure}{0.34\linewidth}
		\includegraphics[width=\linewidth]{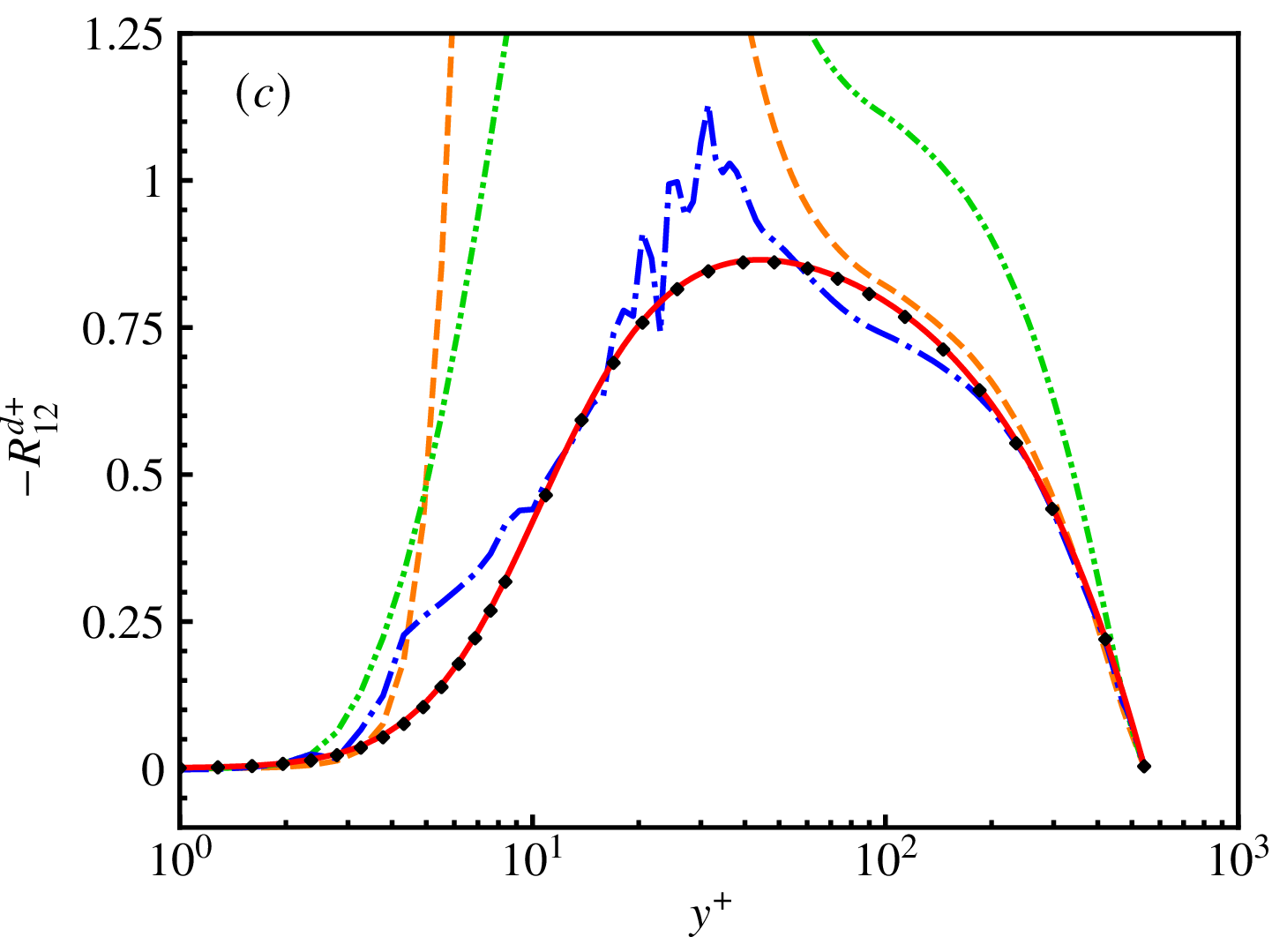}
	\end{subfigure}
	
	\begin{subfigure}{0.32\linewidth}
		\includegraphics[width=\linewidth]{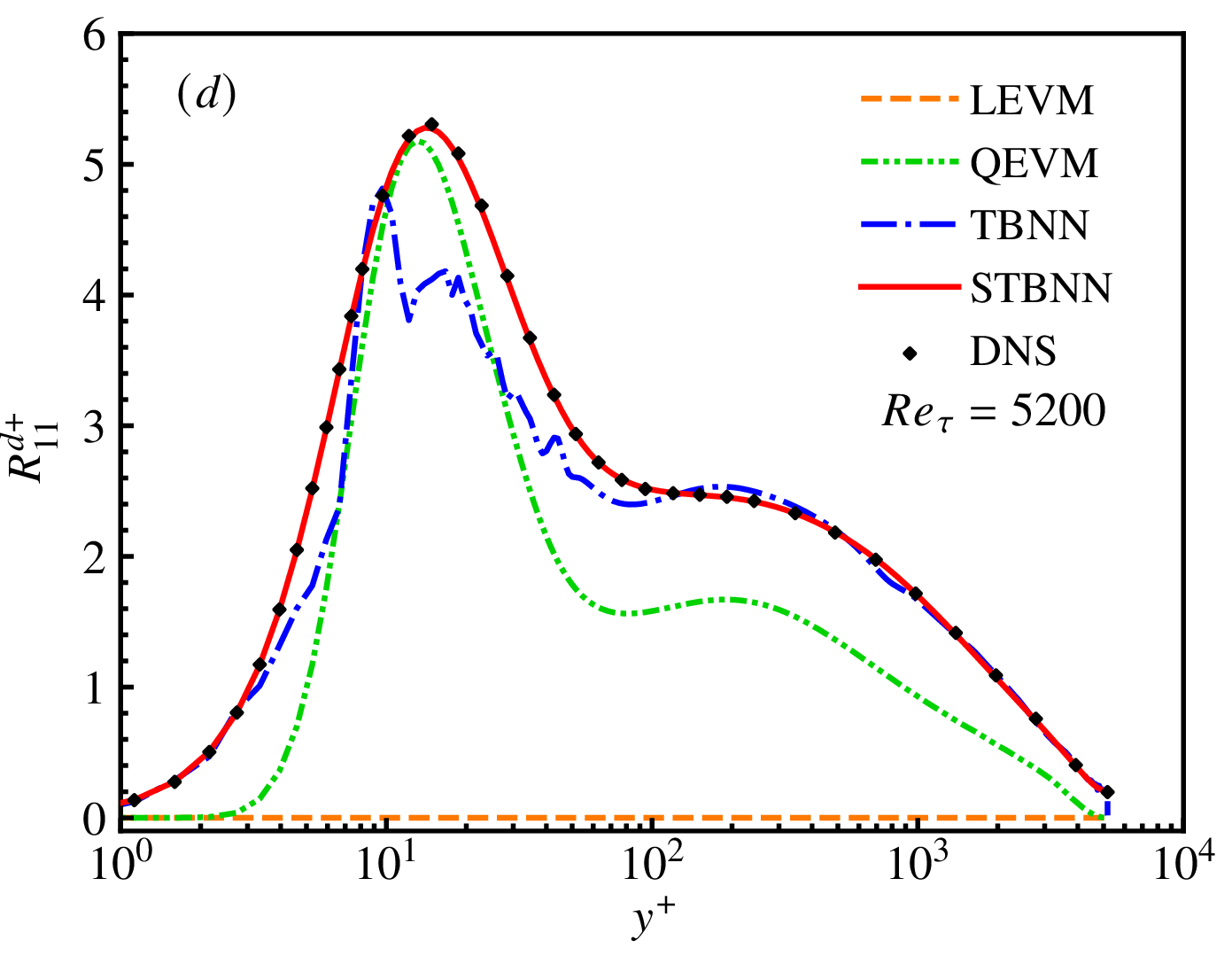}
	\end{subfigure}
	\begin{subfigure}{0.32\linewidth}
		\includegraphics[width=\linewidth]{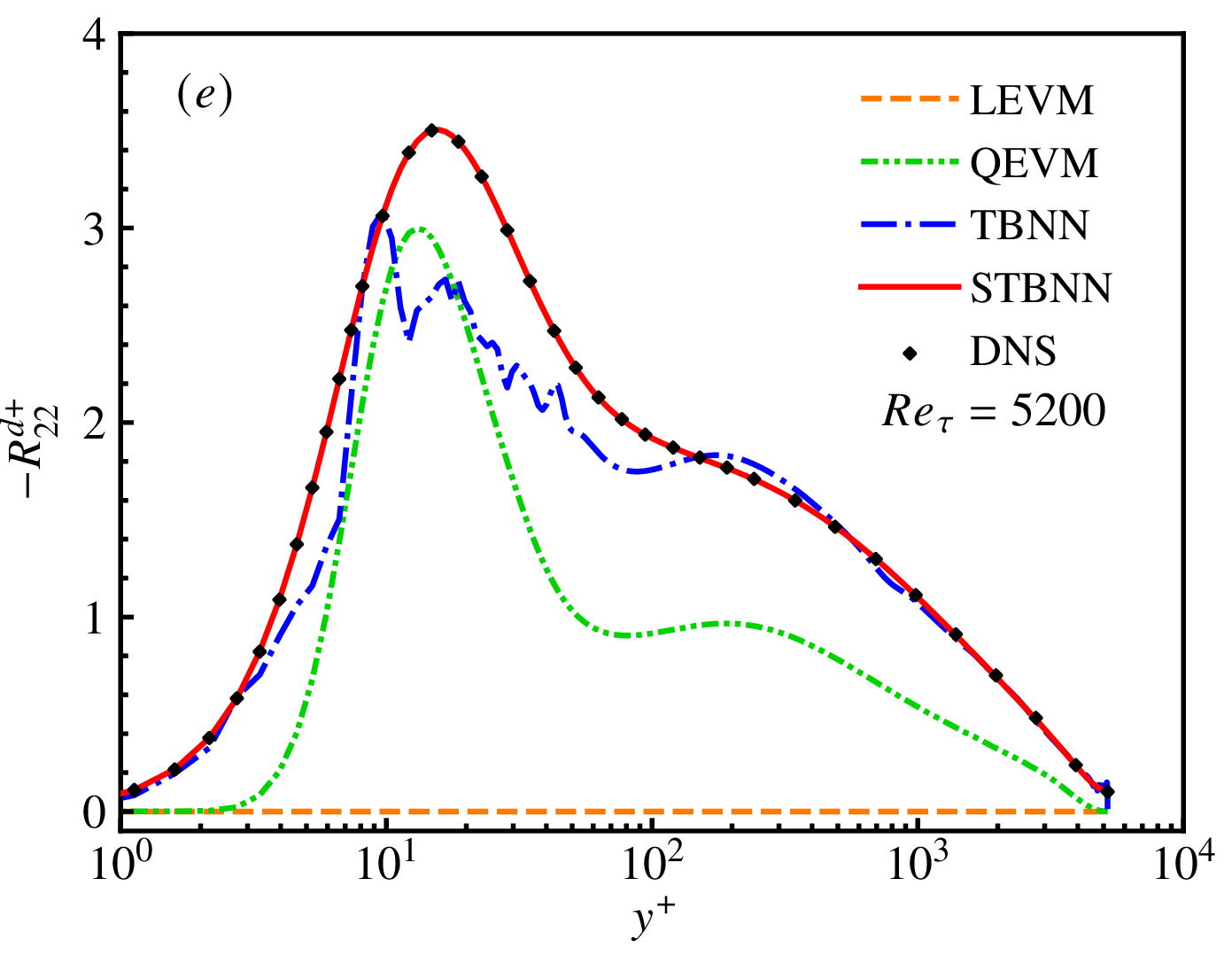}
	\end{subfigure}
	\begin{subfigure}{0.34\linewidth}
		\includegraphics[width=\linewidth]{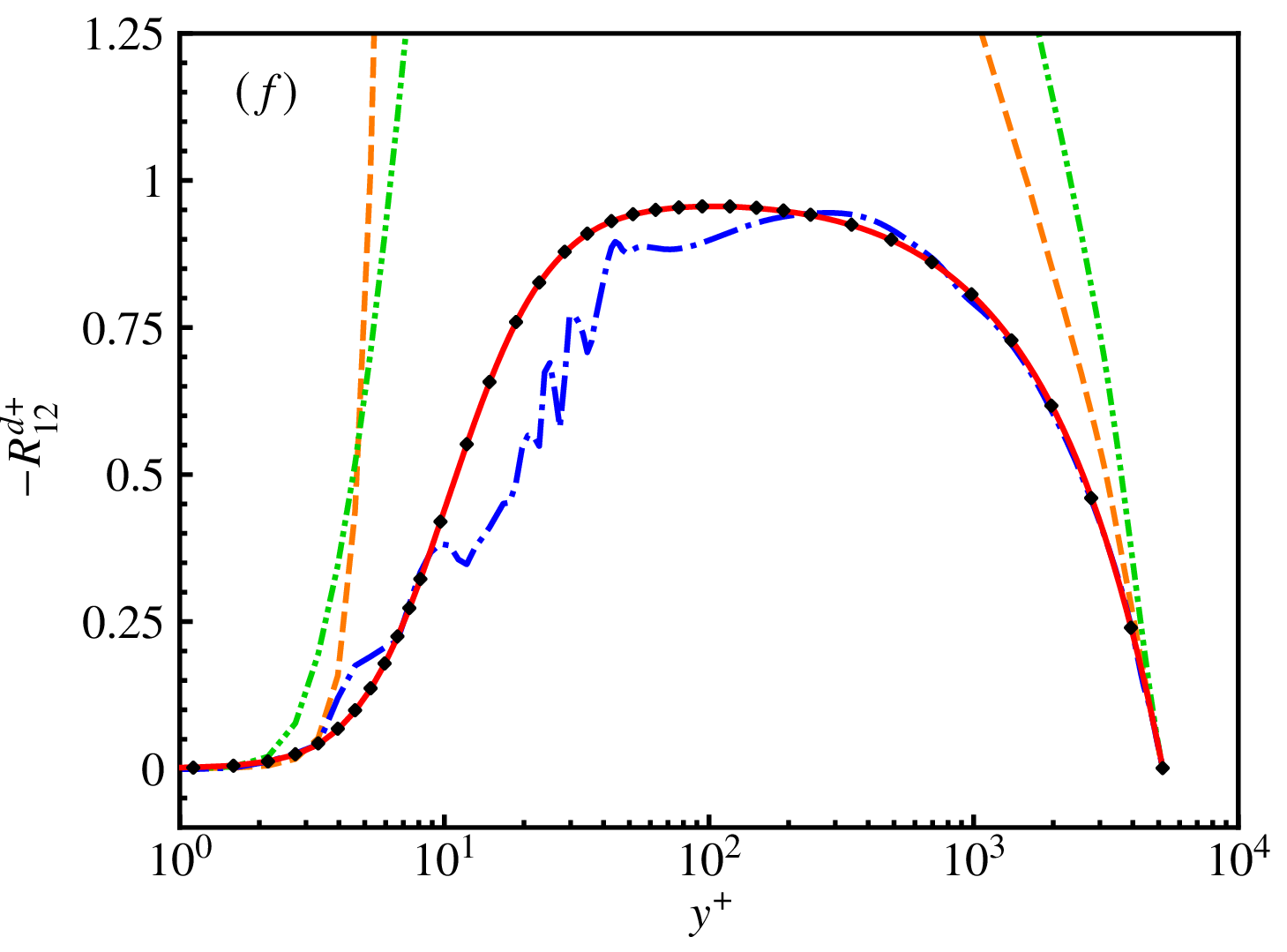}
	\end{subfigure}
	
	\begin{subfigure}{0.32\linewidth}
		\includegraphics[width=\linewidth]{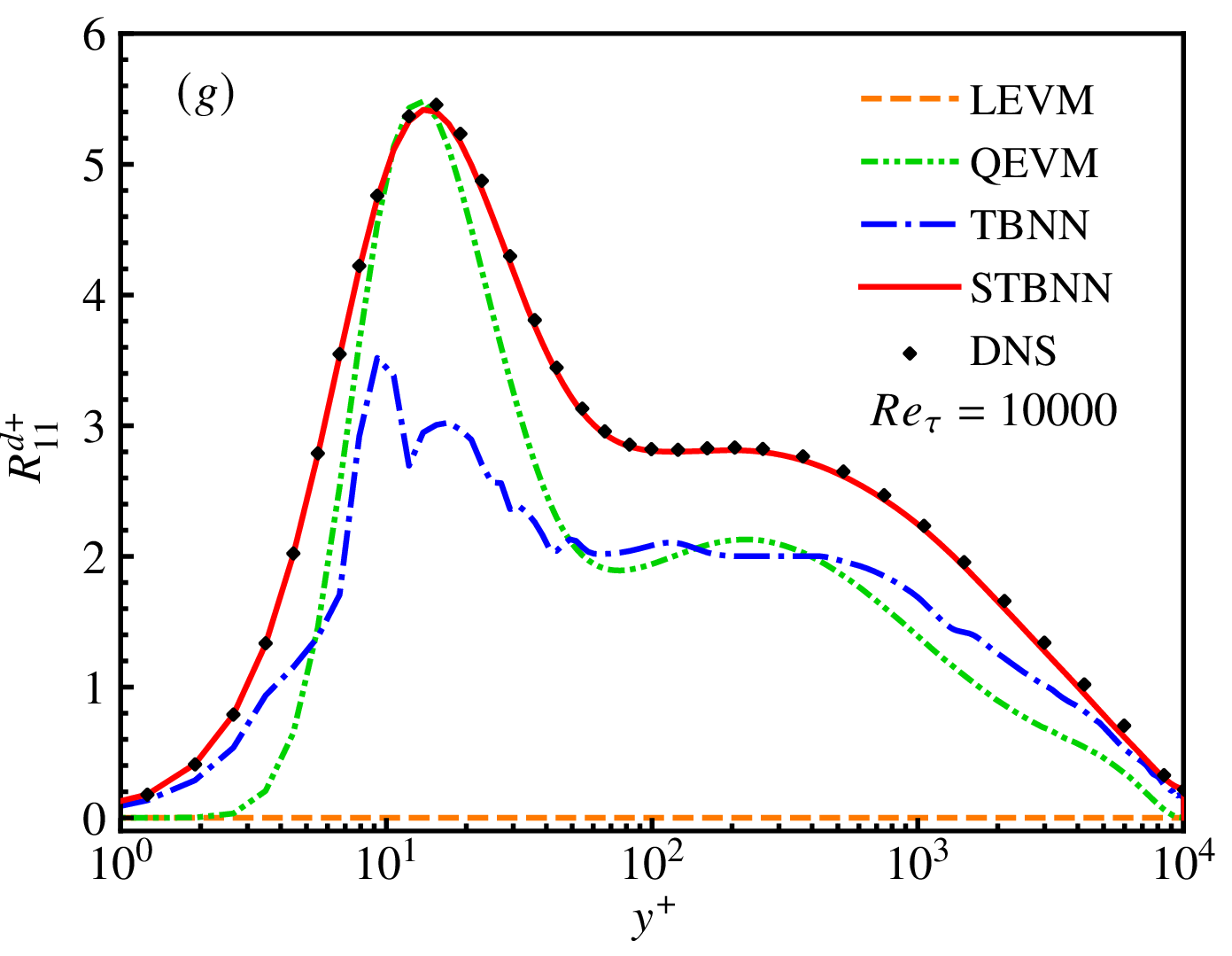}
	\end{subfigure}
	\begin{subfigure}{0.32\linewidth}
		\includegraphics[width=\linewidth]{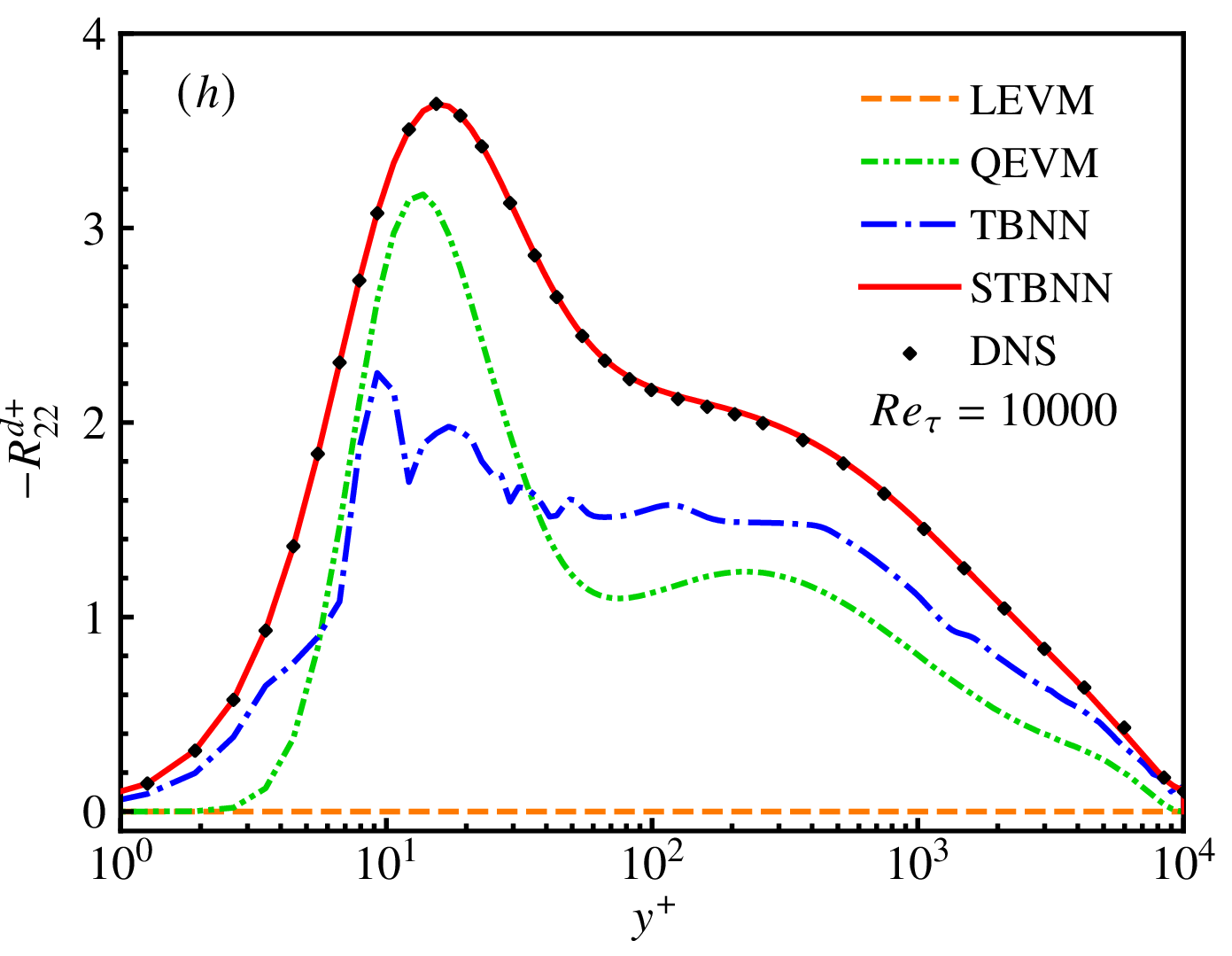}
	\end{subfigure}
	\begin{subfigure}{0.34\linewidth}
		\includegraphics[width=\linewidth]{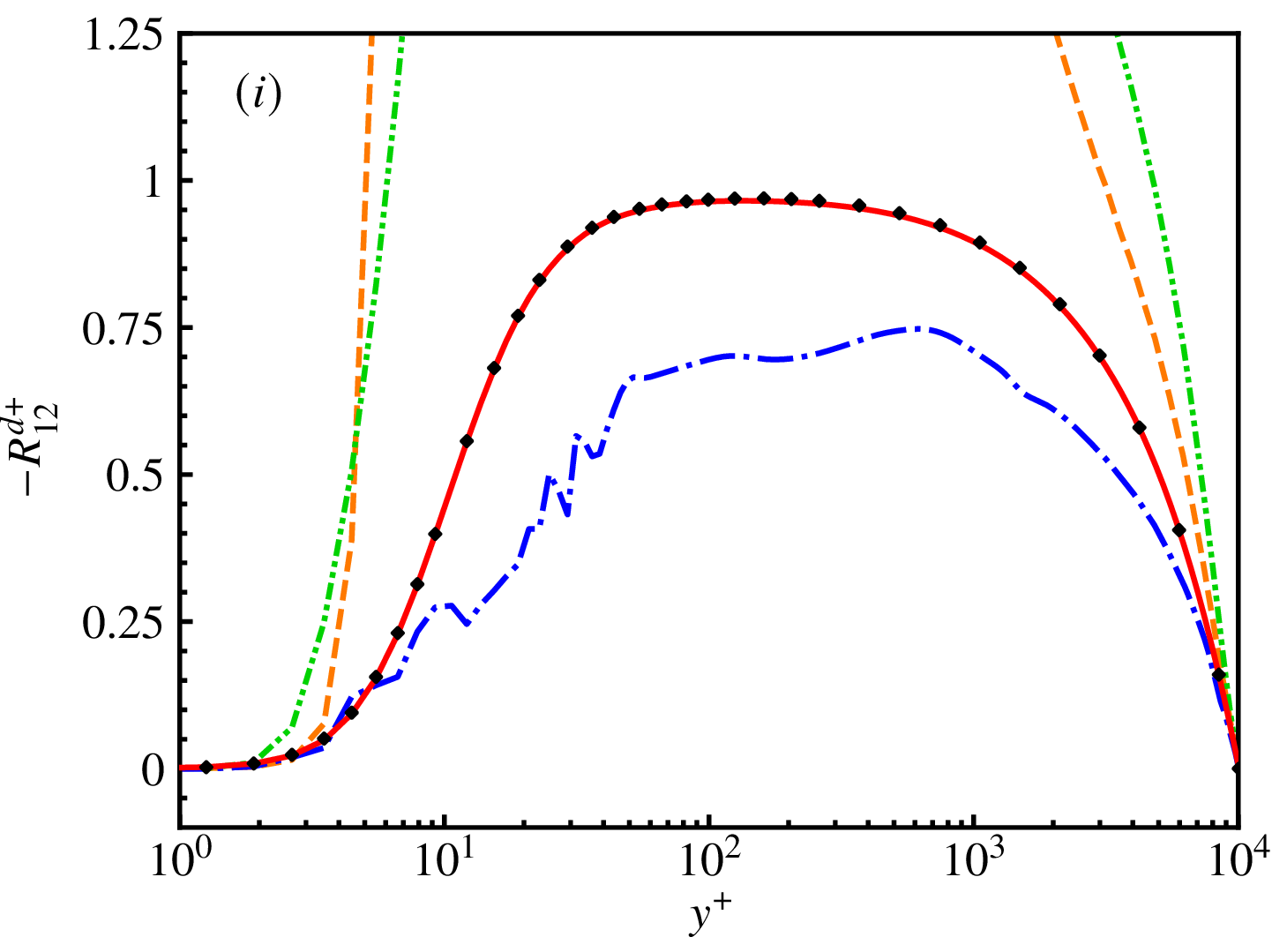}
	\end{subfigure}
	
	\caption{
		Modeled deviatoric Reynolds stress components in plane channel flow at untrained friction Reynolds numbers ${Re}_\tau$.
		The rows correspond to ${Re}_\tau=550$, 5200, and 10000 (top to bottom), while the columns show
		$R_{11}^d$, $R_{22}^d$, and $R_{12}^d$ (left to right).
	}
	\label{fig:5}
\end{figure}

To further examine the local predictive capability, Fig.~\ref{fig:5} presents the wall-normal distributions of the deviatoric Reynolds stress components in wall units $y^+=y u_\tau/\nu$ at untrained friction Reynolds numbers (${ Re}_\tau=550$, 5200 and 10000). Consistent with the statistical metrics, the STBNN predictions closely match the DNS profiles across the entire wall-normal range, accurately reproducing both the near-wall peak and the outer-layer decay for all stress components. In contrast, the QEVM underestimates the peak magnitude and misrepresents the shear stress distribution, while the TBNN exhibits noticeable deviations in the buffer and logarithmic regions, especially at higher Reynolds numbers. The LEVM fails to capture the anisotropic structure of turbulence altogether.
Importantly, the agreement between STBNN and DNS remains robust as $Re_\tau$ increases, indicating that the model preserves the correct wall scaling and outer-layer similarity simultaneously. This confirms that the STBNN can not only improve global statistical accuracy but also reconstruct the correct spatial distribution of Reynolds stress anisotropy, which is essential for reliable \emph{a posteriori} predictions.

\subsection {Periodic hill flows with varying geometries}
In contrast to channel flows, the periodic hill configuration introduces intense mean strain, strong streamline curvature, and large-scale separation, thus providing a substantially more stringent test for the turbulence closures. The \emph{a priori} predictive accuracy for periodic hill flows is summarized in Tables~\ref{tab:5} and~\ref{tab:6}. For the training geometries ($\alpha = 0.5, 1.0$, and $1.5$), the STBNN model yields correlation coefficients exceeding 99\% for all deviatoric stress components, with relative errors below 10\%. More importantly, on the unseen validation geometries ($\alpha = 0.8$ and 1.2), the correlation coefficients remain above 98\%, and the relative errors stay below 20\%, demonstrating robust generalization to untrained hill shapes.

\begin{table}[t]
	\centering
	\caption{Correlation coefficients of the deviatoric Reynolds stress for periodic hill flows.}
	\label{tab:5}
	\begin{tabular}{c l c c c c}
		\hline\hline
		${\rm{Case \slash C}}\left( {R_{ij}^d} \right)$  & Model & $R_{11}^d$ & $R_{22}^d$ & $R_{33}^d$ & $R_{12}^d$ \\
		\hline
		\multirow{4}{*}{%
			\begin{tabular}{c}
				Training set \\[3pt]
				$\alpha \in \left\{ \begin{array}{l}
					0.5,{\mkern 1mu} 1.0,{\mkern 1mu} \\
					1.5
				\end{array} \right\}$
		\end{tabular}}
		& LEVM  & -0.071  & 0.026  & 0       & 0.8683 \\
		& QEVM  & 0.5946  & 0.476  & 0.1949  & 0.922 \\
		& TBNN  & 0.9147  & 0.903  & 0.909   & 0.9574 \\
		& \textbf{STBNN} & \textbf{0.9927}    & \textbf{0.9904} & \textbf{0.9931} & \textbf{0.9917} \\
		\hline
		\multirow{4}{*}{%
			\begin{tabular}{c}
				Validation set \\[3pt]
				$\alpha \in \left\{ \begin{array}{l}
					0.8,{\mkern 1mu} 1.2 \\
				\end{array} \right\}$
		\end{tabular}}
		& LEVM  & -0.022 & 0.022  & 0      & 0.8669 \\
		& QEVM  & 0.623  & 0.4874 & 0.1823 & 0.9191 \\
		& TBNN  & 0.6472 & 0.7185 & 0.4608 & 0.9527 \\
		& \textbf{STBNN} & \textbf{0.9835} & \textbf{0.984} & \textbf{0.9815} & \textbf{0.991} \\
		\hline\hline
	\end{tabular}
\end{table}

\begin{table}[t]
	\centering
	\caption{Relative errors of the deviatoric Reynolds stress for periodic hill flows.}
	\label{tab:6}
	\begin{tabular}{c l c c c c}
		\hline\hline
		${\rm Case \slash Er}(R_{ij}^d)$ & Model & $R_{11}^d$ & $R_{22}^d$ & $R_{33}^d$ & $R_{12}^d$ \\
		\hline
		\multirow{4}{*}{%
			\begin{tabular}{c}
				Training set \\[3pt]
				$\alpha \in \left\{ \begin{array}{l}
					0.5,{\mkern 1mu} 1.0,{\mkern 1mu} \\
					1.5
				\end{array} \right\}$
		\end{tabular}}
		& LEVM  & 1.2072 & 1.2759 & 1 & 0.5939 \\
		& QEVM  & 0.7436 & 0.8853 & 0.9685 & 0.3218 \\
		& TBNN  & 0.2976 & 0.2788 & 0.4272 & 0.2417 \\
		& \textbf{STBNN} & \textbf{0.084} & \textbf{0.0898} & \textbf{0.1139} & \textbf{0.101} \\
		\hline
		\multirow{4}{*}{%
			\begin{tabular}{c}
				Validation set \\[3pt]
				$\alpha \in \left\{ \begin{array}{l}
					0.8,{\mkern 1mu} 1.2 \\
				\end{array} \right\}$
		\end{tabular}}
		& LEVM  & 1.2018 & 1.2787 & 1 & 0.5996 \\
		& QEVM  & 0.7304 & 0.8793 & 0.9853 & 0.3271 \\
		& TBNN  & 0.7719 & 0.5557 & 1.6881 & 0.255 \\
		& \textbf{STBNN} & \textbf{0.128} & \textbf{0.1122} & \textbf{0.1926} & \textbf{0.1047} \\
		\hline\hline
	\end{tabular}
\end{table}

In contrast, the LEVM fails to represent stress anisotropy in separated regions, leading to near-zero or even negative correlations for normal components ($R_{11}^d$, $R_{22}^d$ and $R_{33}^d$). The QEVM improves the normal stresses but still suffers from large errors in the shear stress. The TBNN performs better than both algebraic closures (LEVM and QEVM). However, its accuracy decreases noticeably for unseen geometries, especially in the $R_{33}^d$ component, suggesting limited extrapolation capability in complex separated flows. These results indicate that the self-scaling mechanism in STBNN provides a more consistent mapping between local velocity-gradient invariants and stress anisotropy across varying geometrical configurations.

\begin{figure}\centering
	\begin{subfigure}{0.5\textwidth}
		\centering
		{($a$)}
		\includegraphics[width=0.9\linewidth,valign=t]{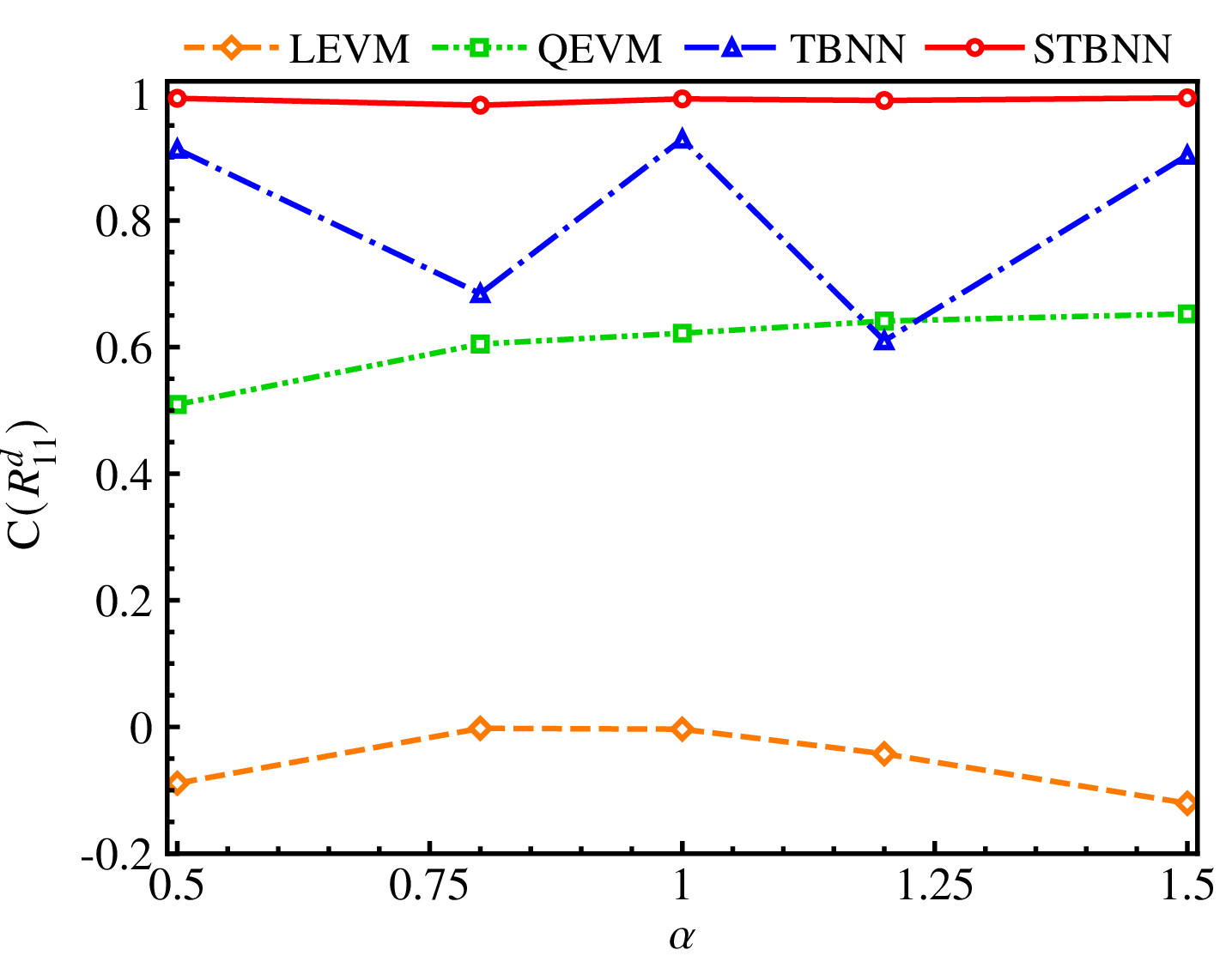}
	\end{subfigure}%
	\begin{subfigure}{0.5\textwidth}
		\centering
		{($b$)}
		\includegraphics[width=0.9\linewidth,valign=t]{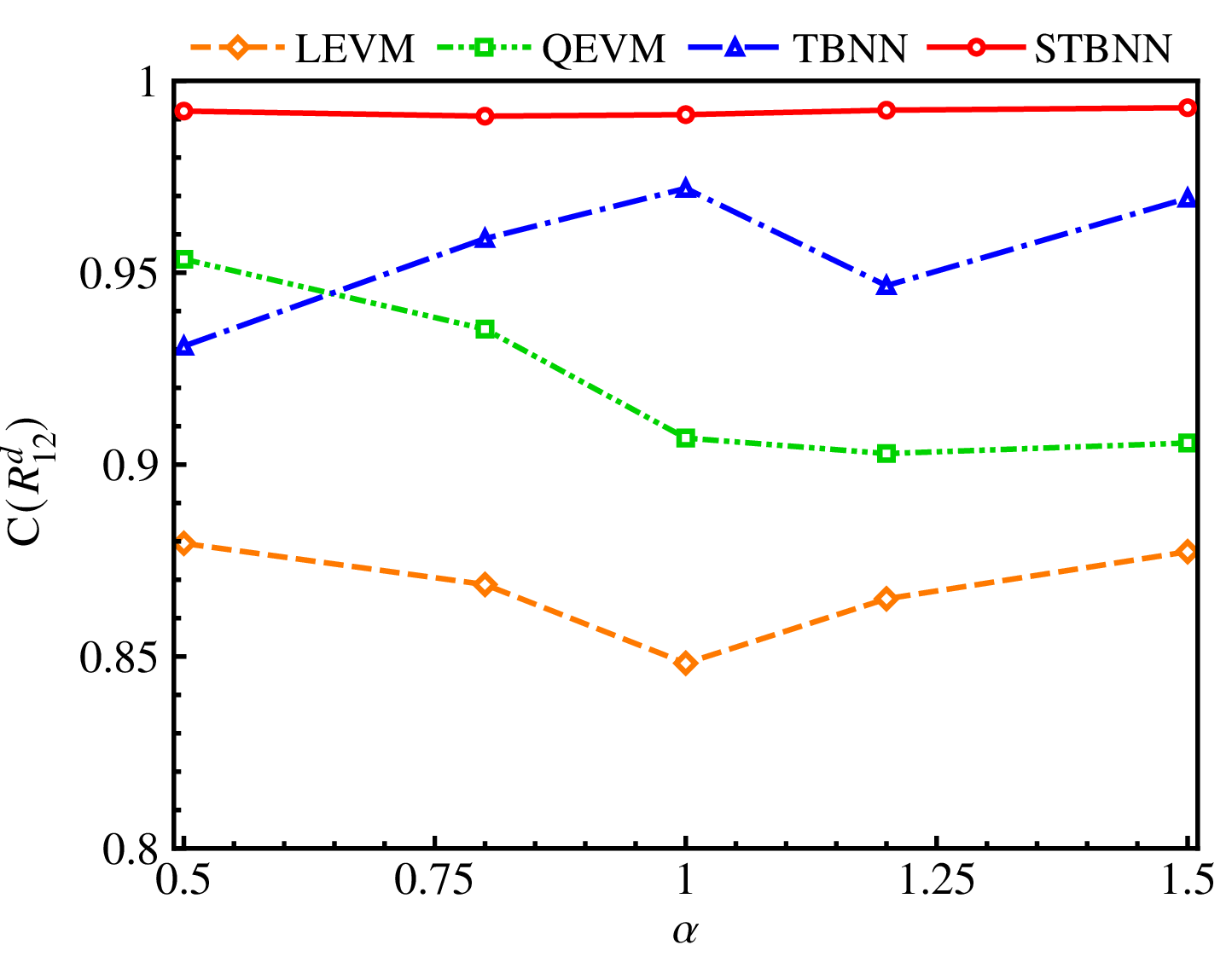}
	\end{subfigure}
	\begin{subfigure}{0.5\textwidth}
		\centering
		{($c$)}
		\includegraphics[width=0.9\linewidth,valign=t]{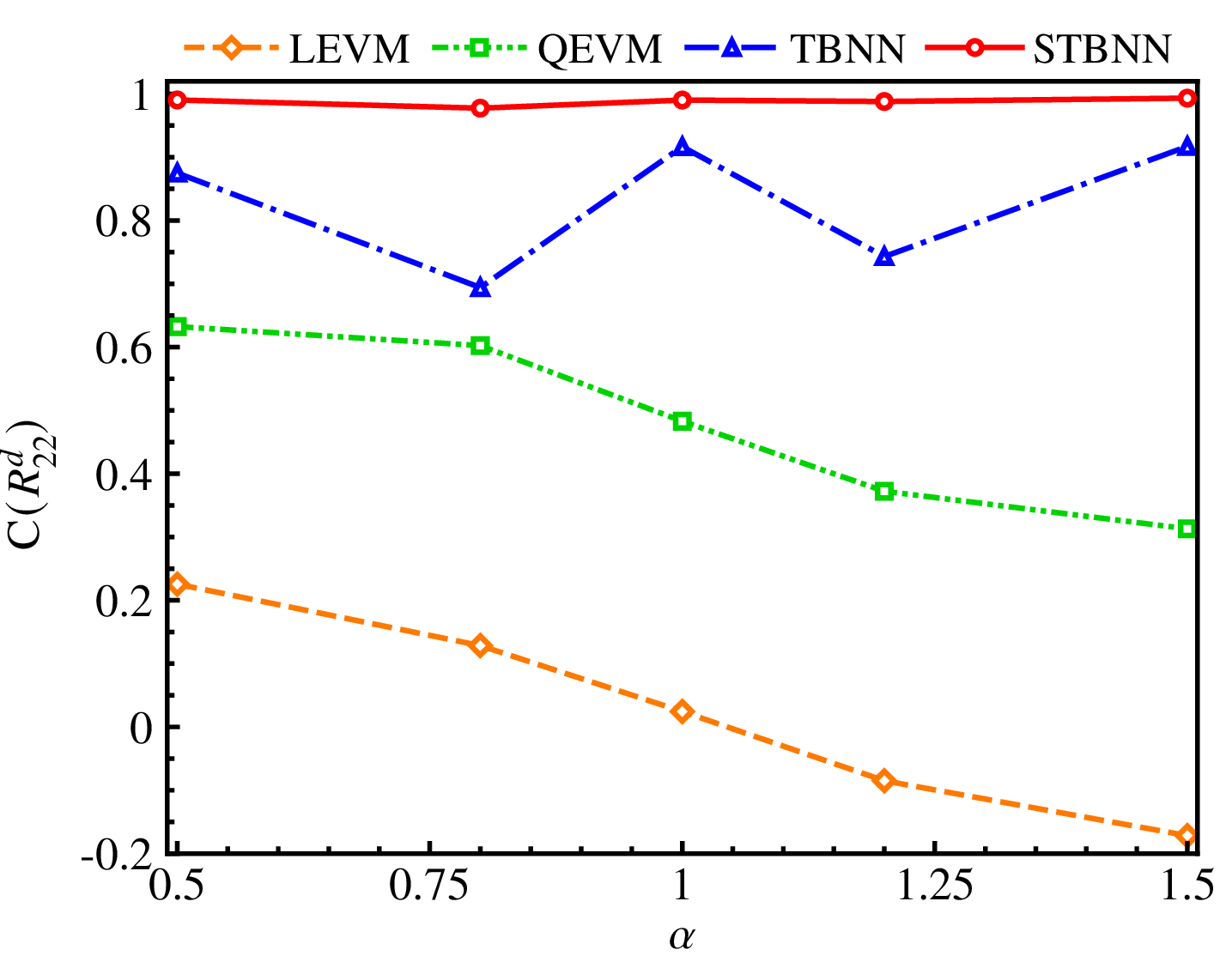}
	\end{subfigure}%
	\begin{subfigure}{0.5\textwidth}
		\centering
		{($d$)}
		\includegraphics[width=0.9\linewidth,valign=t]{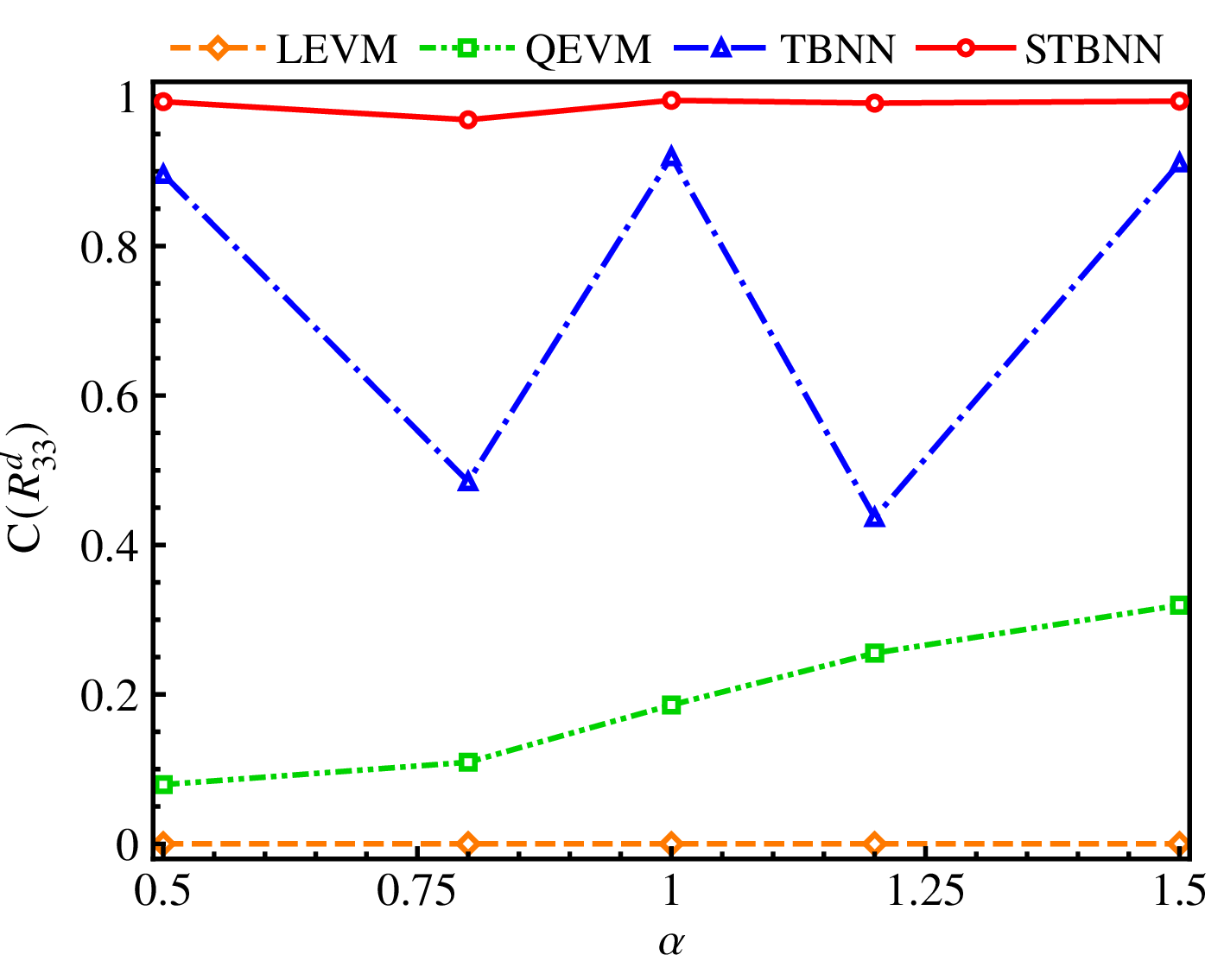}
	\end{subfigure}
	\caption{Correlation coefficients of modeled deviatoric Reynolds stress components in periodic hill flows across steepness ($\alpha$): (a) $R_{11}^d$; (b) $R_{12}^d$; (c) $R_{22}^d$; (d) $R_{33}^d$.}
	
	\label{fig:6}
\end{figure}

\begin{figure}\centering
	\begin{subfigure}{0.5\textwidth}
		\centering
		{($a$)}
		\includegraphics[width=0.9\linewidth,valign=t]{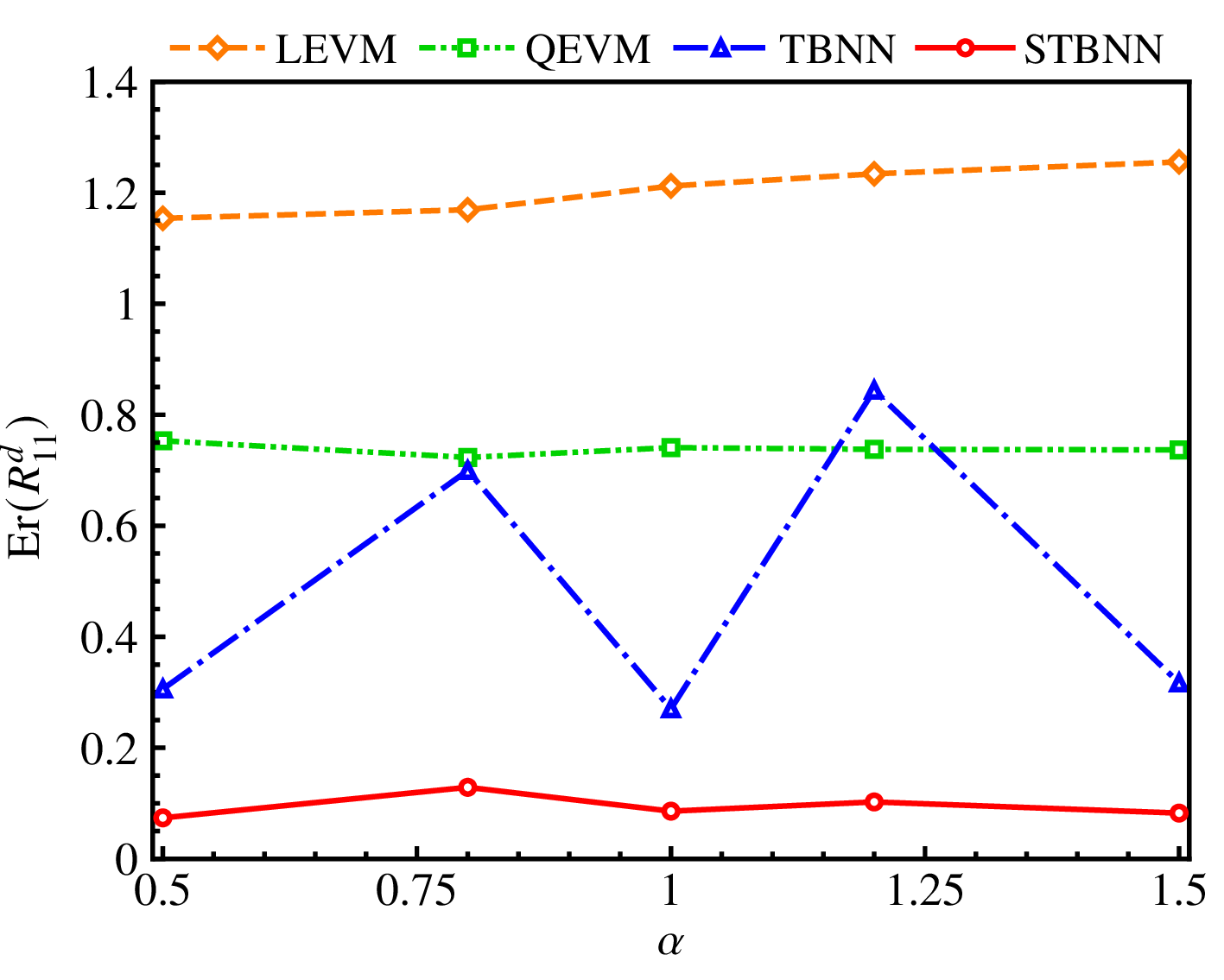}
	\end{subfigure}%
	\begin{subfigure}{0.5\textwidth}
		\centering
		{($b$)}
		\includegraphics[width=0.9\linewidth,valign=t]{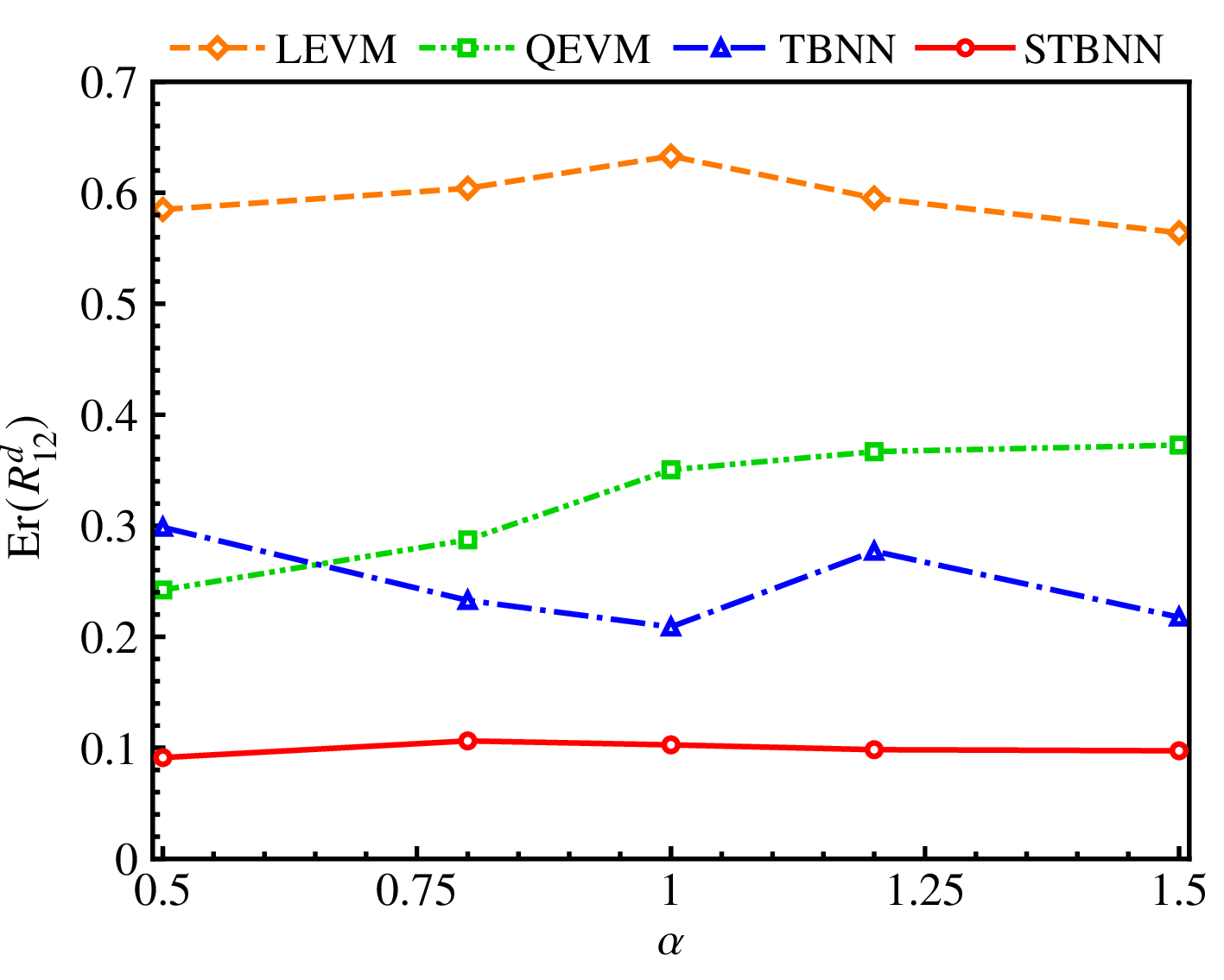}
	\end{subfigure}
	\begin{subfigure}{0.5\textwidth}
		\centering
		{($c$)}
		\includegraphics[width=0.9\linewidth,valign=t]{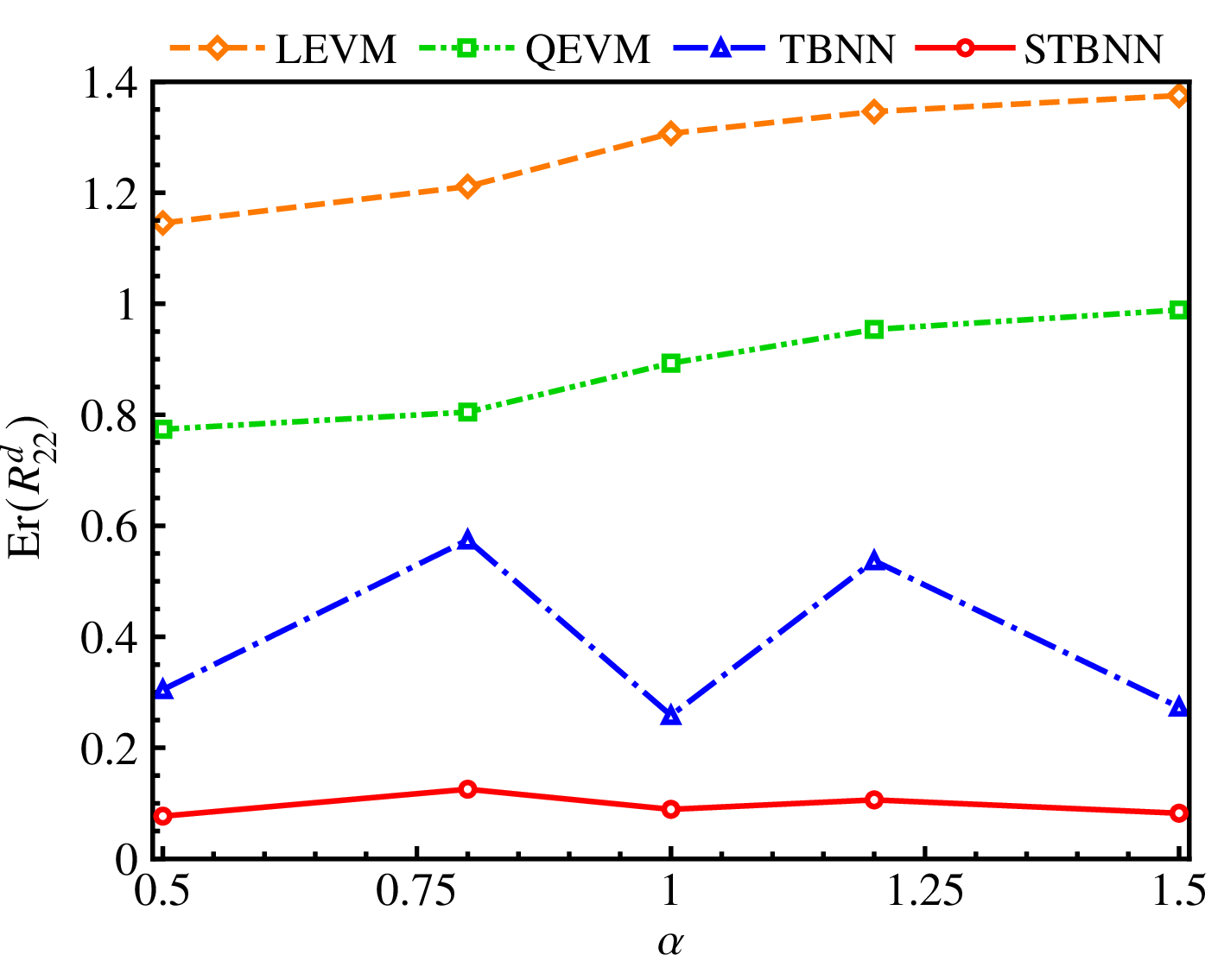}
	\end{subfigure}%
	\begin{subfigure}{0.5\textwidth}
		\centering
		{($d$)}
		\includegraphics[width=0.9\linewidth,valign=t]{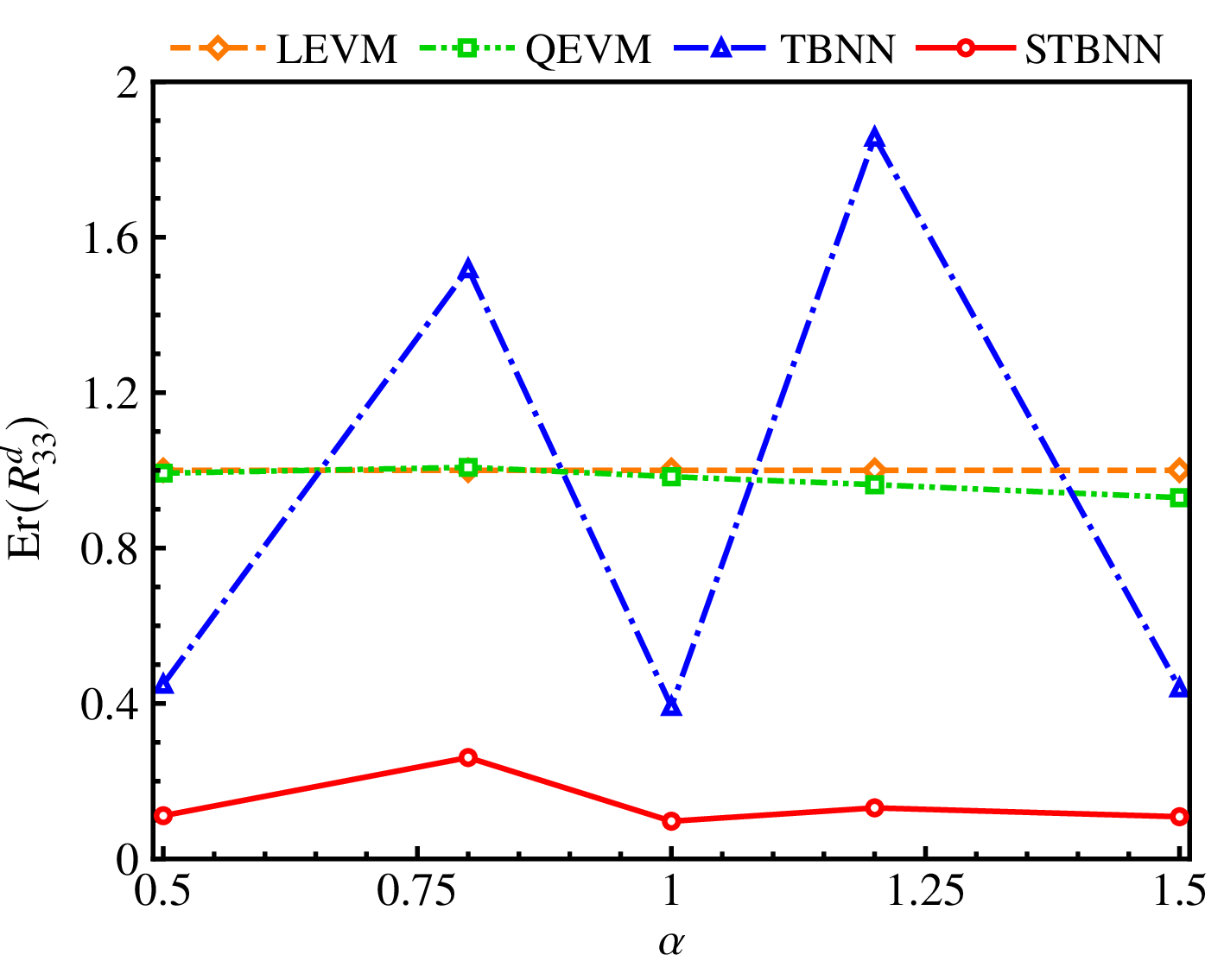}
	\end{subfigure}
	\caption{Relative errors of modeled deviatoric Reynolds stress components in periodic hill flows across across steepness ($\alpha$): (a) $R_{11}^d$; (b) $R_{12}^d$; (c) $R_{22}^d$; (d) $R_{33}^d$.}
	\label{fig:7}
\end{figure}

The variations of correlation coefficients and relative errors with hill steepness are shown in Figs.~\ref{fig:6} and~\ref{fig:7}. The STBNN model exhibits consistently high correlations and low relative errors over the tested range of $\alpha$, indicating reduced sensitivity to geometric variation within the present periodic-hill dataset.

In contrast, the algebraic closures (LEVM and QEVM) show pronounced sensitivity to geometric variations, resulting in degraded performance as the flow regime changes.
The TBNN performs again better than both LEVM and QEVM models. However, its correlations fluctuate noticeably and its errors increase for unseen geometries, indicating limited generalization capability when the flow regime departs from the training configurations.

\begin{figure}
	\centering
	
	\begin{subfigure}{0.45\linewidth}
		\includegraphics[width=\linewidth]{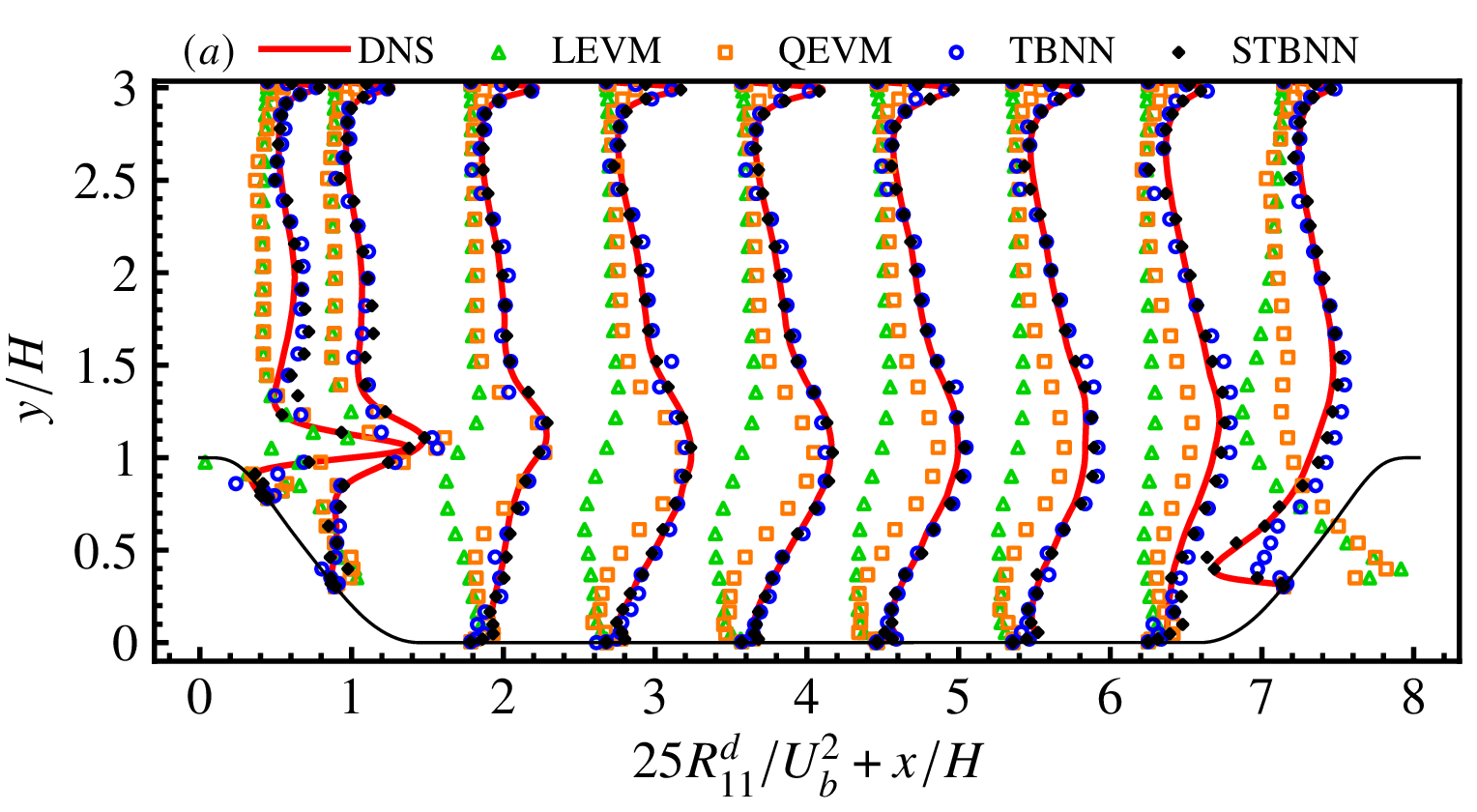}
	\end{subfigure}
	\begin{subfigure}{0.45\linewidth}
		\includegraphics[width=\linewidth]{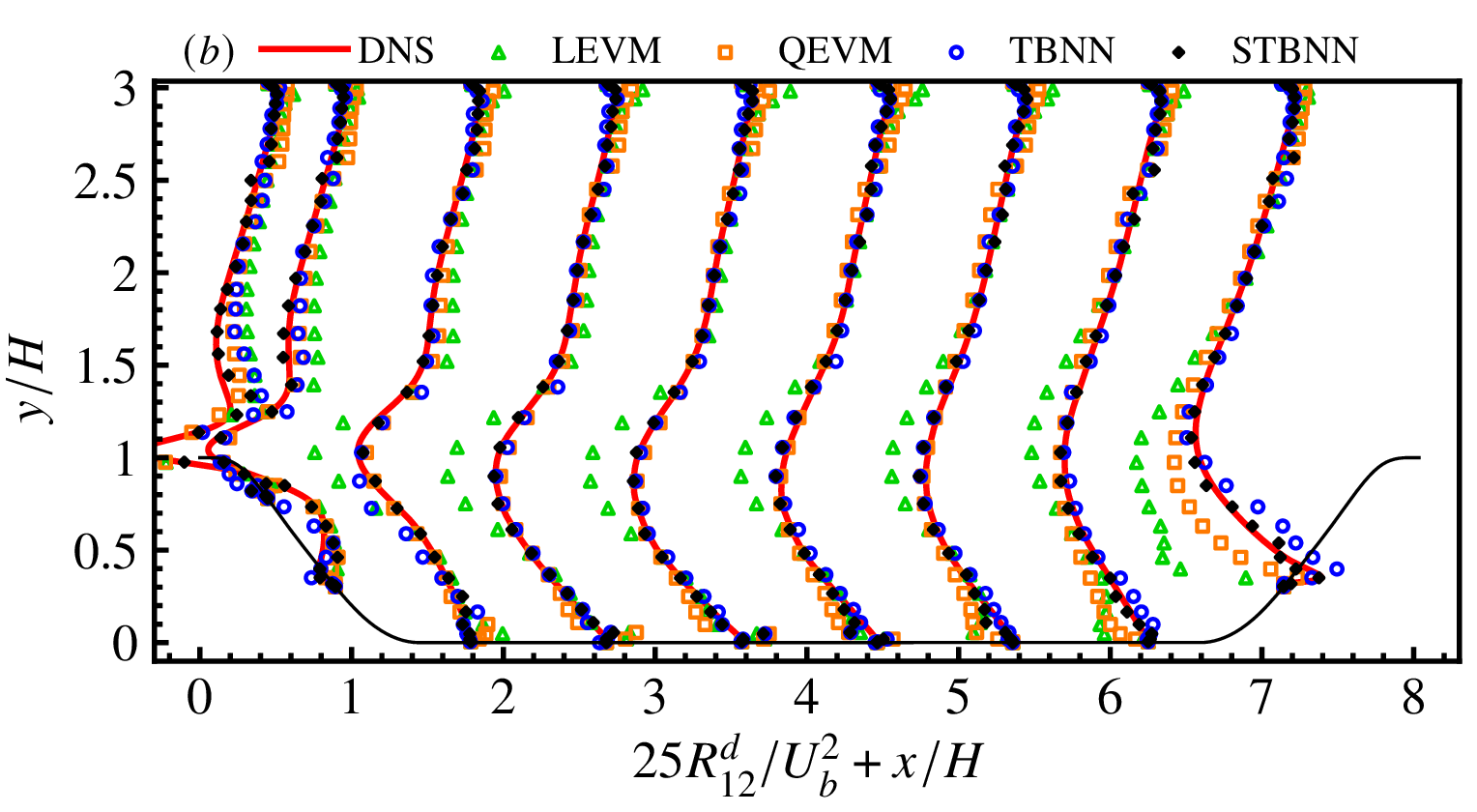}
	\end{subfigure}
	
	\begin{subfigure}{0.45\linewidth}
		\includegraphics[width=\linewidth]{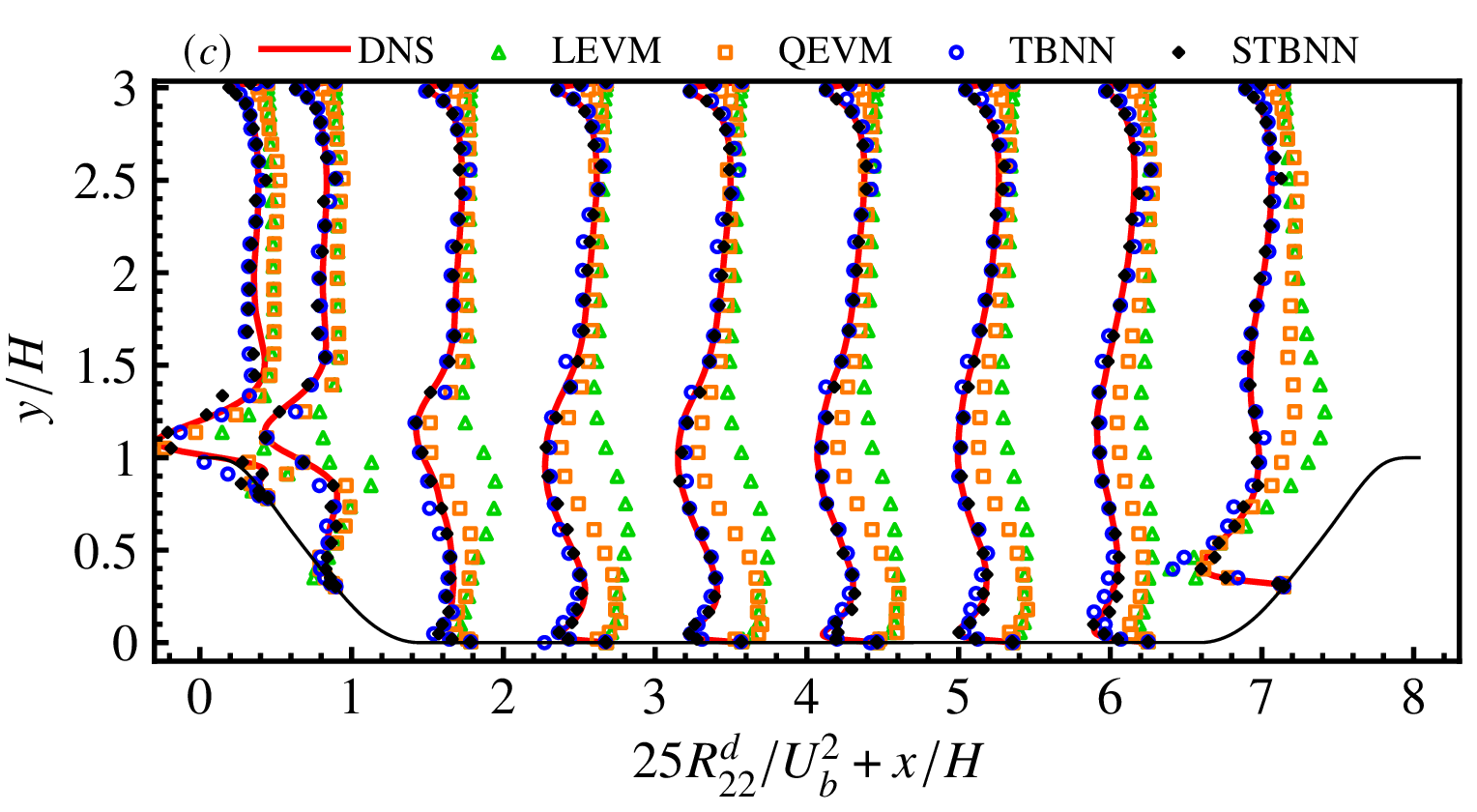}
	\end{subfigure}
	\begin{subfigure}{0.45\linewidth}
		\includegraphics[width=\linewidth]{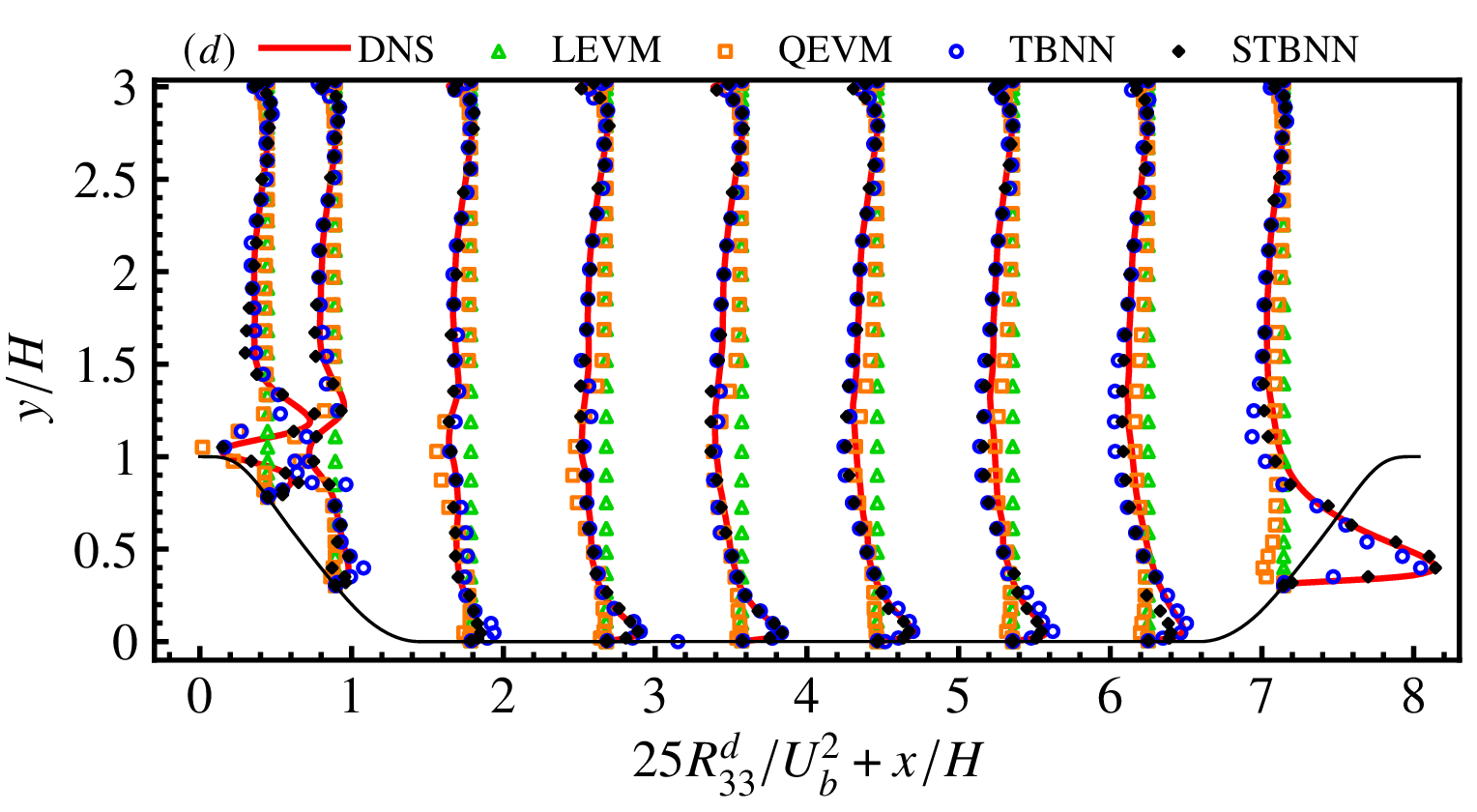}
	\end{subfigure}
	
	\caption{
		Profiles of deviatoric Reynolds stress components in periodic hill flows at untrained steepness ${\alpha}=0.8$:  (a) $R_{11}^d$; (b) $R_{12}^d$; (c) $R_{22}^d$; (d) $R_{33}^d$.
	}
	\label{fig:8}
\end{figure}

The simultaneous stability of both correlation and error metrics suggests that the STBNN captures the relation between local strain-rotation balance and Reynolds-stress anisotropy more consistently than the baseline models within the tested periodic-hill family.
\begin{figure}
	\centering
	
	\begin{subfigure}{0.45\linewidth}
		\includegraphics[width=\linewidth]{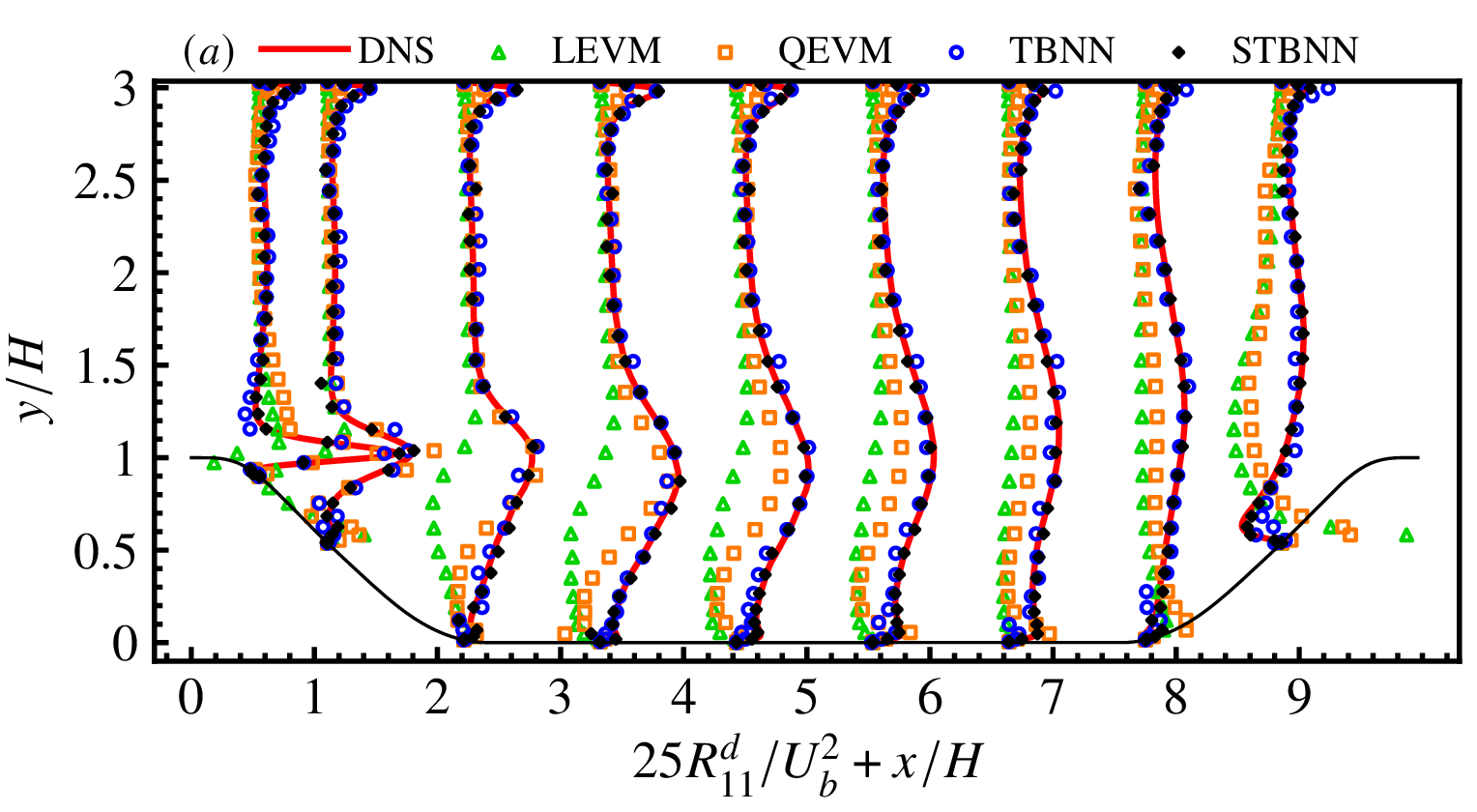}
	\end{subfigure}
	\begin{subfigure}{0.45\linewidth}
		\includegraphics[width=\linewidth]{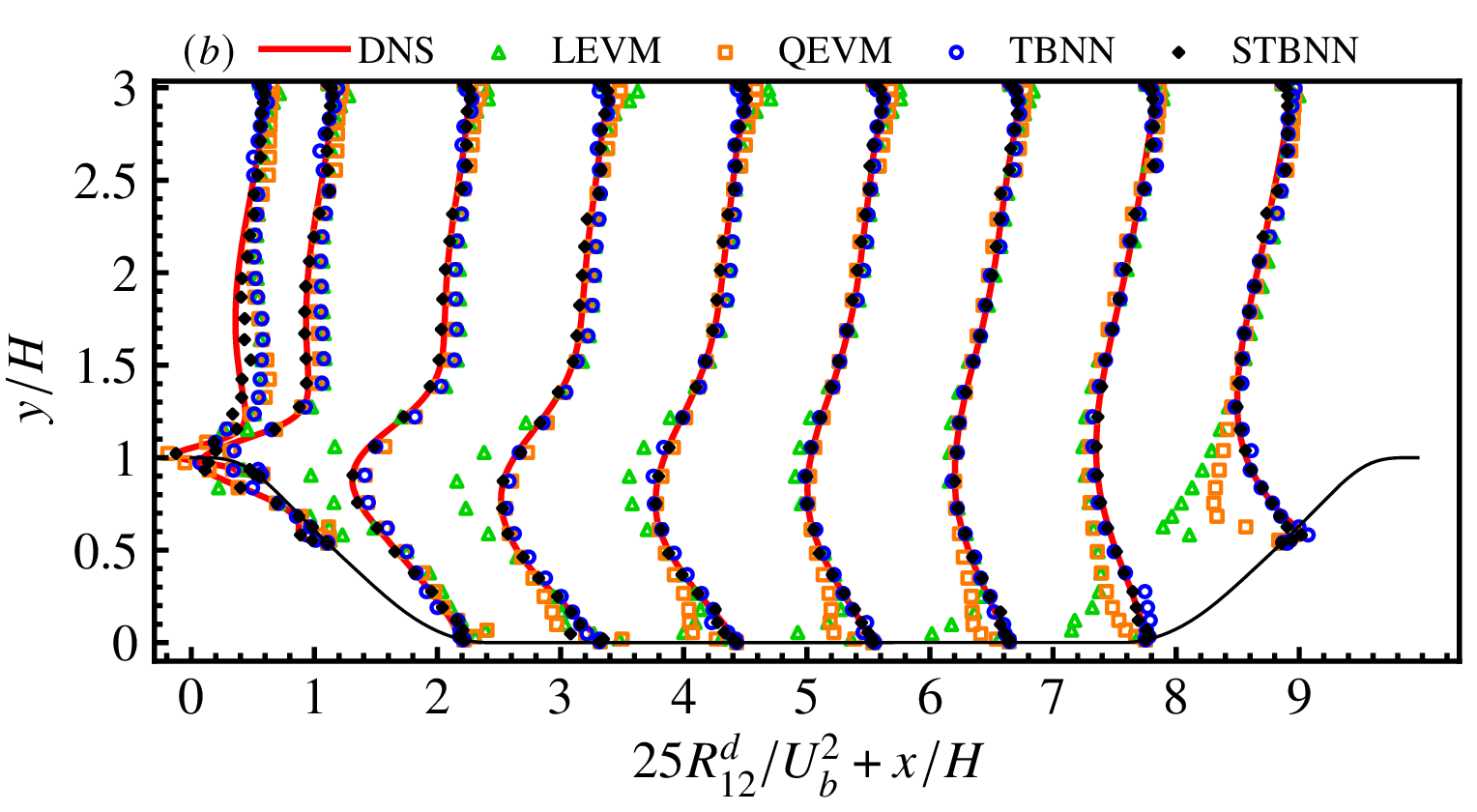}
	\end{subfigure}
	
	\begin{subfigure}{0.45\linewidth}
		\includegraphics[width=\linewidth]{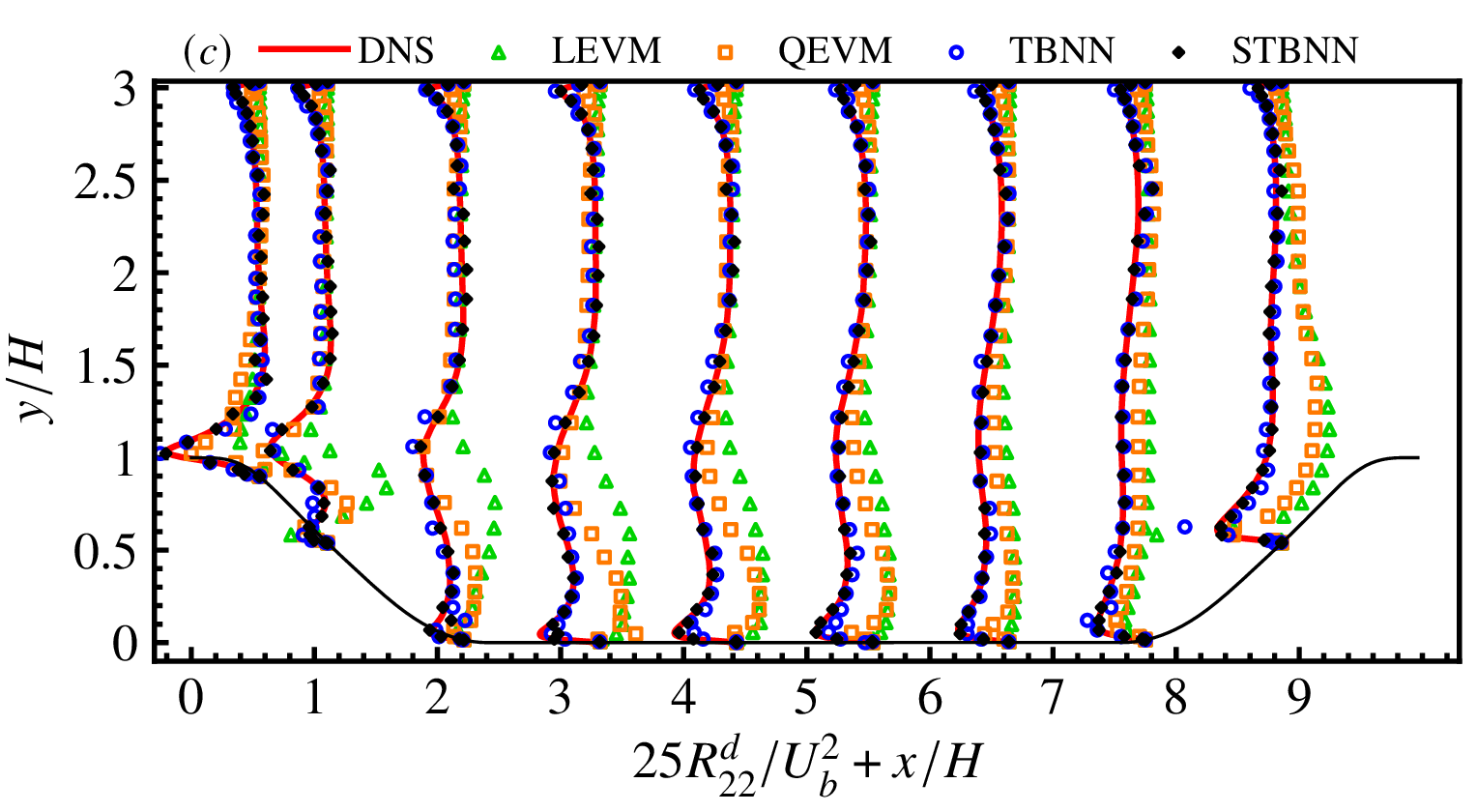}
	\end{subfigure}
	\begin{subfigure}{0.45\linewidth}
		\includegraphics[width=\linewidth]{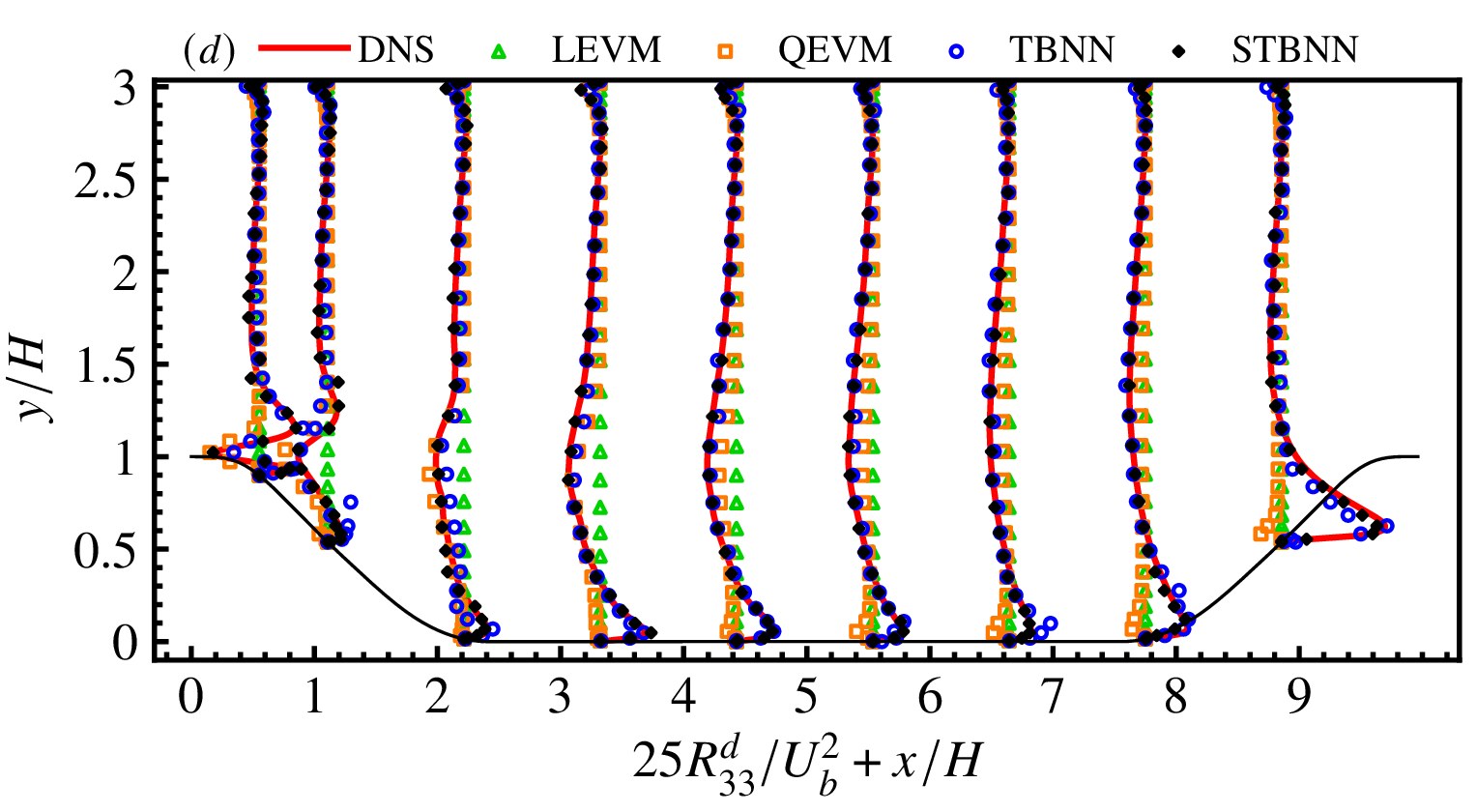}
	\end{subfigure}
	
	\caption{
		Profiles of deviatoric Reynolds stress components in periodic hill flows at untrained steepness ${\alpha}=1.2$:  (a) $R_{11}^d$; (b) $R_{12}^d$; (c) $R_{22}^d$; (d) $R_{33}^d$.
	}
	\label{fig:9}
\end{figure}

\begin{figure}\centering
	\includegraphics[width=0.9\textwidth]{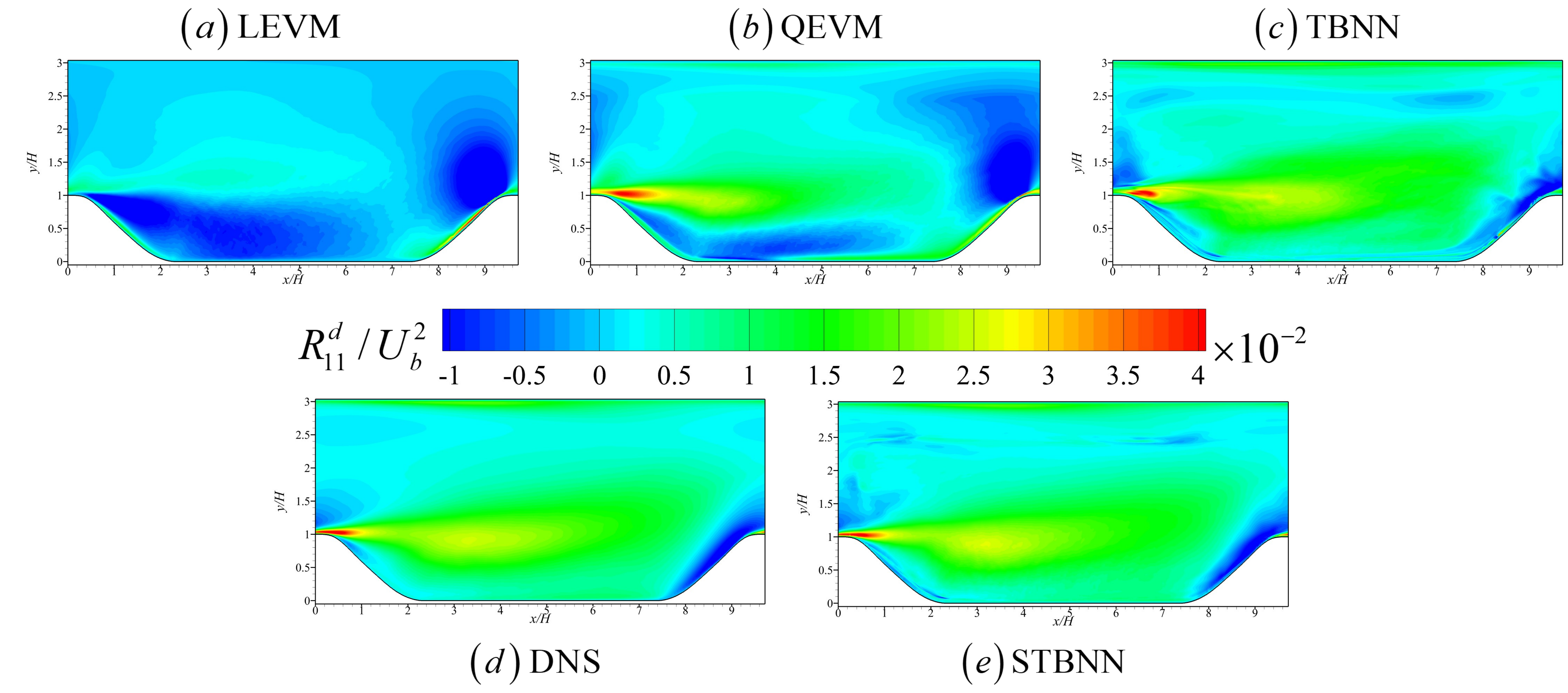}
	\caption{Contours of the normal Reynolds stress component at untrained hill steepness $\alpha=1.2$: (a) LEVM; (b) QEVM; (c) TBNN; (d) DNS; (e) STBNN.}\label{fig:10}
\end{figure}

\begin{figure}\centering
	\includegraphics[width=0.9\textwidth]{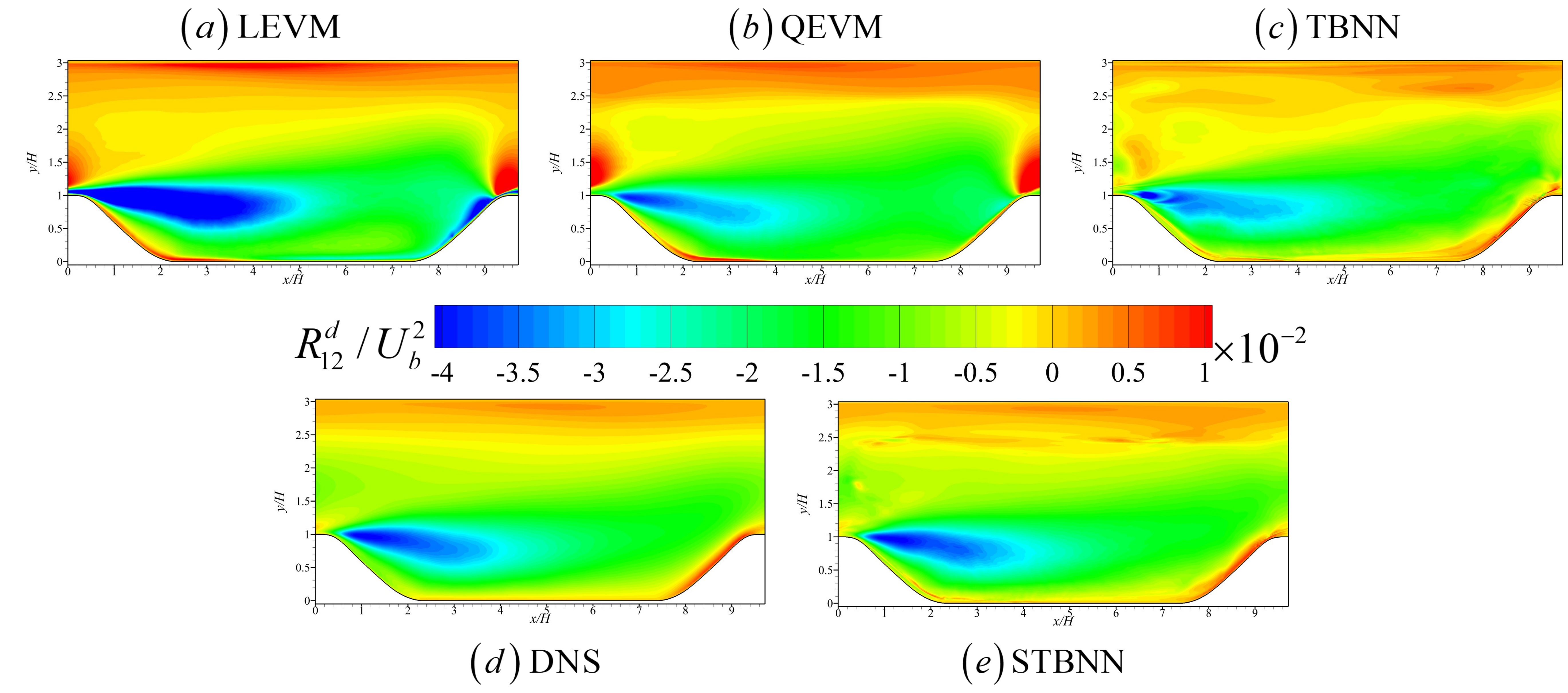}
	\caption{Contours of the shear Reynolds stress component at untrained hill steepness $\alpha=1.2$: (a) LEVM; (b) QEVM; (c) TBNN; (d) DNS; (e) STBNN.}\label{fig:11}
\end{figure}

The predicted Reynolds-stress profiles are shown at streamwise locations
$x/L_x=s/L_{x0}$ with $s \in \{0.5,1,2,3,4,5,6,7,8\}$ based on the reference geometry
$L_{x0}=L_x(\alpha=1)=9$ for the untrained hill steepness
$\alpha=0.8$ and $1.2$ in Figs.~\ref{fig:8} and~\ref{fig:9}. The STBNN predictions closely follow the DNS distributions over the entire domain, including the upstream acceleration region, the separation bubble, the post-separation shear layer and the downstream recovery region. In particular, the model accurately reproduces both the magnitude and the spatial variation of the Reynolds-stress peaks.

The LEVM and QEVM closures exhibit substantial discrepancies in the separated flow. The LEVM severely underpredicts the anisotropy inside the recirculation region and fails to capture the stress redistribution near reattachment, while the QEVM produces overly diffused stress profiles in both the separated shear layer and the recovery region.
The TBNN captures the general trends of the stresses but shows noticeable deviations in the reattachment region near the outlet.

Overall, the close agreement between STBNN and DNS in both separated and recovery regions demonstrates its improved capability to represent strongly non-equilibrium turbulence, whereas the degradation of TBNN on the unseen geometries indicates limited extrapolation ability in complex separated flows.

The contours of the Reynolds stress components for the untrained hill geometry are illustrated in Figs.~\ref{fig:10} and~\ref{fig:11}. The results for $\alpha=0.8$ exhibit similar behavior and are omitted for brevity (see Figs.~\ref{fig:8} and~\ref{fig:9}). The STBNN accurately captures the structure of the separation and recirculation regions. In contrast, LEVM fails to reconstruct the separation zone and yields an overly diffused stress field. QEVM excessively smooths the shear-layer gradients and shifts the stress peak. The TBNN recovers the overall pattern but shows noticeable distortions near the downstream reattachment region, where the stress redistribution is misrepresented. These observations indicate that the STBNN correctly reconstructs the geometry-induced stress topology rather than merely matching local magnitudes, leading to substantially improved predictions in strongly separated flows.

\section {\emph{A posteriori} studies of the STBNN models}\label{sec:level5}
In this section, we perform \emph{a posteriori} analyses to further evaluate the proposed STBNN model in comparison with LEVM, QEVM, and TBNN. The turbulence models are embedded into RANS computations for channel flows at different friction Reynolds numbers and periodic hill flows with varying hill steepness. All \emph{a posteriori} simulations were performed in the open-source solver OpenFOAM v2312 by solving the steady incompressible RANS equations within a finite-volume framework using the semi-implicit method for pressure-linked equations (SIMPLE).\cite{weller1998tensorial} The pressure equation was solved using GAMG, whereas the momentum and turbulence transport equations were solved using PBiCGStab with DILU preconditioning. Gradient terms were discretized using the Gauss linear scheme, diffusive terms using the Gauss linear corrected scheme, and convective terms using a second-order linear-upwind scheme. The residual tolerances were set to $10^{-8}$ for the velocity and turbulence variables and $10^{-6}$ for the pressure. The near-wall mesh resolution of RANS is sufficient to resolve the viscous sublayer, with the first-cell height satisfying $d_w^+<1$ for all cases. For the LEVM and QEVM models, the $k-\varepsilon$ transport equations are solved with low-Reynolds-number wall treatments.\cite{jones1972prediction,launder1974application} For the data-driven closures (TBNN and STBNN), no additional empirical wall modelling is introduced. A $k$-corrective-frozen strategy is adapted,\cite{schmelzer2020discovery, Myklebust2025UnsteadyTBNN} in which the turbulent kinetic energy $k$ and turbulence dissipation $\varepsilon$ is provided by the baseline $k-\varepsilon$ linear eddy-viscosity model.\cite{launder1974application} In this manner, the learned model only replaces the stress anisotropy, ensuring a consistent and closed RANS formulation. Additional details of the grid-convergence assessment are given in Appendix~\ref{sec:appendixC}.

Prior studies have shown that the explicit modeling of Reynolds stresses in the RANS momentum equations often leads to ill-posedness.\cite{wu2019reynolds,brener2021conditioning,brener2025mitigating} In this paper, the modeled deviatoric Reynolds stress is decomposed into linear and nonlinear parts,\cite{wu2018physics,McConkey2025Realisability}
\begin{equation}
R_{ij}^d =  - 2\nu _t^L{S_{ij}} + R_{ij}^ \bot ,\;\;{\rm{where}}\;\;\nu _t^L =  - \frac{1}{2}\frac{{R_{ij}^d{S_{ij}}}}{{{S_{kl}}{S_{kl}}}}.
\label{Rij_decompose}
\end{equation}
The linear part is treated implicitly together with the viscous term, while the nonlinear contribution ($R_{ij}^ \bot $) is incorporated into the RANS momentum equation through its solenoidal component obtained via a Helmholtz–Hodge decomposition.\cite{luther2025NonLinear} Denoting the divergence-free projection by $f_i^{\perp}$,
\begin{equation}
\frac{{\partial R_{ij}^ \bot }}{{\partial {x_j}}} = f_i^ \bot ,\;\;{\rm{and}}\;\;\frac{{\partial f_i^ \bot }}{{\partial {x_i}}} = 0,
\label{div_Rij}
\end{equation}
with the irrotational component absorbed into a modified pressure. This treatment maintains a stable pressure–velocity coupling and avoids the numerical instability commonly observed in explicit Reynolds-stress closures. The average computational cost per iteration in RANS calculations is summarized in Table~\ref{tab:7}. TBNN and STBNN have a computational cost close to that of QEVM, and are about 1.3--1.5 times more expensive than LEVM for both the plane channel and periodic hill flows.

\begin{table}[tbp]
	\begin{center}
		\caption{The average computational cost per iteration for different turbulence models in RANS calculations solved by OpenFOAM.}
		\label{tab:7}%	
			\begin{tabular*}{0.95\textwidth}{@{\extracolsep{\fill}} lcccc}
				\hline\hline
				\small 	
				Model (Plane channel) & LEVM & QEVM & TBNN & STBNN \\ \hline
				t(CPU$\cdot$s)    & 0.154 & 0.222  & 0.202 & 0.228 \\
				t/t$_{\rm{LEVM}}$ & 1 & 1.44 & 1.312 & 1.481 \\ \hline
				Model (Periodic hill) & LEVM & QEVM & TBNN & STBNN \\ \hline
				t(CPU$\cdot$s)     & 0.053 & 0.072 & 0.0718 & 0.0729 \\
				t/t$_{\rm{LEVM}}$  & 1 & 1.357 & 1.354 & 1.374 \\
				\hline\hline
			\end{tabular*}%
	\end{center}
\end{table}%

\subsection {Plane channel flows}

\begin{figure}
	\centering
	
	\begin{subfigure}{0.32\linewidth}
		\includegraphics[width=\linewidth]{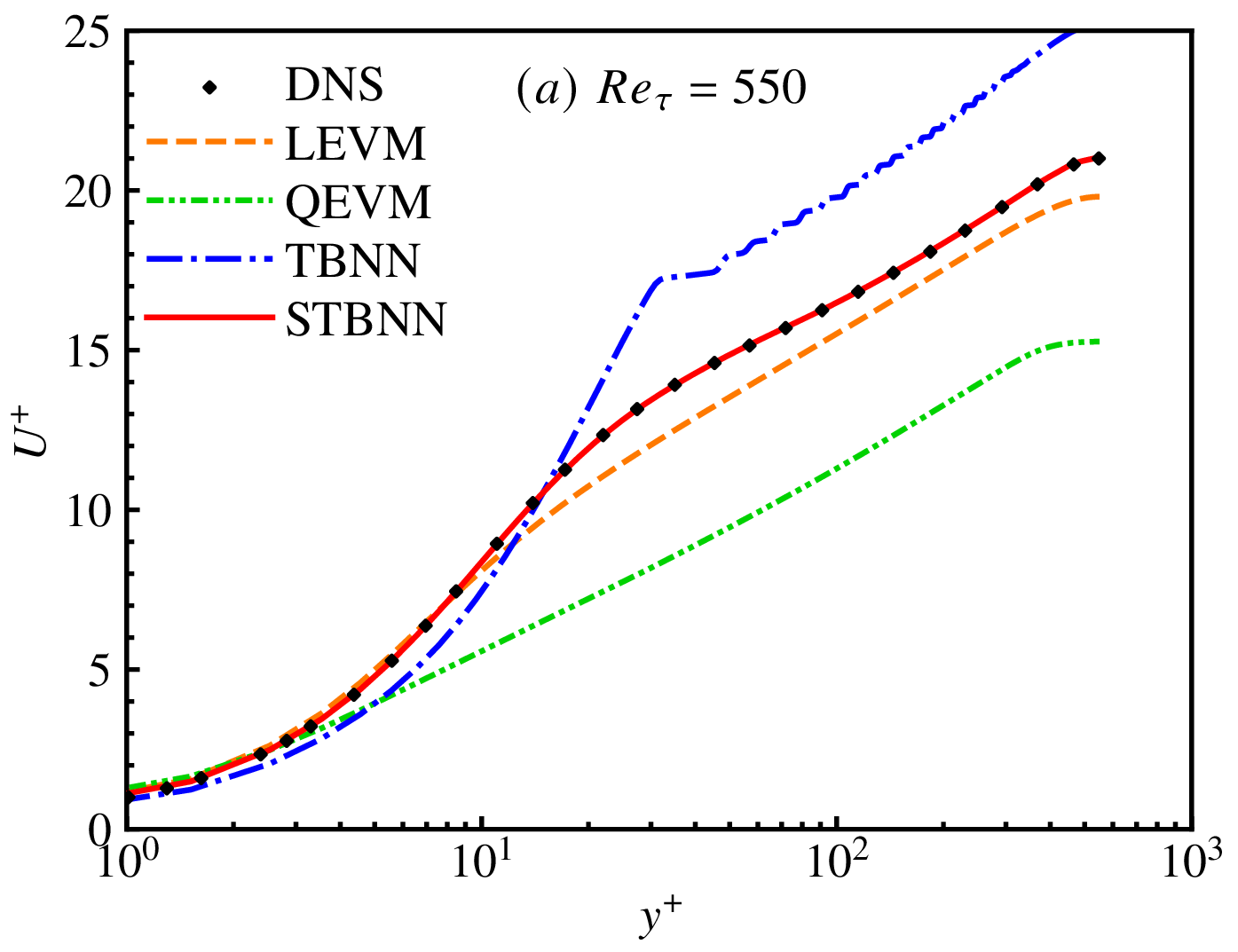}
	\end{subfigure}
	\begin{subfigure}{0.32\linewidth}
		\includegraphics[width=\linewidth]{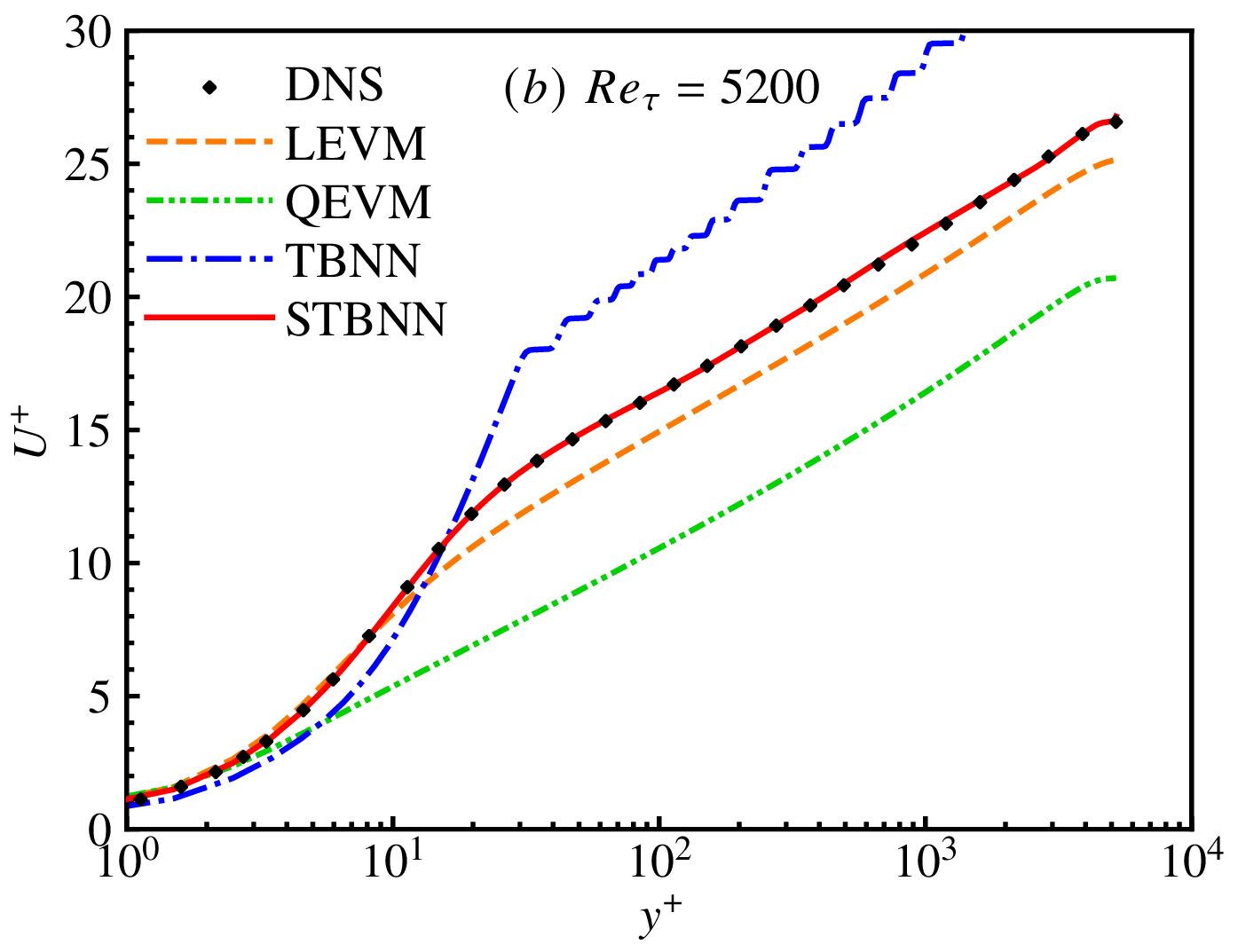}
	\end{subfigure}
	\begin{subfigure}{0.32\linewidth}
		\includegraphics[width=\linewidth]{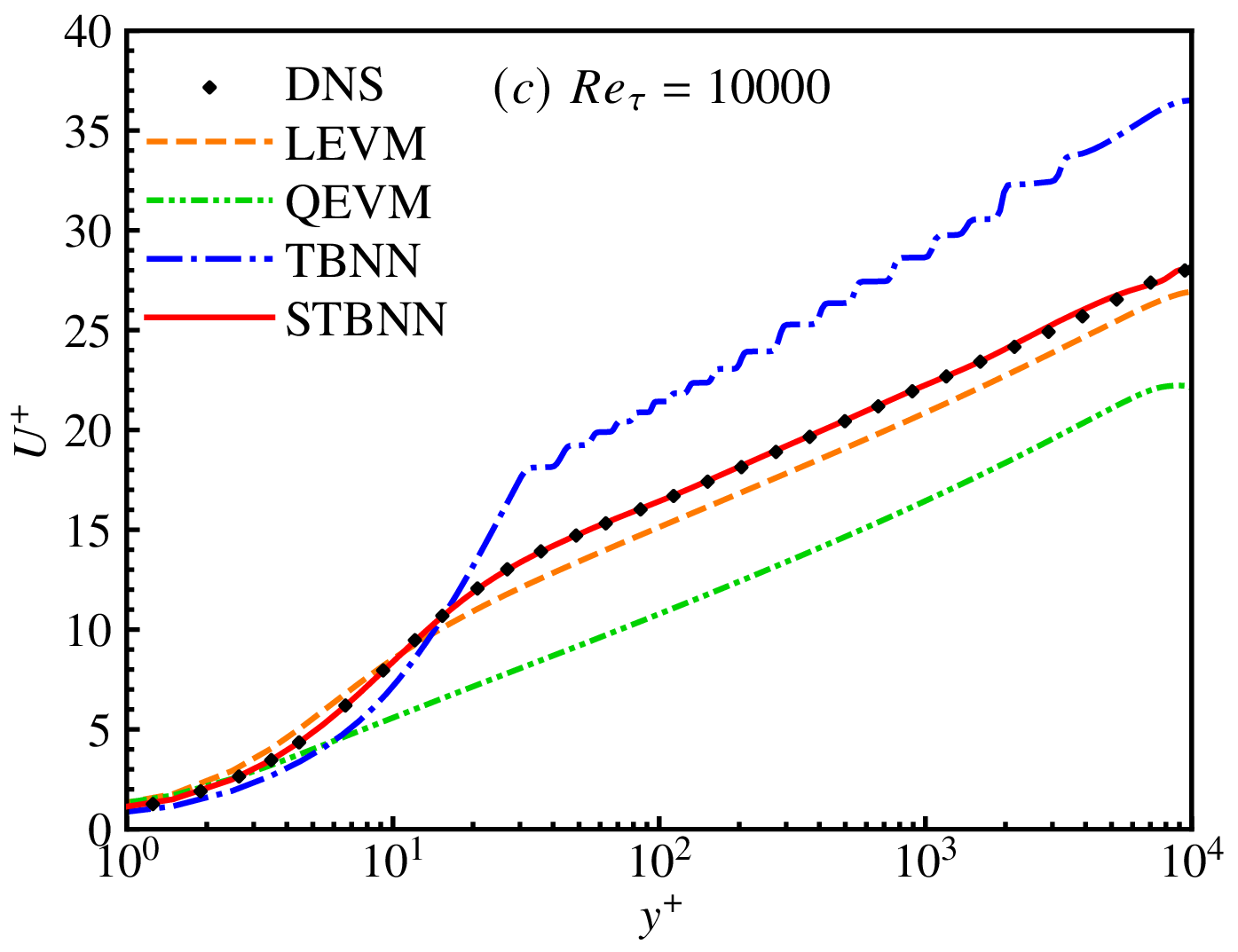}
	\end{subfigure}
	
	\caption{
		Streamwise velocity profiles calculated by different turbulence models at untrained friction Reynolds numbers ${Re}_\tau$:
		(a) ${Re}_\tau = 550$, (b) $5200$, and (c) $10000$.
	}
	\label{fig:12}
\end{figure}

The \emph{a posteriori} performance of the models is first examined in plane channel flows by comparing the mean streamwise velocity profiles normalized by the friction velocity ${U^ + } = {{{\bar u}_1}}/{u_\tau }$. Figure~\ref{fig:12} presents the predicted profiles at friction Reynolds numbers ${Re}_\tau=550$, $5200$, and $10000$, which are not included in the training set. The STBNN predictions agree closely with the high-fidelity DNS across the wall-normal domain, capturing both the near-wall and outer-layer behaviour in cases not used for training. In contrast, LEVM and QEVM exhibit noticeable deviations in the logarithmic region, while the TBNN significantly overpredicts the velocity in the outer layer, leading to an incorrect log-layer slope. The close agreement between STBNN and DNS across the tested Reynolds numbers indicates improved predictive transferability in mean-flow prediction within the present channel-flow cases.

\subsection {Periodic hill flows with varying geometries}
\begin{figure}\centering
	\includegraphics[width=0.9\textwidth]{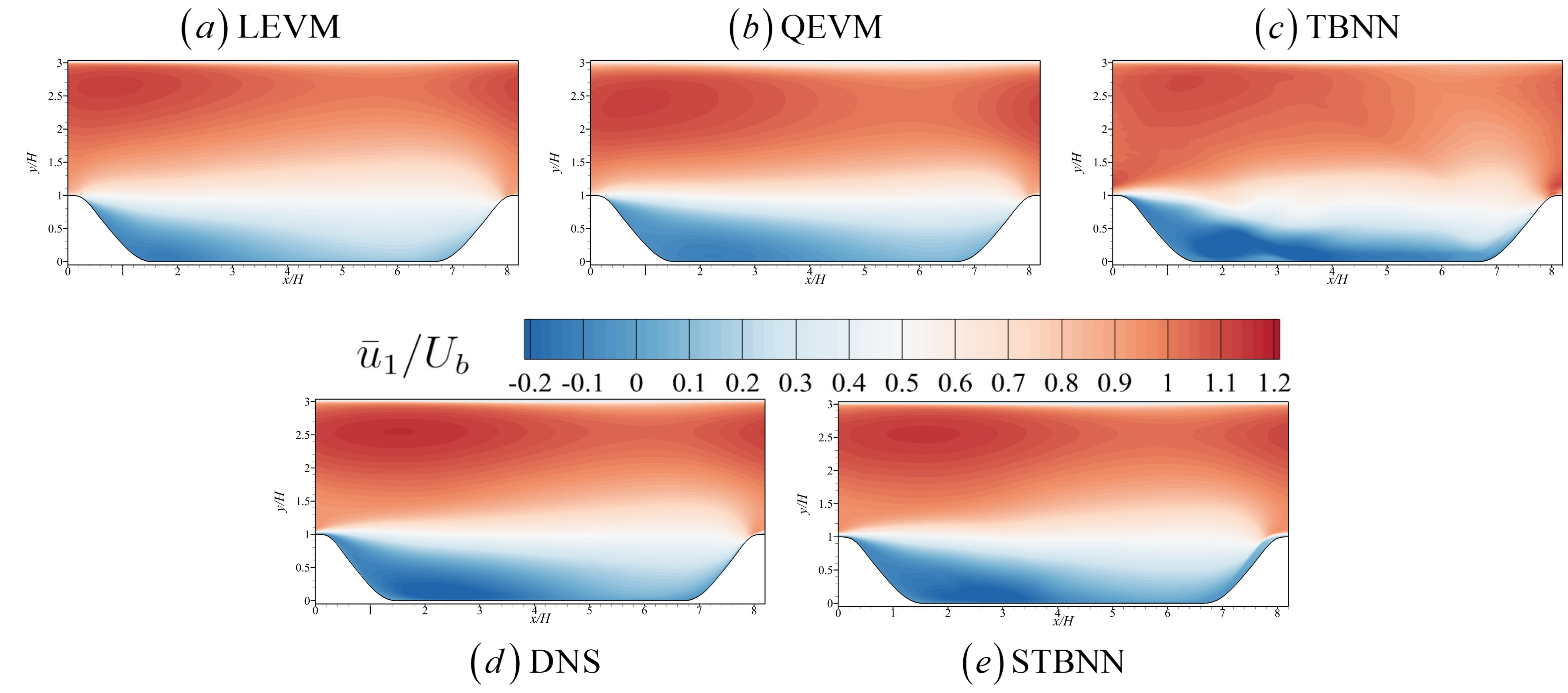}
	\caption{Contours of streamwise velocity at untrained hill steepness $\alpha=0.8$ predicted by different turbulence models:(a) LEVM; (b) QEVM; (c) TBNN; (d) DNS; (e) STBNN.}\label{fig:13}
\end{figure}

\begin{figure}\centering
	\includegraphics[width=0.9\textwidth]{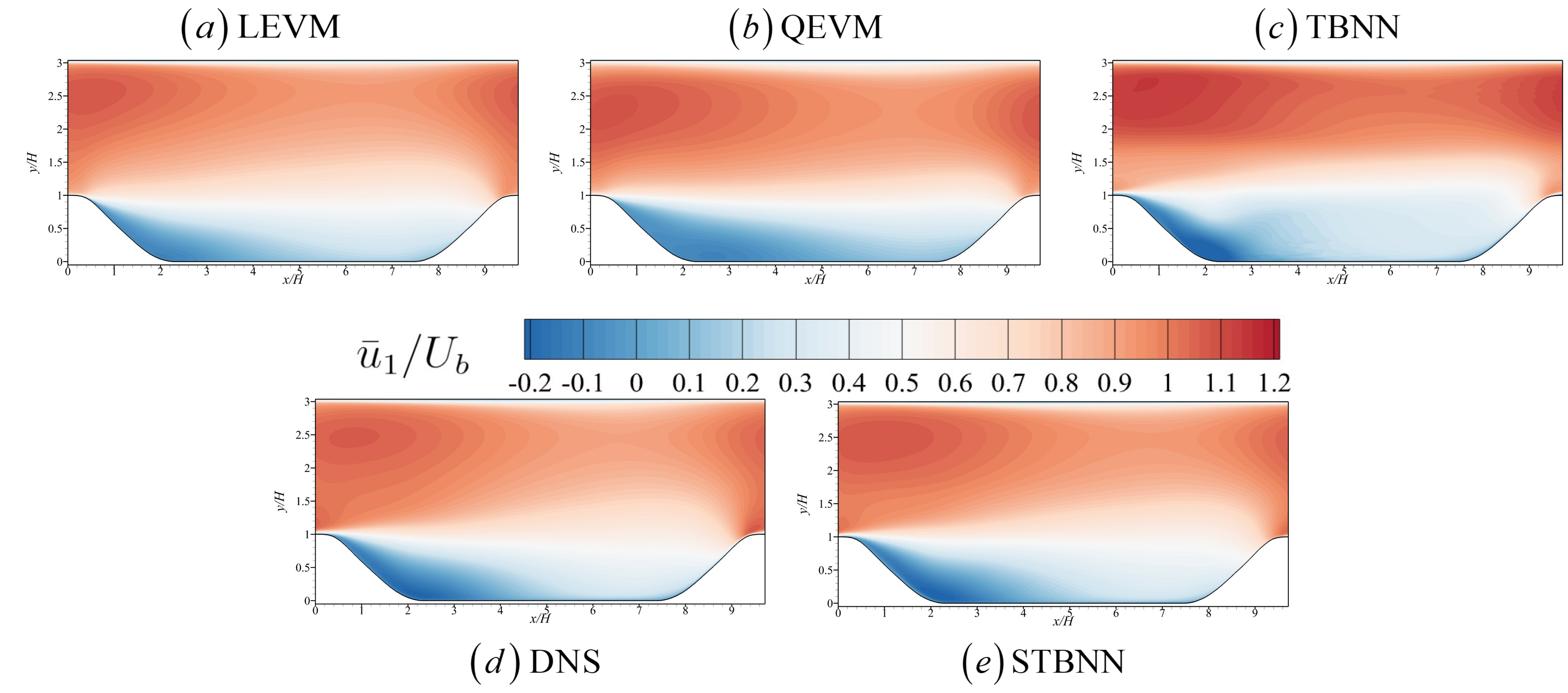}
	\caption{Contours of streamwise velocity at untrained hill steepness $\alpha=1.2$ predicted by different turbulence models:(a) LEVM; (b) QEVM; (c) TBNN; (d) DNS; (e) STBNN.}\label{fig:14}
\end{figure}

The \emph{a posteriori} performance of the models is further examined in periodic hill flows by comparing the mean velocity distributions normalized by the bulk velocity ($U_b$). 
Figs.~\ref{fig:13} and~\ref{fig:14} display the contours of the streamwise velocity for the untrained hill geometries $\alpha=0.8$ and $1.2$, respectively. The LEVM predicts a separation region that is noticeably smaller than in the DNS for both geometries. The QEVM improves the overall flow pattern, but discrepancies persist in the near-wall region. The TBNN predictions exhibit nonphysical fluctuations, particularly near the reattachment region, suggesting limited robustness for untrained geometries. By contrast, the STBNN predicts agree closely with the DNS, accurately reproducing both the separation extent and the downstream recovery.

\begin{figure}
	\centering

\begin{subfigure}{0.45\linewidth}
	\includegraphics[width=\linewidth]{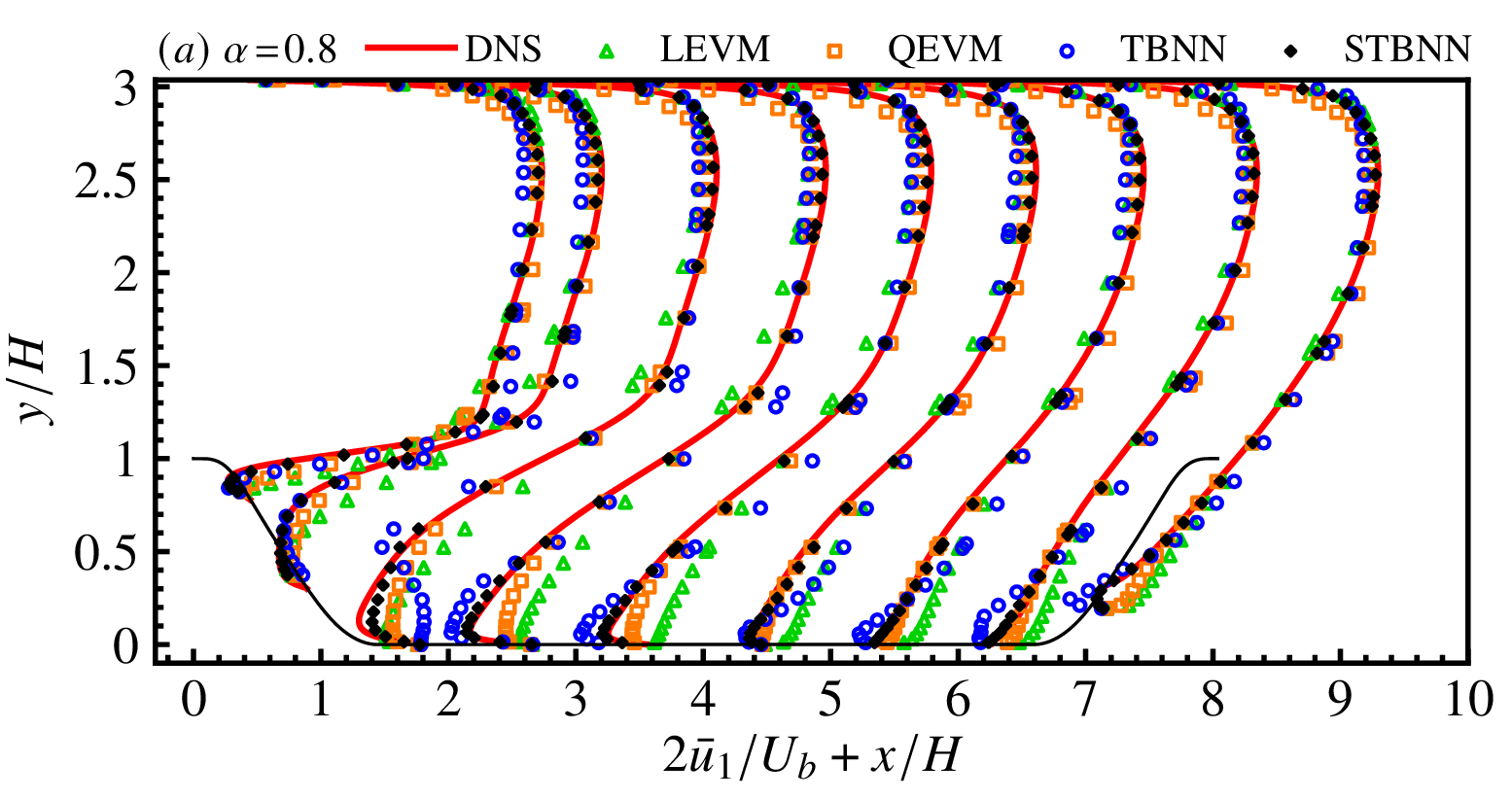}
\end{subfigure}
\begin{subfigure}{0.45\linewidth}
	\includegraphics[width=\linewidth]{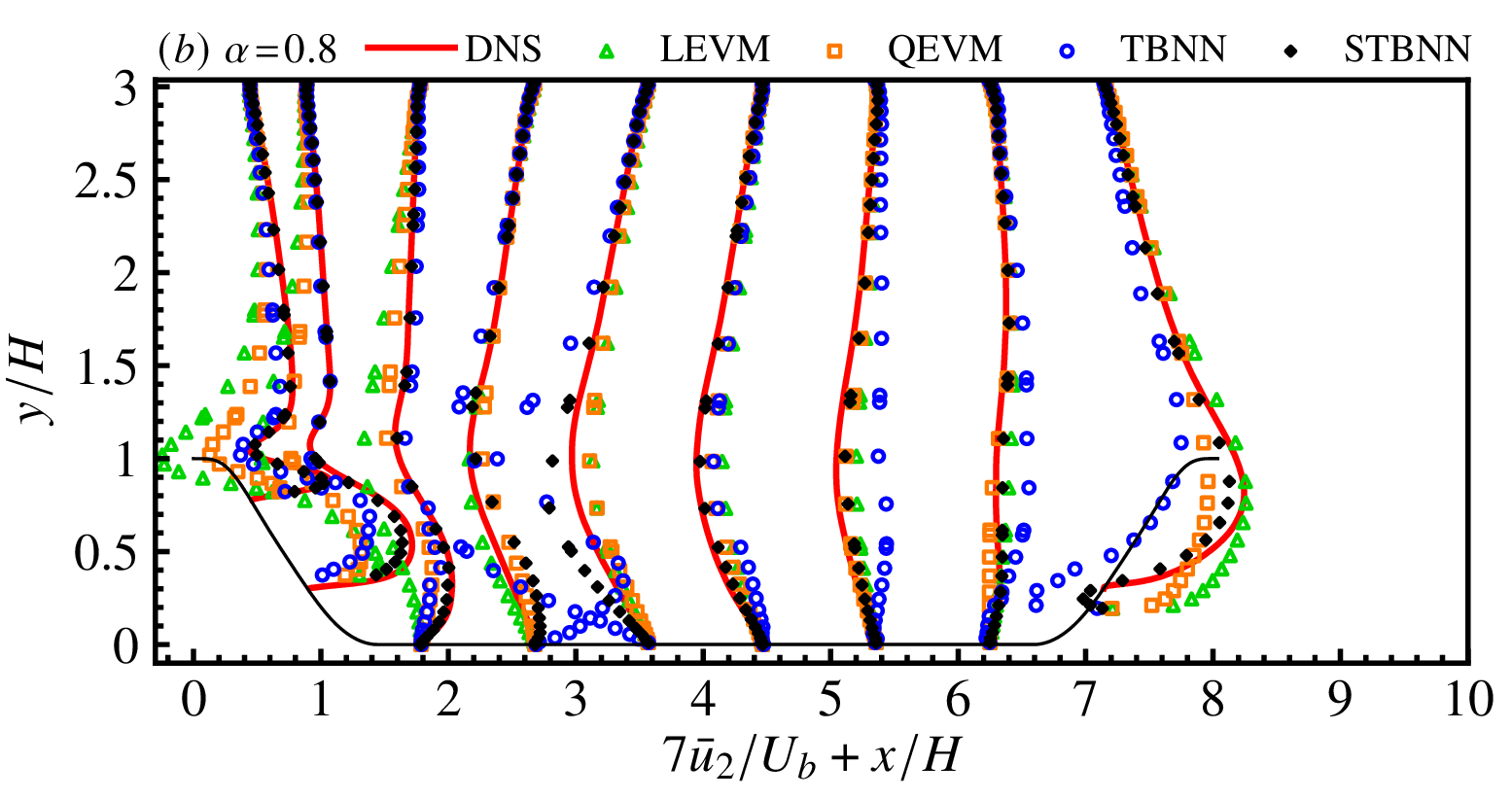}
\end{subfigure}

\begin{subfigure}{0.45\linewidth}
	\includegraphics[width=\linewidth]{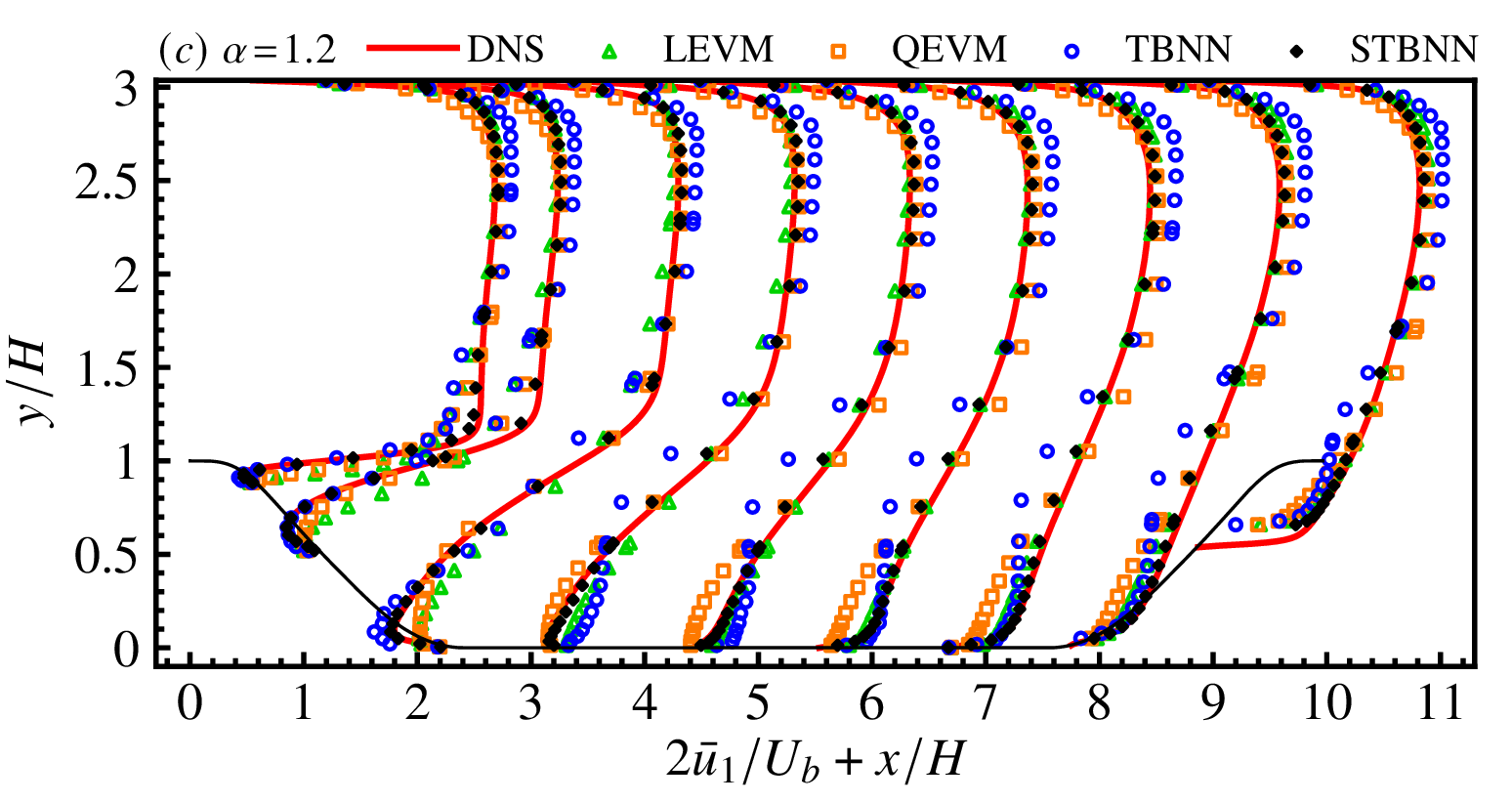}
\end{subfigure}
\begin{subfigure}{0.45\linewidth}
	\includegraphics[width=\linewidth]{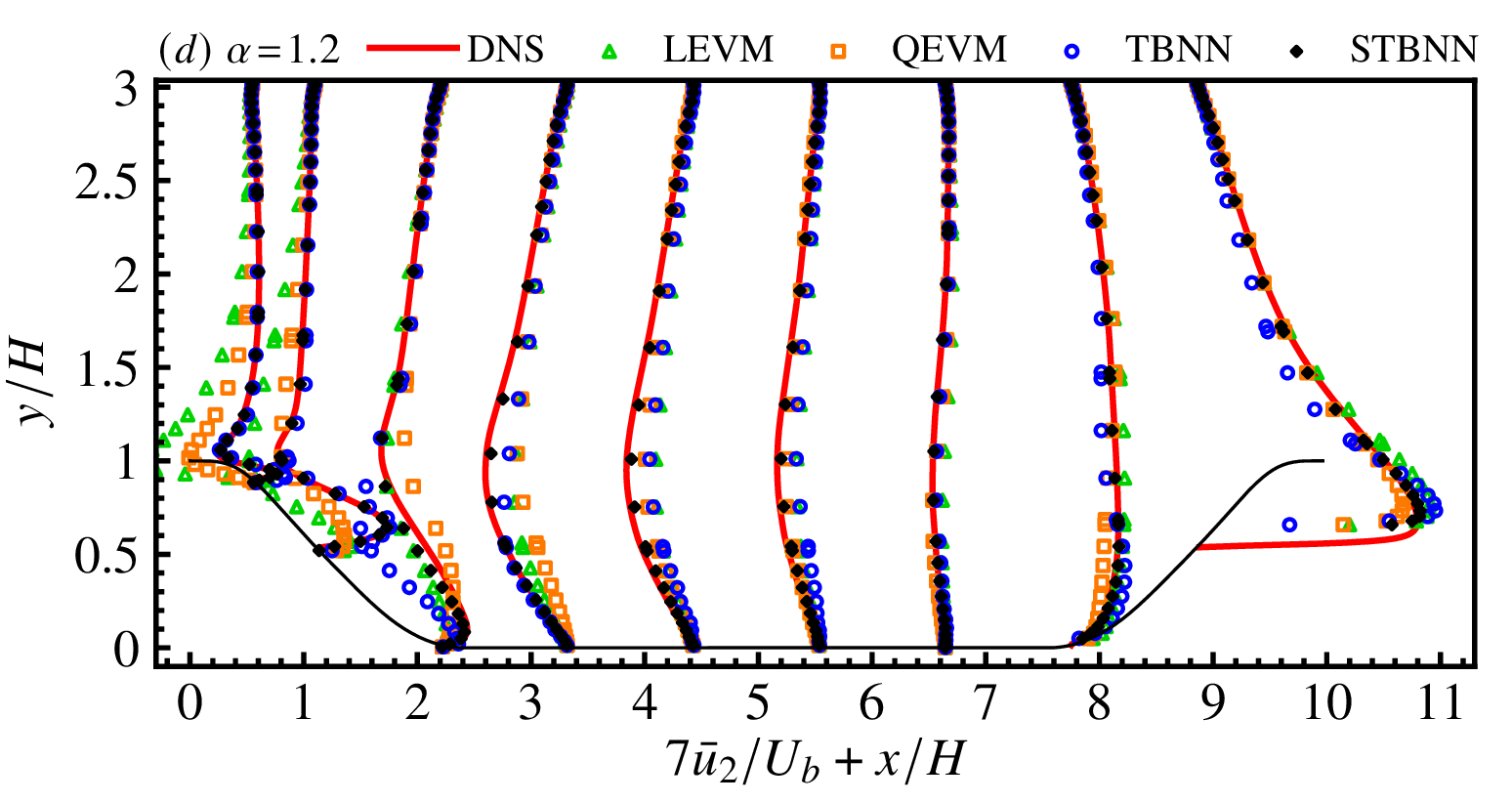}
\end{subfigure}

\caption{
	Velocity profiles calculated by different turbulence models in periodic hill flows with varying geometries.
	The rows correspond to $\alpha=0.8$ and 1.2 (top to bottom), while the columns show
	$\bar u_1$ and $\bar u_2$ (left to right).
}
	\label{fig:15}
\end{figure}

The streamwise and wall-normal velocity profiles at $x/L_x=s/L_{x0}$ with $s\in\{0.5,1,2,3,4,\\5,6,7,8\}$ based on the reference geometry
$L_{x0}=9$ are further illustrated in Fig.~\ref{fig:15}. The baseline closures (LEVM, QEVM and TBNN) show appreciable deviations in the separation bubble and reattachment region, while the STBNN profiles remain in closer agreement with the DNS across most streamwise locations. Overall, the STBNN captures the separated-flow behaviour more reliably than the baseline TBNN model for the untrained periodic-hill cases examined in this study.

\section{Conclusions}\label{sec:level6}

To summarize, a self-scaling tensor-basis neural network (STBNN) framework is proposed for Reynolds-stress modelling of wall-bounded turbulent flows within the RANS paradigm. By introducing an invariant-based normalization constructed from the first two invariants of the velocity-gradient tensor into the tensor-basis formulation, the STBNN provides a scale-consistent representation across the near-wall, logarithmic and outer regions of wall-bounded turbulence, while preserving Galilean and rotational invariance and accounting for the relative contributions of strain and rotation within a tensor-basis representation of the Reynolds-stress anisotropy. 

Extensive \emph{a priori} assessments based on high-fidelity direct numerical simulation (DNS) data of plane channel flow and flow over periodic hills show that the STBNN predicts the Reynolds-stress anisotropy with higher accuracy than the linear eddy-viscosity model (LEVM), the quadratic eddy-viscosity model (QEVM), and the baseline tensor-basis neural network (TBNN). The predictive accuracy remains nearly unchanged with variations in Reynolds number and flow geometry. For channel flows at Reynolds numbers not included in the training set, the correlation coefficients remain above 99\% with relative errors below 1\%. For periodic hill flows with unseen hill steepness, the correlations exceed 98\% and the relative errors remain below 20\%.  By contrast, the baseline TBNN exhibits a noticeable deterioration in accuracy for unseen flow conditions, with significantly increased errors on the validation cases. These results indicate that the STBNN model captures the underlying stress scaling rather than fitting individual flow conditions, leading to more consistent predictive behaviour across the canonical cases examined in this work.

The \emph{a posteriori} RANS calculations provide an integrated assessment of the overall performance of the turbulence closures. For channel flows at $Re_\tau=550,\;5200,$ and $10000$, which are not included in the training dataset, the STBNN remains in close agreement with the DNS velocity profiles, whereas LEVM and QEVM show noticeable deviations and the baseline TBNN exhibits larger outer-layer errors.
In periodic hill flows at $\alpha=0.8$ and $1.2$, also outside the training conditions, the STBNN reproduces the separation bubble and reattachment regions consistently with the DNS, while LEVM underestimates the separation region, QEVM shows near-wall discrepancies, and TBNN displays nonphysical oscillations near reattachment. These results indicate that improved anisotropy modelling leads to more reliable predictions of separated-flow behaviour for the validation periodic-hill geometries examined in the present study.

Overall, the results indicate that the self-scaling formulation alleviates the Reynolds-number sensitivity and near-wall deficiencies present in conventional eddy-viscosity closures and standard tensor-basis neural networks. From a practical perspective, for the present channel and periodic hill flows, the STBNN provides a consistent RANS closure without additional flow-dependent tuning, indicating potential applicability to industrial RANS simulations while maintaining predictive robustness across Reynolds numbers and flow configurations. Further work should consider enforcing additional physical constraints such as realizability and asymptotic near-wall behaviour, extending the framework to more complex three-dimensional separated flows, and developing uncertainty-aware formulations to quantify the reliability of data-driven Reynolds-stress closures. 

\appendix
\section{Effect of activation functions}
\label{sec:appendixA}

In this appendix, we examine the influence of different activation functions on the predictive performance of the proposed STBNN model. Apart from the Gaussian Error Linear Unit (GELU), several commonly used activation functions are considered, including the Rectified Linear Unit (ReLU), Leaky ReLU, and Swish functions.

The activation functions used in this study are defined by
	\begin{equation}
		\mathrm{ReLU}(x) = \max(0, x),
	\end{equation}
	\begin{equation}
		\mathrm{LeakyReLU}(x) =
		\begin{cases}
			x, & x \ge 0, \\
			\alpha x, & x < 0,
		\end{cases}
	\end{equation}
	\begin{equation}
		\mathrm{Swish}(x) = \frac{x}{1 + e^{-x}},
	\end{equation}
	\begin{equation}
		\mathrm{GELU}(x) = x \Phi(x),
	\end{equation}
	where $\alpha=0.01$ is the negative slope coefficient. $\Phi(\cdot)$ denotes the cumulative distribution function of the standard Gaussian distribution.

	The predictive performance of the STBNN framework using different activation functions is summarized in Tables~\ref{tab:appendixA1}--\ref{tab:appendixA4}, including both plane channel and periodic hill flows.

	For the plane channel flow(Tables~\ref{tab:appendixA1} and \ref{tab:appendixA2}), all activation functions yield correlation coefficients that are very close to unity, indicating that the learned mapping is captured with high accuracy in all cases. The relative errors remain at the same order of magnitude across different activation functions, with only minor variations observed.

	For the periodic hill flow(Tables~\ref{tab:appendixA3} and \ref{tab:appendixA4}), a similar trend is obtained. All activation functions provide comparable levels of accuracy, while GELU and Swish show slightly lower relative errors and slightly higher correlation coefficients in most cases. However, the overall differences remain small across all tested activation functions.

	Overall, these results indicate that the predictive performance of the present model is not strongly sensitive to the choice of activation function. The activation primarily influences the optimization process, such as training stability and convergence behavior, while the final prediction accuracy remains comparable across different choices.

\begin{table}[th!]
	\centering
	\caption{Correlation coefficients of the deviatoric Reynolds stress predicted by the STBNN model using different activation functions for plane channel flow.
	}
	\label{tab:appendixA1}
	
		\begin{tabular}{c l c c c c}
			\hline\hline
			${\rm{Case \slash C}}\left( {R_{ij}^d} \right)$  & Activation & $R_{11}^d$ & $R_{22}^d$ & $R_{33}^d$ & $R_{12}^d$ \\
			\hline
			\multirow{4}{*}{%
				\begin{tabular}{c}
					Training set \\[3pt]
					${{Re}}_\tau \in \left\{ \begin{array}{l}
						1000,{\mkern 1mu} 2000,{\mkern 1mu} \\
						4000,{\mkern 1mu} 8000
					\end{array} \right\}$
			\end{tabular}}
			& GELU(present)  & 1.0 & 1.0 & 0.9999 & 1.0 \\
			& ReLU  & 1.0 & 1.0 & 0.9999 & 1.0 \\
			& Leaky ReLU  & 1.0 & 1.0 & 1.0 & 1.0 \\
			& Swish  & 1.0 & 1.0 & 0.9999 & 1.0 \\
			\hline
			\multirow{4}{*}{%
				\begin{tabular}{c}
					Validation set \\[3pt]
					${{Re}}_\tau \in \left\{ \begin{array}{l}
						550,{\mkern 1mu} 5200,{\mkern 1mu} \\
						10000
					\end{array} \right\}$
			\end{tabular}}
			& GELU(present)  & 0.9995 & 0.9999 & 0.996 & 0.9998 \\
			& ReLU  & 0.9995 & 0.9999 & 0.996 & 0.9997 \\
			& Leaky ReLU  & 0.9996 & 0.9999 & 0.9969 & 0.9998 \\
			& Swish  & 0.9995 & 1.0 & 0.9961 & 0.9998 \\
			\hline\hline
		\end{tabular}
	
\end{table}

\begin{table}[th!]
	\centering
	\caption{Relative errors of the deviatoric Reynolds stress predicted by the STBNN model using different activation functions for plane channel flow.}
	\label{tab:appendixA2}
		\begin{tabular}{c l c c c c}
			\hline\hline
			${\rm{Case \slash Er}}\left( {R_{ij}^d} \right)$  & Activation & $R_{11}^d$ & $R_{22}^d$ & $R_{33}^d$ & $R_{12}^d$ \\
			\hline
			\multirow{4}{*}{%
				\begin{tabular}{c}
					Training set \\[3pt]
					${{Re}}_\tau \in \left\{ \begin{array}{l}
						1000,{\mkern 1mu} 2000,{\mkern 1mu} \\
						4000,{\mkern 1mu} 8000
					\end{array} \right\}$
			\end{tabular}}
			& GELU(present)  & 0.0017 & 0.0013 & 0.0052 & 0.001 \\
			& ReLU  & 0.0021 & 0.0022 & 0.0061 & 0.0018 \\
			& Leaky ReLU  & 0.0017 & 0.0019 & 0.0047 & 0.0012 \\
			& Swish  & 0.0026 & 0.0018 & 0.0065 & 0.001 \\
			\hline
			\multirow{4}{*}{%
				\begin{tabular}{c}
					Validation set \\[3pt]
					${{Re}}_\tau \in \left\{ \begin{array}{l}
						550,{\mkern 1mu} 5200,{\mkern 1mu} \\
						10000
					\end{array} \right\}$
			\end{tabular}}
			& GELU(present)  & 0.0251 & 0.0103 & 0.0674 & 0.0103 \\
			& ReLU  & 0.0254 & 0.01 & 0.0679 & 0.0131 \\
			& Leaky ReLU  & 0.0216 & 0.0084 & 0.0617 & 0.0118 \\
			& Swish  & 0.0246 & 0.0087 & 0.0671 & 0.0108 \\
			\hline\hline
		\end{tabular}
\end{table}

\begin{table}[th!]
	\centering
	\caption{Correlation coefficients of the deviatoric Reynolds stress predicted by the STBNN model using different activation functions for  periodic hill flows.}
	\label{tab:appendixA3}
	
		\begin{tabular}{c l c c c c}
			\hline\hline
			${\rm{Case \slash C}}\left( {R_{ij}^d} \right)$  & Activation & $R_{11}^d$ & $R_{22}^d$ & $R_{33}^d$ & $R_{12}^d$ \\
			\hline
			\multirow{4}{*}{%
				\begin{tabular}{c}
					Training set \\[3pt]
					$\alpha \in \left\{ \begin{array}{l}
						0.5,{\mkern 1mu} 1.0,{\mkern 1mu} \\
						1.5
					\end{array} \right\}$
			\end{tabular}}
			& GELU(present)        & 0.9927 & 0.9904 & 0.9931 & 0.9917 \\
			& ReLU        & 0.9899 & 0.9873 & 0.9889 & 0.989 \\
			& LeakyReLU  & 0.9877 & 0.9842 & 0.9861 & 0.9878 \\
			& Swish       & 0.9922 & 0.9902 & 0.9919 & 0.9916 \\
			\hline
			\multirow{4}{*}{%
				\begin{tabular}{c}
					Validation set \\[3pt]
					$\alpha \in \left\{ \begin{array}{l}
						0.8,{\mkern 1mu} 1.2 \\
					\end{array} \right\}$
			\end{tabular}}
			& GELU(present)        & 0.9835 & 0.984 & 0.9815 & 0.991 \\
			& ReLU        & 0.9827 & 0.9821 & 0.9782 & 0.9885 \\
			& LeakyReLU  & 0.9819 & 0.9795 & 0.9759 & 0.9874 \\
			& Swish       & 0.9851 & 0.9854 & 0.9786 & 0.9914 \\
			\hline\hline
		\end{tabular}
\end{table}

\begin{table}[th!]
	\centering
	\caption{Relative errors of the deviatoric Reynolds stress predicted by the STBNN model using different activation functions for periodic hill flows.}
	\label{tab:appendixA4}
		\begin{tabular}{c l c c c c}
			\hline\hline
			${\rm{Case \slash Er}}\left( {R_{ij}^d} \right)$  & Activation & $R_{11}^d$ & $R_{22}^d$ & $R_{33}^d$ & $R_{12}^d$ \\
			\hline
			\multirow{4}{*}{%
				\begin{tabular}{c}
					Training set \\[3pt]
					$\alpha \in \left\{ \begin{array}{l}
						0.5,{\mkern 1mu} 1.0,{\mkern 1mu} \\
						1.5
					\end{array} \right\}$
			\end{tabular}}
			& GELU(present)        & 0.084 & 0.0898 & 0.1139 & 0.101 \\
			& ReLU        & 0.0983 & 0.1031 & 0.1439 & 0.1161 \\
			& LeakyReLU  & 0.1089 & 0.1156 & 0.161 & 0.1233 \\
			& Swish       & 0.0866 & 0.0908 & 0.1229 & 0.1017 \\
			\hline
			\multirow{4}{*}{%
				\begin{tabular}{c}
					Validation set \\[3pt]
					$\alpha \in \left\{ \begin{array}{l}
						0.8,{\mkern 1mu} 1.2 \\
					\end{array} \right\}$
			\end{tabular}}
			& GELU(present)        & 0.128 & 0.1122 & 0.1926 & 0.1047 \\
			& ReLU        & 0.1284 & 0.1206 & 0.2078 & 0.1175 \\
			& LeakyReLU  & 0.1322 & 0.1296 & 0.2185 & 0.1237 \\
			& Swish       & 0.119 & 0.1091 & 0.2084 & 0.1012 \\
			\hline\hline
		\end{tabular}
\end{table}

\section{Effect of network architecture within the STBNN framework}
\label{sec:appendixB}

To further examine the influence of the neural network architecture on the predictive performance, additional comparisons are conducted by replacing the original ANN mapping in the STBNN framework with alternative architectures.

In the present work, the neural network serves as a nonlinear mapping between invariant inputs and tensor-basis coefficients. The baseline ANN adopts a standard fully connected structure, in which nonlinear mappings are approximated through successive affine transformations and activation functions. The forward propagation can be written as
	\begin{equation}
		X_i^{(l)} = \sigma \left( \sum_j W_{ij}^{(l)} X_j^{(l-1)} + b_i^{(l)} \right),
		\label{eq:appendixB_mlp}
	\end{equation}
	where $W_{ij}^{(l)}$, $b_i^{(l)}$ and $\sigma$ denote the weight matrix, bias vector and activation function at layer $l$, respectively. The GELU activation function is adopted in this work, defined as\cite{hendrycks2016gaussian}
	\begin{equation}
		\sigma(x) = x \Phi(x),
		\label{eq:appendixB_gelu}
	\end{equation}
	where $\Phi(\cdot)$ is the cumulative distribution function of the standard Gaussian distribution.

As a representative extension, a residual neural network (ResNet)\cite{he2016deep} is considered, where skip connections are introduced between layers to facilitate the training of deeper networks. Instead of directly learning the mapping, each layer approximates a residual function,
	\begin{equation}
		X^{(l)} = X^{(l-1)} + \sigma \left( \sum_j W_{ij}^{(l)} X_j^{(l-1)} + b_i^{(l)} \right).
		\label{eq:appendixB_resnet}
	\end{equation}

	In addition, a Kolmogorov--Arnold network (KAN)\cite{kalia2025kolmogorov,guo2025physics} is employed, which replaces conventional node-based activations with learnable univariate functional mappings along network edges. The forward KAN mapping can be expressed as
	\begin{equation}
		X_i^{(l)} = \sum_j \sum_{k=0}^{K} a_{ij,k}^{(l)} \, \mathcal{T}_k\!\left( X_j^{(l-1)} \right),
		\label{eq:appendixB_kan}
	\end{equation}
	where $\mathcal{T}_k(\cdot)$ denotes the $k$-th order Chebyshev polynomial, and $a_{ij,k}^{(l)}$ are the corresponding learnable coefficients. In the present work, the expansion is truncated at $K=10$, which is found sufficient for the present mapping accuracy.\cite{guo2025physics}

	For all architectures, the same invariant inputs and tensor-basis coefficient outputs are used, such that the comparison is carried out within a consistent physical formulation. In addition, the depth of neural network  (five hidden layers), number of neurons per layer (twenty), and activation function (GELU) are kept identical across all models to isolate the effect of the mapping architecture.

	Tables~\ref{tab:appendixB1} and \ref{tab:appendixB2} present the correlation coefficients and relative errors of the predicted deviatoric Reynolds stress for plane channel flow. All architectures exhibit nearly identical performance in the training set, indicating that each model is capable of accurately fitting the mapping within the sampled flow conditions. In the validation set, only marginal differences are observed among the three architectures. The KAN variant shows slightly lower errors in certain components, while the ResNet yields comparable correlation levels. Overall, the differences remain small compared to the magnitude of the quantities.

	Similar behaviour is observed for periodic hill flows, as shown in Tables~\ref{tab:appendixB3} and \ref{tab:appendixB4}. All architectures (ANN, ResNet and KAN) yield comparable prediction accuracy across both training and validation cases, with only minor variations among different Reynolds stress components. In particular, no consistent improvement associated with a specific network architecture is observed. In particular, the prediction accuracy across unseen hill geometries remains comparable for all models.

	These results indicate that, under the same invariant-based tensor representation, the choice of neural network architecture has a limited influence on the overall prediction accuracy. Instead, the performance is primarily governed by the physically consistent formulation, in which the invariant inputs and tensor-basis representation constrain the learning process. The robustness observed across different architectures further supports that the proposed STBNN framework is not sensitive to the specific choice of the underlying mapping neural network.
	
\begin{table}[th!]
	\centering
	\caption{Correlation coefficients of the deviatoric Reynolds stress predicted by STBNN with different mapping architectures for plane channel flow.}
	\label{tab:appendixB1}
	
		\begin{tabular}{c l c c c c}
			\hline\hline
			${\rm{Case \slash C}}\left( {R_{ij}^d} \right)$  & Architecture  & $R_{11}^d$ & $R_{22}^d$ & $R_{33}^d$ & $R_{12}^d$ \\
			\hline
			\multirow{3}{*}{%
				\begin{tabular}{c}
					Training set \\[3pt]
					${{Re}}_\tau \in \left\{ \begin{array}{l}
						1000,{\mkern 1mu} 2000,{\mkern 1mu} \\
						4000,{\mkern 1mu} 8000
					\end{array} \right\}$
			\end{tabular}}
			& ANN (present)  & 1.0    & 1.0    & 0.9999 & 1.0    \\
			& ResNet         & 1.0    & 1.0    & 0.9999 & 1.0    \\
			& KAN            & 1.0    & 1.0    & 1.0    & 1.0    \\
			\hline
			\multirow{3}{*}{%
				\begin{tabular}{c}
					Validation set \\[3pt]
					${{Re}}_\tau \in \left\{ \begin{array}{l}
						550,{\mkern 1mu} 5200,{\mkern 1mu} \\
						10000
					\end{array} \right\}$
			\end{tabular}}
			& ANN (present)  & 0.9995 & 0.9999 & 0.996  & 0.9998 \\
			& ResNet         & 0.9995 & 0.9999 & 0.996  & 0.9998 \\
			& KAN            & 0.9995 & 0.9999 & 0.9965 & 0.9998 \\
			\hline\hline
		\end{tabular}
\end{table}

\begin{table}[th!]
	\centering
	\caption{Relative errors of the deviatoric Reynolds stress predicted by STBNN with different mapping architectures for plane channel flow.}
	\label{tab:appendixB2}
	
		\begin{tabular}{c l c c c c}
			\hline\hline
			${\rm{Case \slash Er}}\left( {R_{ij}^d} \right)$  & Architecture  & $R_{11}^d$ & $R_{22}^d$ & $R_{33}^d$ & $R_{12}^d$ \\
			\hline
			\multirow{3}{*}{%
				\begin{tabular}{c}
					Training set \\[3pt]
					${{Re}}_\tau \in \left\{ \begin{array}{l}
						1000,{\mkern 1mu} 2000,{\mkern 1mu} \\
						4000,{\mkern 1mu} 8000
					\end{array} \right\}$
			\end{tabular}}
			& ANN (present)  & 0.0017 & 0.0013 & 0.0052 & 0.001  \\
			& ResNet         & 0.0029 & 0.0023 & 0.0075 & 0.0013 \\
			& KAN            & 0.0014 & 0.0014 & 0.0035 & 0.0008 \\
			\hline
			\multirow{3}{*}{%
				\begin{tabular}{c}
					Validation set \\[3pt]
					${{Re}}_\tau \in \left\{ \begin{array}{l}
						550,{\mkern 1mu} 5200,{\mkern 1mu} \\
						10000
					\end{array} \right\}$
			\end{tabular}}
			& ANN (present)  & 0.0251 & 0.0103 & 0.0674 & 0.0103 \\
			& ResNet         & 0.0242 & 0.0084 & 0.067  & 0.0108 \\
			& KAN            & 0.0247 & 0.0089 & 0.0638 & 0.0094 \\
			\hline\hline
		\end{tabular}
\end{table}			

\begin{table}[th!]
	\centering
	\caption{Correlation coefficients of the deviatoric Reynolds stress predicted by STBNN with different mapping architectures for periodic hill flows.}
	\label{tab:appendixB3}
	
		\begin{tabular}{c l c c c c}
			\hline\hline
			${\rm{Case \slash C}}\left( {R_{ij}^d} \right)$  & Architecture & $R_{11}^d$ & $R_{22}^d$ & $R_{33}^d$ & $R_{12}^d$ \\
			\hline
			\multirow{3}{*}{%
				\begin{tabular}{c}
					Training set \\[3pt]
					$\alpha \in \left\{ \begin{array}{l}
						0.5,{\mkern 1mu} 1.0,{\mkern 1mu} \\
						1.5
					\end{array} \right\}$
			\end{tabular}}
			& ANN (present)  & 0.9927 & 0.9904 & 0.9931 & 0.9917 \\
			& ResNet         & 0.9922 & 0.9909 & 0.9907 & 0.9933 \\
			& KAN            & 0.9895 & 0.9842 & 0.9894 & 0.9984 \\
			\hline
			\multirow{3}{*}{%
				\begin{tabular}{c}
					Validation set \\[3pt]
					$\alpha \in \left\{ \begin{array}{l}
						0.8,{\mkern 1mu} 1.2
					\end{array} \right\}$
			\end{tabular}}
			& ANN (present)  & 0.9835 & 0.984  & 0.9815 & 0.991  \\
			& ResNet         & 0.9868 & 0.9886 & 0.9823 & 0.9936 \\
			& KAN            & 0.9822 & 0.9796 & 0.9793 & 0.9956 \\
			\hline\hline
		\end{tabular}
\end{table}

\begin{table}[th!]
	\centering
	\caption{Relative errors of the deviatoric Reynolds stress predicted by STBNN with different mapping architectures for periodic hill flows.}
	\label{tab:appendixB4}
	
		\begin{tabular}{c l c c c c}
			\hline\hline
			${\rm{Case \slash Er}}\left( {R_{ij}^d} \right)$  & Architecture & $R_{11}^d$ & $R_{22}^d$ & $R_{33}^d$ & $R_{12}^d$ \\
			\hline
			\multirow{3}{*}{%
				\begin{tabular}{c}
					Training set \\[3pt]
					$\alpha \in \left\{ \begin{array}{l}
						0.5,{\mkern 1mu} 1.0,{\mkern 1mu} \\
						1.5
					\end{array} \right\}$
			\end{tabular}}
			& ANN (present)  & 0.084  & 0.0898 & 0.1139 & 0.101  \\
			& ResNet         & 0.0856 & 0.0867 & 0.1315 & 0.0904 \\
			& KAN            & 0.1006 & 0.1162 & 0.1404 & 0.0434 \\
			\hline
			\multirow{3}{*}{%
				\begin{tabular}{c}
					Validation set \\[3pt]
					$\alpha \in \left\{ \begin{array}{l}
						0.8,{\mkern 1mu} 1.2
					\end{array} \right\}$
			\end{tabular}}
			& ANN (present)  & 0.128  & 0.1122 & 0.1926 & 0.1047 \\
			& ResNet         & 0.113  & 0.0951 & 0.1897 & 0.088  \\
			& KAN            & 0.1338 & 0.1281 & 0.2041 & 0.0723 \\
			\hline\hline
		\end{tabular}
\end{table}

\section{Grid-convergence study of \emph{a posteriori} RANS calculation}
\label{sec:appendixC}

A grid-convergence study was carried out for two representative \emph{a posteriori} RANS cases: the plane-channel flow at ${\rm Re}_{\tau}=10000$ and the periodic-hill flow at $\alpha=1.5$. These two cases were selected since they correspond to the highest Reynolds number in the channel-flow set and the longest streamwise domain in the periodic-hill family, respectively.

\begin{figure}[t]
	\centering
	\begin{subfigure}[t]{0.38\textwidth}
		\centering
		\includegraphics[width=\textwidth]{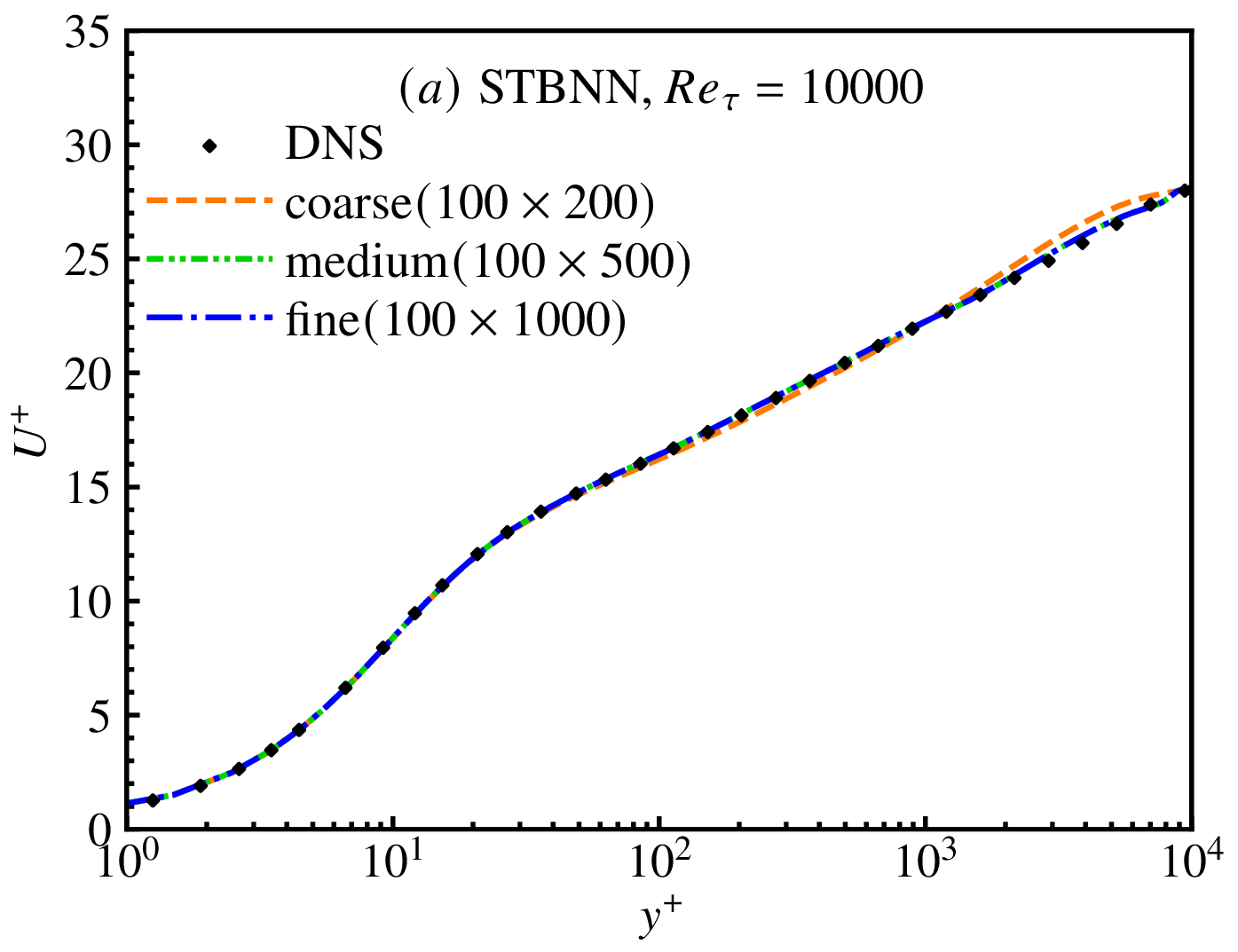}
		\label{fig:C1a}
	\end{subfigure}
	\hfill
	\begin{subfigure}[t]{0.57\textwidth}
		\centering
		\includegraphics[width=\textwidth]{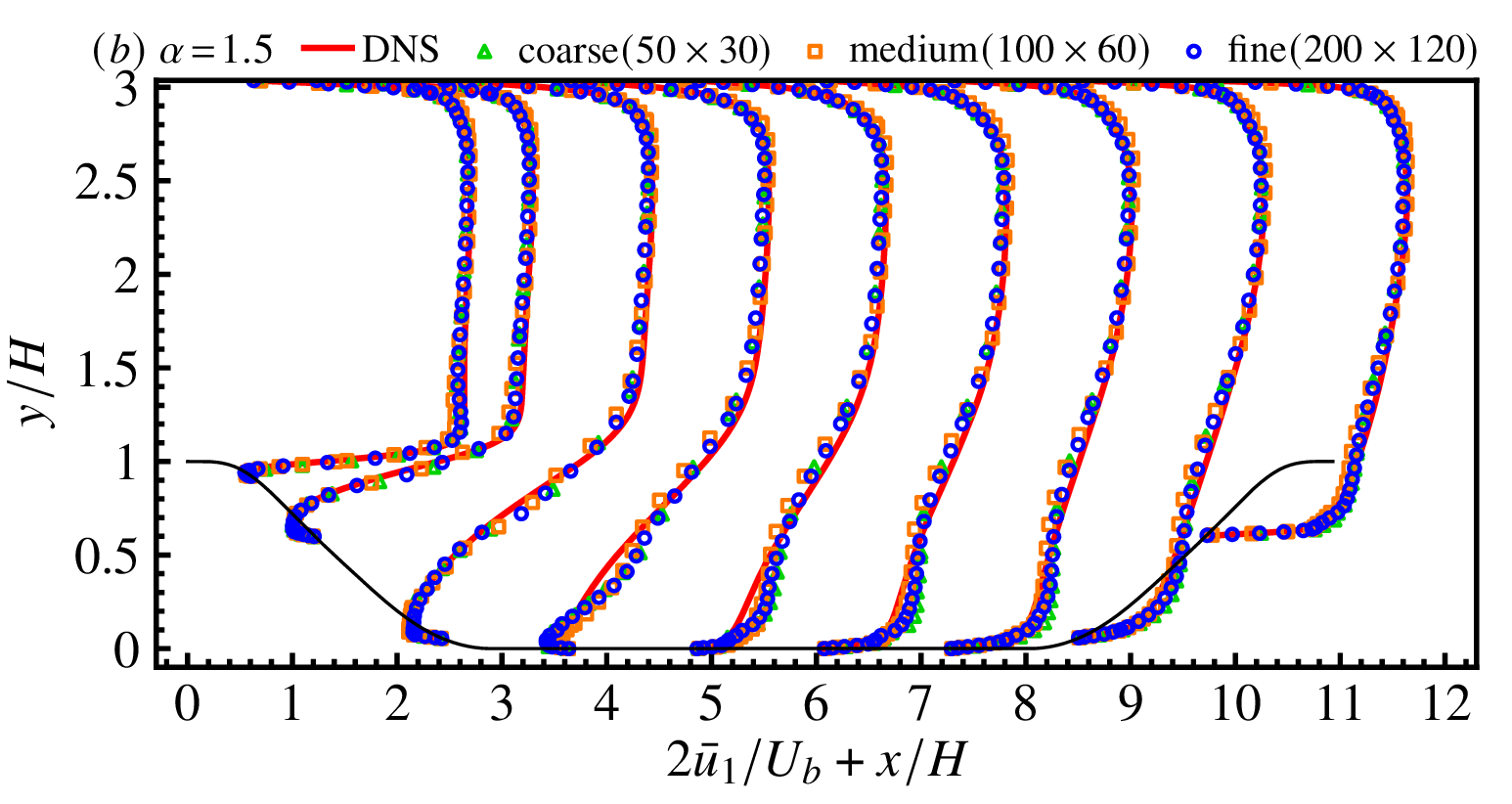}
		\label{fig:C1b}
	\end{subfigure}
	\caption{Mean streamwise velocity profiles for grid-convergence study using STBNN model in \emph{a posteriori} RANS calculations: (a) plane-channel at ${\rm Re}_{\tau}=10000$; (b) periodic-hill flow at steepness $\alpha=1.5$.}
	\label{fig:grid_independence}
\end{figure}

For the plane-channel flow, three meshes, $100\times 200$, $100\times 500$, and $100\times 1000$, were examined. The first-cell height was chosen such that $y^+\approx 1$, ensuring that the viscous sublayer was resolved without wall functions. The corresponding mean streamwise velocity profiles are shown in Fig.~\ref{fig:grid_independence}(a). The difference between the medium and fine meshes is negligible. A small discrepancy is observed only between the coarse mesh and the other two meshes in the outer region.

For the periodic-hill flow at $\alpha=1.5$, three body-fitted meshes, $50\times 30$, $100\times 60$, and $200\times 120$, were considered. Figure~\ref{fig:grid_independence}(b) compares the mean streamwise velocity profiles at several streamwise locations $x/L_x=s/L_{x0}$ with $s\in\{0.5,1,2,3,4,5,6,7,8\}$ based on the reference geometry $L_{x0}=9$. The medium and fine meshes give nearly identical results throughout the domain. The coarse mesh shows slightly larger deviations in the separated-flow region, while the overall profiles remain very close across all three meshes.

These results indicate that further mesh refinement has little effect on the predicted mean velocity for the cases considered here. The mesh resolution used in the present \emph{a posteriori} calculations is therefore sufficient for the assessment in Sec.~\ref{sec:level5}.

\acknowledgments{This research was supported by the Ministry of Education (MOE), Singapore, under the MOE AcRF Tier 2 project POSEIDON: Predicting cOaStal brEakIng waves with advanced Data-driven turbulence mOdelliNg (Grant No. MOE-T2EP50222-0009). }

\section*{Conflict of Interest}

The authors have no conflicts to disclose.

\section*{Author Contributions}

\textbf{Zelong Yuan:} Conceptualization (equal); Data curation (equal); Formal analysis (lead); Investigation (equal); Methodology (equal); Validation (equal); Visualization (equal); Writing--original draft (equal).

\textbf{Yuzhu Pearl Li:} Conceptualization (equal); Data curation (equal); Funding acquisition (lead); Methodology (equal); Project administration (lead); Resources (lead); Software (equal); Supervision (lead); Writing--review \& editing (lead).

\section*{DATA AVAILABILITY}
The data that support the findings of this study are available upon request.

%\nocite{*}
\bibliography{reference}
\end{document}